\documentclass[apj]{emulateapj}
 \usepackage{rotating}

\def\ltsima{$\; \buildrel < \over \sim \;$}
\def\simlt{\lower.5ex\hbox{\ltsima}}
\def\gtsima{$\; \buildrel > \over \sim \;$}
\def\simgt{\lower.5ex\hbox{\gtsima}}
\def\kms{{\rm\,km\,s^{-1}}}

\def\kpc{{\rm\,kpc}}

\def\msun{{\rm\,M_\odot}}
\def\lsun{{\rm\,L_\odot}}

\def\pc{{\rm\,pc}}

\def\deg{^\circ}
\def\s{\ifmmode \widetilde \else \~\fi}
\def\={\overline}

\def\spose#1{\hbox to 0pt{#1\hss}}

\def\lta{\mathrel{\spose{\lower 3pt\hbox{$\mathchar"218$}}
     \raise 2.0pt\hbox{$\mathchar"13C$}}}
\def\gta{\mathrel{\spose{\lower 3pt\hbox{$\mathchar"218$}}
     \raise 2.0pt\hbox{$\mathchar"13E$}}}
\def\Dt{\spose{\raise 1.5ex\hbox{\hskip3pt$\mathchar"201$}}}    % upper case
\def\dt{\spose{\raise 1.0ex\hbox{\hskip2pt$\mathchar"201$}}}    % lower case

\def\dotsfill{\leaders\hbox to 1em{\hss.\hss}\hfill}

\newcommand{\ud}{\mathrm{d}}

\shorttitle{PAndAS properties of 23 M31 dSph}
\shortauthors{N. F. Martin et al.}

\begin{document}


\title{The PAndAS view of the Andromeda satellite system -- II. Detailed properties of 23 M31 dwarf spheroidal galaxies}


\author{Nicolas F. Martin$^{1,2}$, Rodrigo A. Ibata$^1$, Geraint F. Lewis$^3$, Alan McConnachie$^4$, Arif Babul$^5$, Nicholas F. Bate$^6$, Edouard Bernard$^7$, Scott C. Chapman$^8$, Michelle M. L. Collins$^9$, Anthony R. Conn$^3$, Denija Crnojevi\'c$^{10}$, Mark A. Fardal$^{11}$, Annette M. N. Ferguson$^{12}$, Michael Irwin$^6$, A. Dougal Mackey$^{13}$, Brendan McMonigal$^3$, Julio F. Navarro$^5$, R. Michael Rich$^{14}$}
\email{nicolas.martin@astro.unistra.fr}

\altaffiltext{1}{Observatoire astronomique de Strasbourg, Universit\'e de Strasbourg, CNRS, UMR 7550, 11 rue de l'Universit\'e, F-67000 Strasbourg, France}
\altaffiltext{2}{Max-Planck-Institut f\"ur Astronomie, K\"onigstuhl 17, D-69117 Heidelberg, Germany}
\altaffiltext{3}{Sydney Institute for Astronomy, School of Physics A28, The University of Sydney, NSW 2006, Australia}
\altaffiltext{4}{NRC Herzberg Institute of Astrophysics, 5071 West Saanich Road, Victoria, BC, V9E 2E7, Canada}
\altaffiltext{5}{Department of Physics and Astronomy, University of Victoria, Victoria, BC, Canada}
\altaffiltext{6}{Institute of Astronomy, University of Cambridge, Madingley Road, Cambridge CB3 0HA}
\altaffiltext{7}{Laboratoire Lagrange UMR 7293, UniversitŽ de Nice Sophia-Antipolis, CNRS, Observatoire de la C™te d'Azur, CS 34229, 06304, Nice Cedex 04, France}
\altaffiltext{8}{Department of Physics and Atmospheric Science, Dalhousie University, Coburg Road, Halifax, NS B3H 1A6, Canada}
\altaffiltext{9}{Department of Physics, University of Surrey, Guildford, GU2 7XH, Surrey, UK}
\altaffiltext{10}{Physics Department, Texas Tech University, Lubbock, TX 79409, USA}
\altaffiltext{11}{Department of Astronomy, University of Massachusetts, Amherst, MA 01003, USA}
\altaffiltext{12}{Institute for Astronomy, University of Edinburgh, Blackford Hill, Edinburgh EH9 3HJ}
\altaffiltext{13}{Research School of Astronomy and Astrophysics, Australian National University, Canberra, ACT 2611, Australia}
\altaffiltext{14}{Department of Physics and Astronomy, University of California, Los Angeles, PAB, 430 Portola Plaza, Los Angeles, CA 90095-1547, USA}

\begin{abstract}
We present a comprehensive analysis of the structural properties and luminosities of the 23 dwarf spheroidal galaxies that fall within the footprint of the Pan-Andromeda Archaeological Survey (PAndAS). These dwarf galaxies represent the large majority of Andromeda's known satellite dwarf galaxies and cover a wide range in luminosity ($-11.6\lta M_V\lta-5.8$ or $10^{4.2}\lta L \lta10^{6.5}\lsun$) and surface brightness ($25.1\lta\mu_0\lta29.3$ mag/arcsec$^2$). We confirm most previous measurements, but find And~XIX to be significantly larger than before ($r_h=3065^{+1065}_{-935}\pc$, $M_V=-10.1^{+0.8}_{-0.4}$) and cannot derive parameters for And~XXVII as it is likely not a bound stellar system. We also significantly revise downward the luminosities of And~XV and And~XVI, which are now $M_V\sim-7.5$ or $L\sim10^5\lsun$. Finally, we provide the first detailed analysis of Cas~II/And~XXX, a fairly faint system ($M_V=-8.0^{+0.4}_{-0.3}$) of typical size ($r_h=270\pm50\pc$), located in close proximity to the two bright elliptical dwarf galaxies NGC 147 \& 185. Combined with the set of homogeneous distances published in an earlier contribution, our analysis dutifully tracks all relevant sources of uncertainty in the determination of the properties of the dwarf galaxies from the PAndAS photometric catalogue. We further publish the posterior probability distribution functions of all the parameters we fit for in the form of MCMC chains available online; these inputs should be used in any analysis that aims to remain truthful to the data and properly account for covariance between parameters.
\end{abstract}

%Finally, the abstract could be beefed up a bit to include some details; for example, the range of dSph luminosities, surface brightnesses, and radii probed, and how they compare with other Local Group objects in the same luminosity range would be welcome.


\section{Introduction}

The stellar content of dwarf galaxies, their dynamical, kinematic, and chemical properties, their spatial distribution around and their rate of infall onto their host can all be linked to the faint end of galaxy formation in a cosmological model \citep[e.g.][]{mateo98,grebel04,gilmore07,sales07a,tolstoy09,mcconnachie12,weisz14a,pawlowski14b}. However, for a long time, the small number of dwarf galaxies known to inhabit the Local Group, where these dim systems can be studied with the most detail, limited the extent of insight they could provide.

At the turn of the century, panoramic photometric CCD surveys opened up the realm of low surface brightness stellar structures. Not only did these surveys provide the means to discover much dimmer dwarf galaxies (see, e.g., \citealt{mcconnachie12} and \citealt{belokurov13} for reviews, and \citealt{willman12} for some of the questions these new systems raise), but the homogeneity of these surveys provided the means to build samples of dwarf-galaxy properties free of the systematics that usually stem from observing different stellar systems with different instruments and studying them with different analytic techniques. The reader is referred to, e.g., \citet{martin07a}, \citet{simon07}, \citet{dejong08a}, or \citet{weisz14} for both spectroscopic and photometric studies that strive to take such global approaches to infer the properties of Milky Way dwarf galaxies and, therefore, limit the impact of systematics. One such example that is particularly relevant to this paper is the work of \citet[][hereafter M08]{martin08b}, who presented a comprehensive study of the structural and luminosity-related properties of all the Milky Way dwarf galaxies found in the Sloan Digital Sky Survey at the time. This study further proposed to eschew binning and smoothing and rely instead on the information carried by every star present in the photometric catalogue of a dwarf galaxy through the expression of the likelihood of families of structural models.

Around the Andromeda galaxy (M31), new dwarf galaxies were revealed by successive photometric surveys with increasing coverage. Such surveys include a dedicated SDSS stripe along the major axis of M31 \citep[e.g.,][]{zucker04b} and a survey of the inner halo of the galaxy with the INT/WFC \citep{ferguson02,irwin08}. These efforts culminated in a systematic, deep surveying of the region within $150\kpc$ of M31 for the Pan-Andromeda Archaeological Survey (PAndAS; \citealt{mcconnachie09,ibata14a}) that led to the discovery of 16~new dwarf galaxies \citep{martin06b,ibata07, mcconnachie08,martin09,richardson11,martin13b}. Beyond the PAndAS footprint, additional surveys continue to reveal new systems \citep{bell11,slater11,martin13a,martin13c}, yet without reaching the depth achieved by PAndAS. In total, 24 proposed dwarf spheroidal galaxies ($-12\lta M_V\lta-6$) lie within the PAndAS footprint. As such, the homogeneous view of the M31 surroundings provided by this survey brings forth a unique opportunity to build a homogeneous set of properties for a large majority of the known dwarf galaxy satellites of the Andromeda galaxy. \citet{aconn11,aconn12} have initiated this endeavor by determining tip-of-the-red-giant-branch (TRGB) distances for all the Andromeda satellites in the PAndAS footprint. We now extend this effort to provide a complete list of structural parameters and luminosity-related properties for all these dwarf galaxies.

The paper is structured as follows: section~2 presents the PAndAS data used for the analysis of the dwarf galaxies, section~3 explains our statistical framework to infer the properties of the dwarf galaxies, and section~4 describes our results and compares them with those of past studies. The paper is summarized in section~5.

This contribution is the second in a series of PAndAS papers focussing on dwarf galaxies in the PAndAS footprint. Paper~I focussed on an automated search for M31's dwarf galaxies in the PAndAS data \citep{martin13b} while Paper~III derives the dwarf galaxy search completeness limits within the PAndAS footprint (N. F. Martin et al., in preparation).

\section{Data}

The specifics of the design, acquisition, reduction, and calibration of the PAndAS survey have been explained in detail in previous contributions from our collaboration \citep[e.g.,][]{ibata14a} and we shall not revisit these here. It is, however, worth mentioning that this Canada-France-Hawaii Telescope (CFHT) Large Program was built on two previous CFHT surveys led by R. Ibata and A. McConnachie \citep{ibata07,mcconnachie08}, which were further complemented by $\sim230$ hours of observation with the same wide-field imager, MegaCam, in order to cover a contiguous region of $\sim390$ deg$^2$ within $\sim150\kpc$ of M31 and $\sim50\kpc$ of its companion M33. Each one of the 1-deg$^2$ MegaCam fields was observed for at least 45 minutes in both the MegaCam $g$ and $i$ bands, under exquisite conditions (median seeing of $0.67''$ and $0.60''$, respectively, for the two bands). The final ($5\sigma$) depths of the survey are 26.0 and 24.7, respectively, with some variations from field to field \citep{ibata14a}. In the analysis, we only use the catalogue of aperture magnitudes and these are de-reddened following equation (1) of \citet{ibata14a} when needed. We also always use the stellarity flags to keep only star-like objects, as defined in the same publication.

\begin{figure}
\begin{center}
\includegraphics[width=0.85\hsize,angle=270]{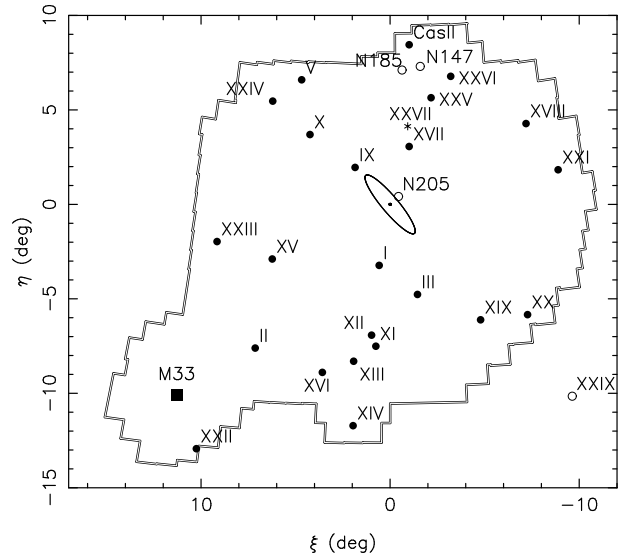}
\caption{\label{PAndAS_map}Distribution of dwarf galaxies around M31, whose disk is represented by the central ellipse. North is to the top and East to the left. M33 is represented by the large square in the bottom-left corner. The polygon represents the PAndAS footprint and all dwarf galaxies studied in this paper are shown as filled circles, except for the peculiar And~XXVII, located by the asterisk. Other dwarf galaxies are shown as hollow circles.}
\end{center}
\end{figure}

The dwarf galaxy sample we focus on here comprises all the dwarf spheroidal galaxies that fall in the PAndAS footprint, as seen in Figure~\ref{PAndAS_map}: And~I, And~II, And~III, And~V, And~IX, And~X, And~XI, And~XII, And~XIII, And~XIV, And~XV, And~XVI, And~XVII, And~XVIII, And~XIX, And~XX, And~XXI, And~XXII, And~XXIII, And~XXIV, And~XXV, And~XXVI, and the latest PAndAS discovery Cas~II/And~XXX. The brightest dwarf galaxies satellite of M31 (M32, NGC~205, NGC~147, and NGC~185) are beyond the scope of this paper as their complexity, the crowding they suffer from, and/or their location in the survey render their study difficult and require dedicated analyses. The reader is referred to \citet{crnojevic14} for a detailed study of NGC~147 and NGC~185. The case of And~XXVII, which was originally included in the current analysis, is discussed in detail in Section~\ref{AndXXVII} as our results lead us to the conclude that it is likely not a bound dwarf galaxy.

\begin{figure*}
\begin{center}
\includegraphics[width=0.8\hsize]{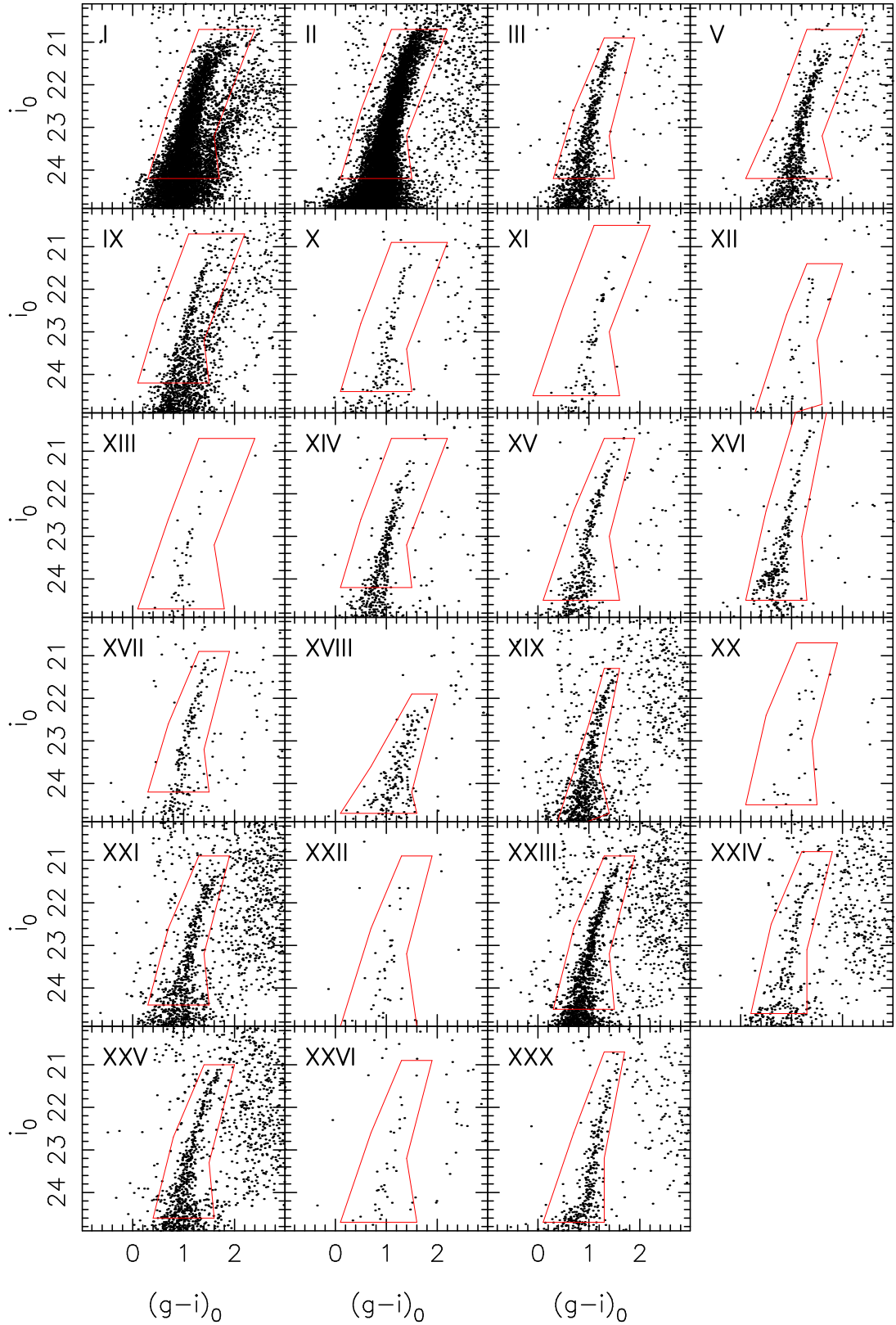}
\caption{\label{CMDs}CMD of all M31 dwarf galaxies studied in this paper for a region within two half-light radii of a dwarf galaxy's centroid except for And XIX, for which the region has a smaller major axis of $12'$. All dwarf galaxies show a more or less populated RGB, depending on the absolute luminosity of the system. The main contamination stems from red foreground dwarf stars that belong to the Milky Way disk or from nearby M31 halo stellar substructure in the case of And~I and And~IX. The polygons shown in red correspond to the CMD selection boxes used to isolate potential member stars on which the algorithms are ran.}
\end{center}
\end{figure*}

Figure~\ref{CMDs} presents the color-magnitude diagrams (CMD) of all the dwarf galaxies in the sample. All the CMDs include stars within two elliptical half-light radii of a galaxy's centroid\footnote{In the case of the very large And~XIX, we limit ourselves to a region within $12'$ of the centroid to avoid clutter.} as inferred through our algorithm described below. The great variety of M31 dwarf galaxies is readily visible in this Figure, with the brightest systems at the top corresponding to discoveries made from photographic plate studies (And~I, And~II, And~III, And~V; \citealt{vandenbergh72,armandroff98}), followed by systems discovered as overdensities of red giant branch (RGB) stars with CCD surveys and photometry: And~IX and And~X in the SDSS \citep{zucker04b,zucker07}, And~XIV in a series of isolated M31 outer halo fields \citep{majewski07}, And~XVII in the INT/WFC survey that preceded PAndAS \citep{irwin08}, and all the other ones from pre-PAndAS or PAndAS photometry \citep{martin06b,ibata07, mcconnachie08,martin09,richardson11}. Some of these systems have very well-defined RGBs (e.g. And~XIV, And~XXI, And~XXIII) while others have RGBs that are barely visible, with only $\sim20$ stars (e.g. And~XII, And~XX, And~XXII, And~XXVI). The varying amount of contamination from foreground MW stars is also evidence for the varying spatial extent of these systems. Finally, these dwarf galaxies are sometimes projected on other M31 halo stellar structures, which explains the complex RGB features present in the CMDs of  And~I and And~IX that both overlap the metal-rich M31 Giant Stream or its associated NE Shelf debris \citep{ibata01c,ferguson02,fardal07}.

Figures~\ref{map_I_II} to \ref{map_XXX}, in Appendix~\ref{maps}, show the spatial distribution of PAndAS stars in the regions used for the inference around each dwarf galaxy, along with the favored $2r_h$ region.

\section{Methods}
\subsection{Structure}
\label{structure}
The probability distribution functions (PDFs) of the dwarf galaxies' structural parameters are determined with an updated version of the M08 algorithm that now accounts for spatially incomplete data sets (such as the PAndAS observations that sometimes suffer from chip gaps and holes created by saturated foreground stars) and relies on Markov Chain Monte Carlo samplings of the posterior PDFs. We take this opportunity to describe this latest version of the algorithm.

For a sky region, $\mathcal{A}$, around a given dwarf galaxy, we work with a set of $n$ CMD-selected PAndAS data points, $\mathcal{D}_n = \{\overrightarrow{d_k}\}_{1\leq k\leq n}$. Each datum is defined by its spatial coordinates on the sky, which can be simplified to the coordinates on the plane tangent to the sky at the dwarf galaxy's centroid assuming literature values\footnote{Note that we infer updated centroid positions as two of the model parameters correspond to offsets from these literature values, which only serve as starting points.}, $\overrightarrow{d_k}=\{x_k,y_k\}$. The likelihood that these data points follow a specific radial density model, defined by the set of parameters $\mathcal{P}=\{p_1,p_2,\dots,p_j\}$, is then defined as

\begin{equation}
\label{eqn:Ptot}
P_\mathrm{tot}\left(\mathcal{D}_n|\mathcal{P}\right) = \prod_k P_k\left(\overrightarrow{d_k}|\mathcal{P}\right),
\end{equation}

\noindent where $P_k\left(\overrightarrow{d_k}|\mathcal{P}\right)$ is the likelihood of datum $k$ to be generated from the chosen model. For the problem at hand, this likelihood can be expressed as the stellar surface density of the model, $\rho_\mathrm{model}$, normalized to the number of stars expected to be in region $\mathcal{A}$. In other words,

\begin{equation}
\label{eqn:Pk}
P_k\left(\overrightarrow{d_k}|\mathcal{P}\right) = \frac{\rho_\mathrm{model}\left(\overrightarrow{d_k}|\mathcal{P}\right)}{\int_\mathcal{A}\rho_\mathrm{model}\ud\mathcal{A}}.
\end{equation}

As described in M08, the family of radial density dwarf galaxy models at radius $r$, $\rho_\mathrm{model}(r)$, are defined by an offset $(x_0,y_0)$ of the centroid from the initial literature value, an ellipticity $\epsilon$ defined as $\epsilon=1-b/a$, with $b/a$ the minor-to-major-axis ratio of the system, the position angle of the major axis $\theta$, defined East of North, the half-light radius\footnote{What we really determine here is the half-density radius since we study the distribution of stars on the sky, irrespective of their luminosity. However, under the assumption that a system harbors uniform stellar populations and does not suffer from mass segregation, this quantity is equivalent to the more common half-light radius. $r_h$ is related to the exponential radius, $r_e$, by $r_h = 1.68r_e$.} of its assumed exponential radial profile, $r_h$, and the number of stars in the system for the chosen CMD selection\footnote{For every dwarf galaxy, this CMD selection box is different and tailored around the visible RGB, as shown by the red polygons in Figure~\ref{CMDs}. It is wide enough in color to include all RGB stars, down to a depth that is usually above the HB (located around $i_0\sim25.0$ in most cases) so as not to be sensitive to field-to-field depth variations. The low density of faint dwarf galaxies like And~XII, And~XXII, or And~XXVI forces us to dig deeper into the data, down to the faint limit. These dwarf galaxies also happen to be the smallest ones, which limits the impact of field-to-field variations since they are completely included in a given MegaCam field.}, $N^*$, such that

\begin{equation}
\label{eqn:rho_dw}
\rho_\mathrm{dwarf}(r) = \frac{1.68^2}{2\pi r_h^2(1-\epsilon)}N^*\exp\left(-1.68 r/r_h\right).
\end{equation}

\noindent Further assuming a constant contamination level\footnote{In cases such as those of And~X and ~XIX (\S~\ref{section_X} and \ref{section_XIX} below) for which there is suspicion of a non-uniform background due to M31 stellar halo substructures, we also tested the use of smoothly varying background models is the form $\Sigma_b = ax+by+cxy+d$, with $a$, $b$, and $c$ parameters of the model and $d$ determined via its normalization. However, it does not yield significantly better results than with a flat background and comes at the cost of a more flexible model that does not converge as easily without added priors. We therefore restrict ourselves to a simple, flat background model and warn the reader in case of doubt about this assumption.}, $\Sigma_b$, around the studied stellar system yields

\begin{equation}
\label{eqn:rho_mod}
\rho_\mathrm{model}(r) = \rho_\mathrm{dwarf}(r) + \Sigma_b.
\end{equation}

\noindent In this equation, $r$ is the elliptical radius, which relates to projected sky coordinates $(x,y)$ via

\begin{eqnarray}
\label{eqn:r}
r=\bigg(  \Big(\frac{1}{1-\epsilon} ((x-x_0)\cos\theta - (y-y_0)\sin\theta)\Big)^2\nonumber\\
+ \Big((x-x_0)\sin\theta +(y-y_0)\cos\theta\Big)^2  \bigg)^{1/2}.
\end{eqnarray}

As in M08, we normalize the likelihood function by further enforcing that the model has the same number of stars in region $\mathcal{A}$, $n$, as is found in the PAndAS data, which sets the value of the background level,

\begin{equation}
\label{eqn:Sigma}
\Sigma_b = \left(n - \int_\mathcal{A}\rho_\mathrm{dwarf} \ud\mathcal{A}\right)/\int_\mathcal{A}\ud\mathcal{A}.
\end{equation}

\noindent Note that, contrary to M08, we do not assume that $\mathcal{A}$ is continuous. In effect, it can include regions with no data (chip gaps, bright stars, etc) and, consequently, the integration of equation (\ref{eqn:Sigma}) is performed numerically over a grid with pixels no larger than 5 times smaller than the inferred half-light radius.

Together, Equations (\ref{eqn:rho_dw}), (\ref{eqn:rho_mod}), (\ref{eqn:r}), and (\ref{eqn:Sigma}) entirely define the likelihood of a data point as described in equation (\ref{eqn:Pk}) and, from there, the likelihood of the data given the model, $P_\mathrm{tot}$, with equation~(\ref{eqn:Ptot}). Following Bayes' rule, $P_\mathrm{tot}(\mathcal{D}_n|\mathcal{P})$ is proportional to what one is ultimately concerned with, $P(\mathcal{P}|\mathcal{D}_n)$, the probability of the model given the data, via multiplicative priors. Since the PAndAS data is generally powerful enough that any set of sensible priors yields similar results, we use flat priors for all parameters, making sure they nevertheless remain physical: 

\begin{itemize}
\item $0\leq\epsilon<1.0$,
\item $\theta$ included in an interval of $180\deg$,
\item $0<r_h\leq r_\mathrm{max}$,
\item $N^*>0$.
\end{itemize}

In most cases, $r_\mathrm{max} = +\infty$, except for the faintest dwarf galaxies for which the algorithm can diverge and attempt to fit a model that is as wide as possible, which, in effect, means a simple increase of the contamination level and the suppression of the dwarf galaxy in the model. However, since we know from spectroscopy and deeper photometry that an M31 stellar overdensity is present in every field, we fix $r_\mathrm{max}$ to a value that is clearly larger than the size of the stellar overdensity in these few cases. This choice will be detailed in section~\ref{results} when relevant.

Finally, we use a home-made, simple Metropolis-Hastings algorithm to sample the posterior distribution functions. The algorithm was typically ran for hundreds of thousands of iterations to achieve visually appealing PDFs in the figures, which ensured convergence. The outcomes of the fits are presented later with a detailed discussion of the results for all 23 dwarf galaxies in the sample (Figures~\ref{posterior_pdf_AndI} to \ref{posterior_pdf_AndXXX} and Section~4). The special case of And~XXVII is discussed in sub-section~\ref{AndXXVII}. The structural-parameter inferences are also summarized in tabular form in Table~\ref{table:param_list}. The figures show the marginalized one-dimensional posterior PDFs for the 6 parameters of the inference. The favored parameters are measured as the modes of these 1d PDFs and uncertainties are bound by the values of the parameters for which the probability is 61 percent of the peak value\footnote{This definition is equivalent to a $\pm1\sigma$ deviation for a Gaussian distribution and mathematically equivalent to the commonly used `$\Delta\chi^2=1$' for such a distribution.}. The figures further show, for every dwarf galaxy, a comparison of the most probable exponential profile, compared to the data binned following the most probable centroid, ellipticity, and position angle.

It should however be emphasized that a truthful representation of the PDFs, including covariance between parameters and a proper sampling of the PDFs are given in electronic form (see Appendix~\ref{chains}) and distributed with this paper. Any analysis that wishes to use the results presented here should strive to use these outputs of the MCMC algorithm instead of the distribution moments summarized in Table~\ref{table:param_list}.

\begin{figure*}
\begin{center}
\includegraphics[width=0.35\hsize,angle=270]{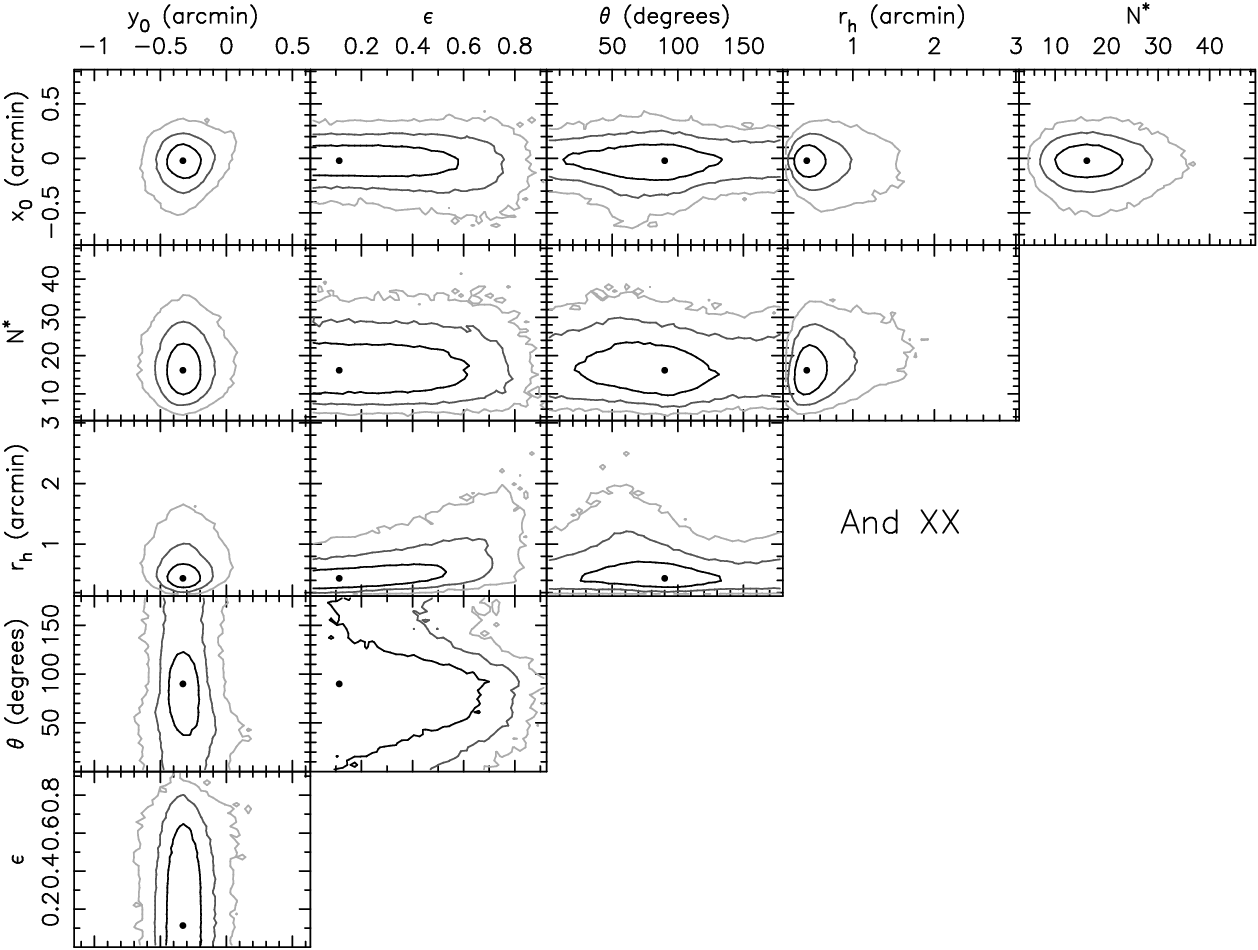}
\hspace{1cm}
\includegraphics[width=0.35\hsize,angle=270]{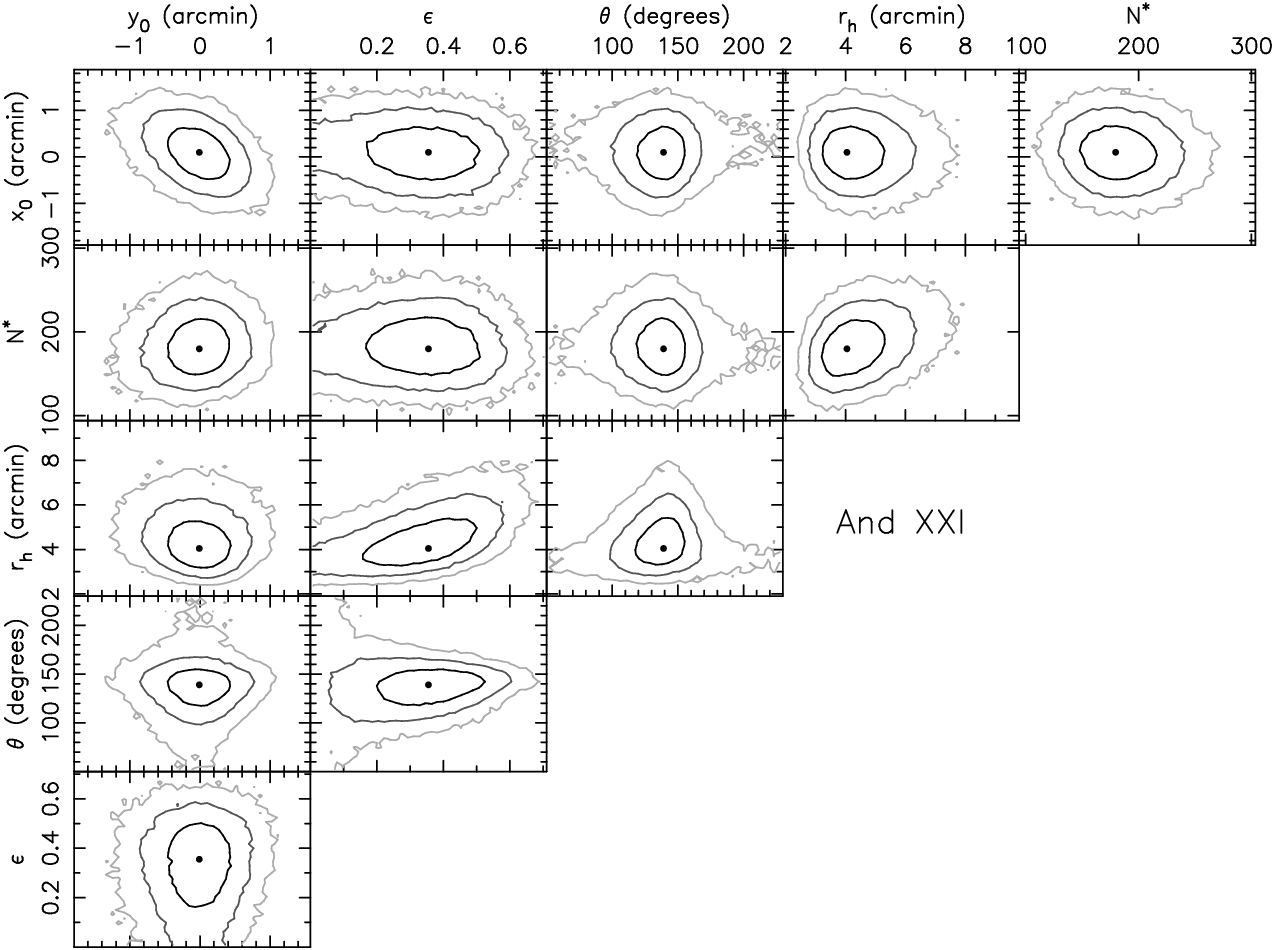}
\caption{\label{2d_plots} Two-dimensional PDFs of all the parameters for the And~XX (left) and And~XXI (right) inference. In each panel, the dot represents the favored model whereas the contours correspond to 1, 2, and $3\sigma$ confidence intervals assuming Gaussian distributions. Of course, these are only indicative in the case of non-Gaussian distributions.}
\end{center}
\end{figure*}

Figure~\ref{2d_plots} presents the 2-dimensional PDFs of all the parameters for a fainter (And~XX) and a brighter (And~XXI) galaxy in the sample. These plots reveal the typical correlations one might expect between the various parameters. Similar figures can easily be constructed from the provided chains for the other dwarf galaxies.

Deriving the physical half-light radius of the stellar systems from the angular half-light radius of the model requires an estimate of the distance to the dwarf galaxies. These have been determined homogeneously by \citet{aconn12} from the same PAndAS data set. Therefore, for each step of our MCMC chain, we randomly draw a distance from their heliocentric distance PDFs for the relevant dwarf galaxy and use this distance to calculate the physical half-light radius. This way, we fold in the distance uncertainties in our assessment of the physical sizes of the systems. Although this has the consequence of sometimes giving poor constraints (e.g. the physical $r_h$ PDF of And~XIV is double-peaked because its distance PDF is double-peaked), we favor the homogeneity of the measurements by using only PAndAS data. Nevertheless, since we also provide the angular sizes, it will be straightforward to recalculate the physical sizes from a future set of more accurate distances.

As a final note, we wish to point out that we chose exponential functions to characterize the radial density profile of the dwarf galaxies because these have been commonly used to characterize dwarf galaxies \citep[starting from][]{faber83} and correspond to a generic phase of dwarf spheroidal evolution \citep[e.g.,][]{read05}. In addition, they are easier to deal with analytically than the also common Plummer profile and are parametrized by one fewer parameter than King profiles, which leads to easier convergence. Moreover, neither of the three types of profiles actually accounts for the complexity of the brighter dwarf galaxies that often contain multiple stellar populations with different spatial distributions. This is commonly seen in the better known Milky Way dwarf galaxies \citep[e.g.][]{tolstoy04} but has also been shown to be the case for And~II, one of the most studied Andromeda companions \citep{mcconnachie07a,ho15}. It remains that exponential profiles are a simple way to characterize dwarf galaxies, especially over a range of luminosity and, likely, stellar content, even if the latter is difficult to constrain for the fainter systems.

\subsection{Absolute magnitudes}

One of the outcomes of the structural parameter fit is that we now have samplings of the PDF of the number of stars, $N^*$, that a dwarf galaxy hosts for the chosen CMD selection box, corrected for chip gaps and other holes in the data. In and of itself, $N^*$ is not very meaningful since the selection boxes change with the dwarf galaxy, the distance to it, and the depth of the data in this particular region of the survey. This number can, however, be converted into the absolute magnitude of the dwarf galaxy by sampling an artificial CMD representative of the stellar system (M08). We use this methodology below to calculate the absolute magnitude of the dwarf galaxies after determining the (in)completeness function of the data at any location in the PAndAS survey.

\subsubsection{The PAndAS completeness model}
\label{section_completeness}
We start by determining the completeness function of the survey data in the region around And~XIV, for which we have deep and wide Subaru/SuprimeCam data. These SuprimeCam data were observed on the night of August 22, 2009 as part of an ongoing program to gather deep photometric follow-up for all M31 dwarf galaxies discovered post-2004 with wide-field imagers on 8-meter class telescopes \citep[e.g.][]{brasseur11a}. The data comprise $3\times400$s and $9\times220$s dithered exposures in the SuprimeCam $g$ and $i$ bands, respectively, obtained under exquisite conditions (image quality $\sim0.4''$).

\begin{figure*}
\begin{center}
\includegraphics[width=0.4\hsize,angle=270]{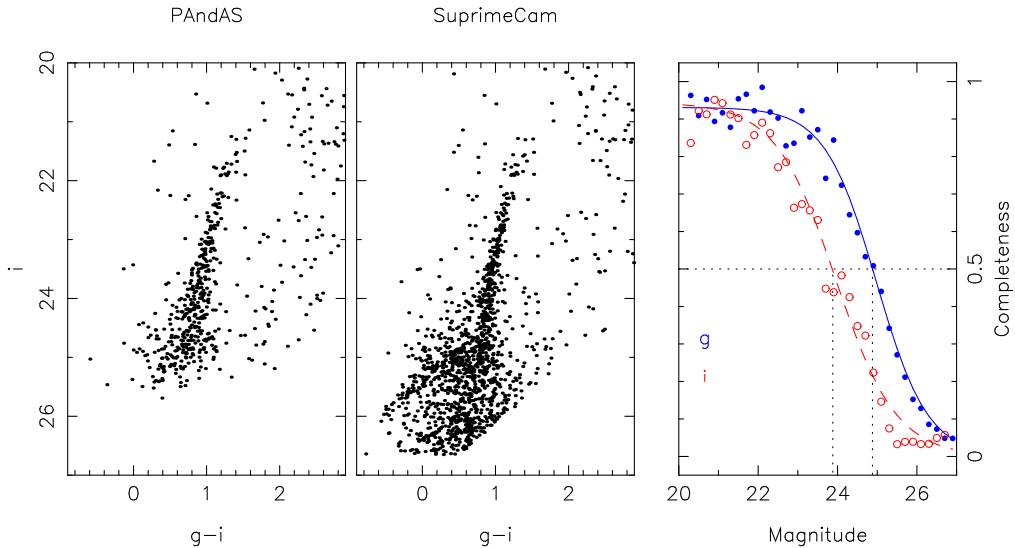}
\caption{\label{completeness} Comparison of the CMD of stars within $5'$ of And~XIV from PAndAS data (left) and from the SuprimeCam data used to derive the completeness function of the PAndAS data (middle). Neither of the two CMDs are de-reddened. The right-hand panel shows the fraction of SuprimeCam stars present in the PAndAS data for $g$-band magnitudes by blue, filled circles and $i$-band magnitudes by red, hollow circles. The full blue line represents the best completeness model in the $g$ band, as defined by equation~(\ref{eqn:completeness}), while the dashed red line corresponds to the best model in the $i$ band. The thin dotted lines indicate the values of 50-percent completeness, $m_{50}$.}
\end{center}
\end{figure*}

The images are reduced and stacked and the photometry is performed in the usual way, using a version of the CASU pipeline \citep{irwin01} updated to work on SuprimeCam data. The resulting photometric catalogue is calibrated onto the PAndAS photometry and the two left-most panels of Figure~\ref{completeness} show a comparison of the PAndAS and SuprimeCam CMDs for the region within $5'$ of And~XIV's centroid; the SuprimeCam data are obviously much deeper than PAndAS.

After cross-identifying the PAndAS and the SuprimeCam catalogues over the full SuprimeCam footprint minus a small region near the center of And~XIV that could suffer from crowding, and minus the chip gaps and the halos of bright stars that are present in either of the two data sets, we calculate the fraction, $\eta$, of SuprimeCam stellar-like objects\footnote{Stellar-like objects have CASU classification flags of $-1$ or $-2$ in both the $g$- and $i$-band observations.} that are present in the stellar-like PAndAS catalogue for a given magnitude bin\footnote{By doing so, we assume that the SuprimeCam data are 100\% complete.}. These fractions are represented by the filled and hollow circles in the right-most panel of Figure~\ref{completeness} for the $g$ and $i$ band, respectively. Note that we do not require an object to be observed in the two bands at this stage as we aim to build two independent $g$- and $i$-band completeness functions, which, when combined, will yield the completeness of the full CMD.

These completeness data points are used to constrain models of the form

\begin{equation}
\label{eqn:completeness}
\eta(m) = \frac{A}{1+\exp\left(\frac{m-m_\mathrm{50}}{\rho}\right)},
\end{equation}

\noindent where $m$ is either the $g$-band or the $i$-band magnitude. For the $g$ band, we find that the favored values are $A = 0.94$, $m_\mathrm{50} = 24.88$, and $\rho = 0.65$, whereas, for the $i$ band, we get $A = 0.93$, $m_\mathrm{50} = 23.88$, and $\rho = 0.74$. The curves in the right-most panel of Figure~\ref{completeness} represent these best models.

Since the depth of the data varies as a function of position in the survey, we further need to shift these completeness models to brighter/fainter magnitudes when the PAndAS data is shallower/deeper. In order to do so, we use the median magnitude of stars whose photometric uncertainties are in the range 0.09--0.11, $m_\mathrm{ref}$, as a reference point (i.e., the local $10\sigma$ depth of the data). For a given MegaCam field $k$ in PAndAS, the completeness model of the $m$ band, with $m=g$ or $i$, therefore becomes $\eta(m + m_\mathrm{ref, XIV} - m_{\mathrm{ref,}k}$), where $m_\mathrm{ref, XIV}$ is the reference magnitude of the field that contains And~XIV.

An additional test with a shallower SuprimeCam field around And~V yields model parameters that are consistent, within the uncertainties, with the model constrained on the And~XIV data. We however favor the model built from the And~XIV as it corresponds to the deepest SuprimeCam data we have at our disposal.

\subsubsection{Calculating $M_V$}

For a given dwarf galaxy, we start by drawing a distance modulus from the distance MCMC chains of \citet{aconn12}. We then model the PDF of its member stars in the CMD as a 13-Gyr \textsc{Parsec} isochrone \citep{bressan12} of the appropriate metallicity\footnote{Whenever possible, we use the spectroscopic metallicities compiled by \citet{collins13}, with the exception of And~I for which we use the photometric metallicities derived by \citet{kalirai10}, And~II for which we use the more recent spectroscopic value from \citet{ho15}, and And~X for which we use the photometric metallicity derived by \citet{brasseur11b}.}, with the associated luminosity function (assuming the default \citet{chabrier01} IMF), shifted to this distance modulus, reddened by the relevant amount of extinction measured from \citet{schlegel98} as recalibrated by \citet{schlafly11}, and convolved by the photometric uncertainties (see \citealt{martin13b} for more details).\footnote{As we did in \citet{martin13b}, we further convolve the color-magnitude PDF of these single stellar populations with an additional, empirically-determined 0.03-magnitude term that is added in quadrature to the photometric uncertainties. This has the consequence of producing slightly wider PDFs and accounts for the dwarf galaxies hosting more than single stellar populations. The impact of this widening is however almost insignificant in the current analysis.} This model PDF is defined down to the hydrogen burning limit.

A random drawing from the structural parameter MCMC chain of the dwarf galaxy yields the target $N^*_i$ that we wish to reproduce for a particular realization $i$ of the dwarf galaxy's CMD. We then sample the color-magnitude PDF and flag sample stars that fall in the CMD selection box that was used to determine the structural parameters. We further randomly determine if that star falls below both completeness functions for its $g$ and $i$ magnitudes and flag them accordingly. We repeat this process until we have accumulated $N^*_i$ flagged stars. Summing up the flux of all the stars generated by this procedure (flagged or not) yields the apparent $g$-band and $i$-band magnitude of that realization of the dwarf galaxy, which we convert to the $V$ band with the color equations given in section~2.2 of \citet{ibata07} and correct for the extinction. In addition to the apparent magnitude, $m_{V,0}$, correcting from the distance modulus assumed earlier yields the $V$-band absolute magnitude, $M_{V,0}$, of this realization of the dwarf galaxy.

In order to account for uncertainties that stem from the (in)completeness model, we determine the impact on the And~I magnitude values of varying the $g$- and $i$-band models by up to $\pm1\sigma$ from the favored values determined in sub-section~\ref{section_completeness} above. These typically yield uncertainties of $\pm0.2$ and we therefore add a Gaussian random deviate of width 0.2 to the values of $m_{V,0}$ determined through the procedure above.

One may wonder about the impact of our choice of CMD selection boxes driving the uncertainties on $N^*$ and, from there, those on $m_{V,0}$. Our framework naturally takes these into account as a wider and therefore more contaminated selection box will yield a more uncertain value for $N^*$, which will drive a more uncertain mesure of the system's magnitude.

As we will see below, barring cases for which a direct comparison is not warranted (And~XV, XVI, XVIII, and XIX), the revised magnitude values differ from the latest literature values by being, overall, $0.29\pm0.11$~magnitudes fainter, which could be a consequence of our choice of a very old stellar population. New results show that (at least some) Andromeda dwarf galaxies contain a significant fraction of intermediate age stars, younger than the 13~Gyr assumed above \citep{weisz14b,skillman16}. A test performed by assuming that And~I is instead entirely composed of an 8-Gyr stellar population does lead to magnitudes that are systematically brighter by $\sim0.2$ magnitudes. Without systematic star-formation histories inferred from deep photometry for all the dwarf galaxies presented here, it is however too soon to use more complex stellar populations for the current study.

For every dwarf galaxy, we show the $m_{V,0}$ and $M_{V,0}$ PDFs resulting from 500 iterations of this procedure in the two right-most panels of Figures~\ref{posterior_pdf_AndI} to~\ref{posterior_pdf_AndXXX}. Table~\ref{table:luminosity_list} summarizes the luminosity-related properties of the dwarf galaxies and lists the median of the distributions and the limits of their central 68\% confidence interval. As for the structural parameters, we list random samples of the $m_{V,0}$ and $M_{V,0}$ PDFs in the MCMC chains distributed with this paper (see Appendix~\ref{chains}).

\subsection{Central surface brightness}

Following equation~(6) of M08, the central surface brightness of a dwarf galaxy, $\mathcal{F}_0$, expressed in units of flux per angular area, can be calculated as

\begin{equation}
\mathcal{F}_0 = \frac{\mathcal{F}}{2\pi r_e^2(1-\epsilon)},
\end{equation}

\noindent where $\mathcal{F}$ is the total apparent flux of the system, which is easily calculated from the apparent magnitude $m_{V,0}$ determined in the previous section. The quantity $\mathcal{F}_0$ can then easily be transformed back into the more usual central surface brightness $\mu_0$ expressed in mag/arcsec$^2$. These values are calculated for all dwarf galaxies and every iteration of the chains and the resulting PDFs are summarized in Table~\ref{table:luminosity_list}.

\begin{table*}
\begin{center}
\caption{\label{table:param_list}Derived structural parameters of the dwarf galaxies}
\begin{tabular}{l|llllll}
\hline
Name & $\alpha$ (J2000) & $\delta$ (J2000) & $\epsilon$ & $\theta$ (deg) & $r_h$ (arcmin) & $r_h$ (pc)\footnote{Calculated using the posterior heliocentric distance PDFs calculated by \citet{aconn12}.}\\
\hline
And~I & $0^\mathrm{h}45^\mathrm{m}39.7^\mathrm{s} \pm 0.3^\mathrm{s}$ & $+38\deg02'15'' \pm 6''$ & $0.28\pm0.03$ & $30\pm4$ & $3.9\pm0.1$ & $815\pm40$\\
And~II & $1^\mathrm{h}16^\mathrm{m}26.8^\mathrm{s} \pm 0.4^\mathrm{s}$ & $+33\deg26'07'' \pm 6''$ & $0.16\pm0.02$ & $31\pm5$ & $5.3 \pm 0.1$ & $965\pm45$\\
And~III & $0^\mathrm{h}35^\mathrm{m}30.9^\mathrm{s} \pm 0.4^\mathrm{s}$ & $+36\deg29'56'' \pm 8''$ & $0.59\pm0.04$ &$140\pm3$ & $2.0\pm0.2$ & $405\pm35$\\
And~V & $1^\mathrm{h}10^\mathrm{m}17.5^\mathrm{s} \pm 0.4^\mathrm{s}$ & $+47\deg37'42'' \pm 6''$ & $0.26^{+0.09}_{-0.07}$ & $54\pm10$ & $1.6^{+0.2}_{-0.1}$ & $345\pm40$\\
And~IX & $0^\mathrm{h}52^\mathrm{m}53.4^\mathrm{s} \pm 0.7^\mathrm{s}$ & $+43\deg11'57'' \pm 8''$ & $0.00^{+0.16}_{-0.00}$ & $41\pm65$ & $2.0^{+0.3}_{-0.2}$ & $360^{+60}_{-50}$\\
And~X\footnote{Substructure in the field may impact the structural parameters.}\footnote{Stricter priors were used to ensure convergence. See the text for more detail.} & $1^\mathrm{h}06^\mathrm{m}35.4^\mathrm{s} \pm 0.6^\mathrm{s}$ & $+44\deg48'27'' \pm 10''$ & $0.10^{+0.34}_{-0.10}$ & $30^{+20}_{-12}$ & $1.1^{+0.4}_{-0.2}$ & $210^{+70}_{-35}$\\
And~XI & $0^\mathrm{h}46^\mathrm{m}19.7^\mathrm{s} \pm 0.6^\mathrm{s}$ & $+33\deg48'10'' \pm 8''$ & $0.19^{+0.28}_{-0.19}$ & $54\pm30$ & $0.6\pm0.2$ & $120^{+53}_{-44}$\\
And~XII$^\mathrm{c}$ & $0^\mathrm{h}47^\mathrm{m}28.3^\mathrm{s}\pm 1.3^\mathrm{s}$ & $+34\deg22'38'' \pm 37''$ & $0.61^{+0.16}_{-0.48}$ & $16^{+12}_{-36}$ & $1.8^{+1.2}_{-0.7}$ & $420^{+280}_{-200}$\\
And~XIII & $0^\mathrm{h}51^\mathrm{m}51.0^\mathrm{s} \pm 0.5^\mathrm{s}$ & $+33\deg00'16'' \pm 13''$ & $0.61^{+0.14}_{-0.20}$ & $-20^{+9}_{-12}$ & $0.8^{+0.4}_{-0.3}$ & $130^{+80}_{-62}$\\
And~XIV & $0^\mathrm{h}51^\mathrm{m}35.0^\mathrm{s} \pm 0.5^\mathrm{s}$ & $+29\deg41'23'' \pm 10''$ & $0.17^{+0.16}_{-0.17}$ & $-4\pm14$ & $1.5\pm0.2$ & $\sim265$\\
And~XV & $1^\mathrm{h}14^\mathrm{m}18.3^\mathrm{s} \pm 0.5^\mathrm{s}$ & $+38\deg07'11'' \pm 7''$ & $0.24\pm0.10$ & $38\pm15$ & $1.3\pm0.1$ & $230^{+35}_{-25}$\\
And~XVI & $0^\mathrm{h}59^\mathrm{m}30.3^\mathrm{s} \pm 0.4^\mathrm{s}$ & $+32\deg22'34'' \pm 4''$ & $0.29\pm0.08$ & $98\pm9$ & $1.0\pm0.1$ & $130^{+30}_{-15}$\\
And~XVII & $0^\mathrm{h}37^\mathrm{m}06.3^\mathrm{s} \pm 0.6^\mathrm{s}$ & $+44\deg19'23'' \pm 6''$ & $0.50\pm0.10$ & $110\pm9$ & $1.4\pm0.3$ & $285^{+55}_{-45}$\\
And~XVIII & $0^\mathrm{h}02^\mathrm{m}15.6^\mathrm{s} \pm 0.5^\mathrm{s}$ & $+45\deg05'28'' \pm11''$ & $0.03^{+0.28}_{-0.03}$ & $42^{+40}_{-84}$ & $0.8\pm0.1$ & $265\pm50$\\
And~XIX$^\mathrm{b}$ & $0^\mathrm{h}19^\mathrm{m}34.5^\mathrm{s}\pm3.0^\mathrm{s}$ & $+35\deg02'41'' \pm53''$ & $0.58^{+0.05}_{-0.10}$ & $34\pm5$ & $14.2^{+3.4}_{-1.9}$ & $3065^{+935}_{-1065}$\\
And~XX & $0^\mathrm{h}07^\mathrm{m}30.6^\mathrm{s} \pm 0.4^\mathrm{s}$ & $+35\deg07'37'' \pm 6''$ & $0.11^{+0.41}_{-0.11}$ & $90^{+20}_{-44}$ & $0.4^{+0.2}_{-0.1}$ & $90^{+35}_{-20}$\\
And~XXI & $23^\mathrm{h}54^\mathrm{m}47.9^\mathrm{s} \pm 1.6^\mathrm{s}$ & $+42\deg28'14'' \pm 20''$ & $0.36^{+0.10}_{-0.13}$ & $139\pm13$ & $4.1^{+0.8}_{-0.4}$ & $1005\pm175$\\
And~XXII/Tri~I & $1^\mathrm{h}27^\mathrm{m}40.4^\mathrm{s} \pm 0.6^\mathrm{s}$ & $+28\deg05'25'' \pm 7''$ & $0.61^{+0.10}_{-0.14}$ & $114\pm10$ & $0.9^{+0.3}_{-0.2}$ & $225\pm75$\\
And~XXIII & $1^\mathrm{h}29^\mathrm{m}21.0^\mathrm{s} \pm 0.8^\mathrm{s}$ & $+38\deg43'26'' \pm 13''$ & $0.41^{+0.05}_{-0.06}$ & $138\pm5$ & $5.4\pm0.4$ & $1190\pm100$\\
And~XXIV & $1^\mathrm{h}18^\mathrm{m}32.7^\mathrm{s} \pm 1.8^\mathrm{s}$ & $+46\deg22'13'' \pm 18''$ & $0.10^{+0.31}_{-0.10}$ & $90\pm34$ & $2.6^{+1.0}_{-0.5}$ & $680^{+250}_{-140}$\\
And~XXV & $0^\mathrm{h}30^\mathrm{m}09.9^\mathrm{s} \pm 0.9^\mathrm{s}$ & $+46\deg51'41'' \pm 16''$ & $0.03^{+0.16}_{-0.03}$ & $-16\pm30$ & $2.7^{+0.4}_{-0.2}$ & $545^{+95}_{-65}$\\
And~XXVI$^\mathrm{c}$ & $0^\mathrm{h}23^\mathrm{m}46.3^\mathrm{s} \pm 1.0^\mathrm{s}$ & $+47\deg54'43'' \pm 16''$ & $0.35^{+0.33}_{-0.35}$ & $50\pm90$ & $1.0^{+0.6}_{-0.5}$ & $150^{+140}_{-80}$\\
Cas~II/And~XXX & $0^\mathrm{h}36^\mathrm{m}34.6^\mathrm{s} \pm 0.6^\mathrm{s}$ & $+49\deg38'49'' \pm5''$ & $0.43^{+0.10}_{-0.12}$ & $110\pm9$ & $1.5\pm0.2$ & $270\pm50$\\
\end{tabular}
\end{center}
\end{table*}

\begin{table*}
\begin{center}
\caption{\label{table:luminosity_list}Derived luminosity-related properties of the dwarf galaxies}
\begin{tabular}{lllll}
\hline
Name & $m_{V,0}$ & $M_{V,0}$\footnote{Calculated using the posterior heliocentric distance PDFs calculated by \citet{aconn12}.} & $\log_{10}(L_V/\lsun$)$^\mathrm{a}$ & $\mu_0$ (mag/arcsec$^2$)\\
\hline
And~I & $13.1\pm0.2$ & $-11.2\pm0.2$ & $6.38\pm0.10$ & $25.4\pm0.2$\\
And~II & $12.4\pm0.2$ & $-11.6\pm0.2$ & $6.54\pm0.09$ & $25.6\pm0.2$\\
And~III & $14.8\pm0.2$ & $-9.5\pm0.3$ & $5.68\pm0.10$ & $25.1\pm0.3$\\
And~V & $15.1\pm0.2$ & $-9.3\pm0.2$ & $5.61\pm0.10$ & $25.6\pm0.3$\\
And~IX & $15.6\pm0.3$ & $-8.5\pm0.3$ & $5.3\pm0.1$ & $26.7\pm0.3$\\
And~X\footnote{Substructure in the field may impact the structural parameters.}\footnote{Stricter priors were used to ensure convergence. See the text for more detail.} & $16.7\pm0.3$ & $-7.4\pm0.3$ & $4.9\pm0.1$ & $26.6^{+0.6}_{-0.5}$\\
And~XI & $18.0\pm0.4$ & $-6.3^{+0.6}_{-0.4}$ & $4.4\pm0.2$ & $26.8\pm0.6$\\
And~XII$^\mathrm{c}$ & $17.7\pm0.5$ & $-7.0^{+0.7}_{-0.5}$ & $4.7\pm0.2$ & $28.7^{+0.7}_{-0.9}$\\
And~XIII & $17.8\pm0.4$ & $-6.5^{+0.7}_{-0.5}$ & $4.5^{+0.2}_{-0.3}$ & $26.6\pm0.6$\\
And~XIV & $15.8\pm 0.3$ & $-8.5^{+0.4}_{-0.3}$ & $5.3^{+0.1}_{-0.2}$ & $26.3\pm0.3$\\
And~XV & $16.0\pm0.3$ & $-8.0^{+0.3}_{-0.4}$ & $5.1^{+0.2}_{-0.1}$ & $26.1\pm0.3$\\
And~XVI & $16.1\pm0.3$ & $-7.3\pm0.3$ & $4.8\pm0.1$ & $25.5\pm0.3$\\
And~XVII & $16.6\pm0.3$ & $-7.8\pm0.3$ & $5.0\pm0.1$ & $26.4^{+0.4}_{-0.3}$\\
And~XVIII & $16.2\pm0.4$ & $-9.2^{+0.3}_{-0.4}$ & $5.6\pm0.2$ & $25.2^{+0.4}_{-0.5}$\\
And~XIX$^\mathrm{b}$ & $14.5\pm0.3$ & $-10.0^{+0.8}_{-0.4}$ & $5.9^{+0.1}_{-0.3}$ & $29.3\pm0.4$\\
And~XX & $18.0\pm0.4$ & $-6.4^{+0.5}_{-0.4}$ & $4.4\pm0.2$ & $25.8\pm0.7$\\
And~XXI & $15.5\pm0.3$ & $-9.1\pm0.3$ & $5.5\pm0.1$ & $28.0\pm0.3$\\
And~XXII/Tri~I & $18.0\pm0.4$ & $-6.7^{+0.7}_{-0.5}$ & $4.6^{+0.2}_{-0.3}$ & $26.9^{+0.6}_{-0.5}$\\
And~XXIII & $14.6\pm0.2$ & $-9.8^{+0.2}_{-0.3}$ & $5.8\pm0.1$ & $27.5\pm0.2$\\
And~XXIV & $16.3\pm0.3$ & $-8.4\pm0.4$ & $5.3\pm0.2$ & $28.2\pm0.4$\\
And~XXV & $15.3^{+0.3}_{-0.2}$ & $-9.0\pm0.3$ & $5.5\pm0.1$ & $27.1\pm0.3$\\
And~XXVI$^\mathrm{c}$ & $18.5^{+0.7}_{-0.5}$ & $-5.8^{+0.9}_{-1.0}$ & $4.2\pm0.4$ & $28.0\pm1.0$\\
Cas~II/And~XXX & $16.0^{+0.3}_{-0.2}$ & $-8.0^{+0.4}_{-0.3}$ & $5.1\pm0.1$ & $26.1\pm0.3$\\
\end{tabular}
\end{center}
\end{table*}

\section{Results}
This section presents the detailed comparison of our results for each dwarf galaxy with literature values from the latest and/or deepest analyses. Whenever available, we will focus on the angular half-light radius and the apparent magnitude, as opposed to the physical half-light radius and the absolute magnitude since these latter parameters could be significantly affected by changes in the distance estimates. These were already covered in the detailed analysis of \citet{aconn12} that relied entirely on PAndAS data.

\label{results}
\subsection{And I}
\begin{figure*}
\begin{center}
\includegraphics[width=0.4\hsize,angle=270]{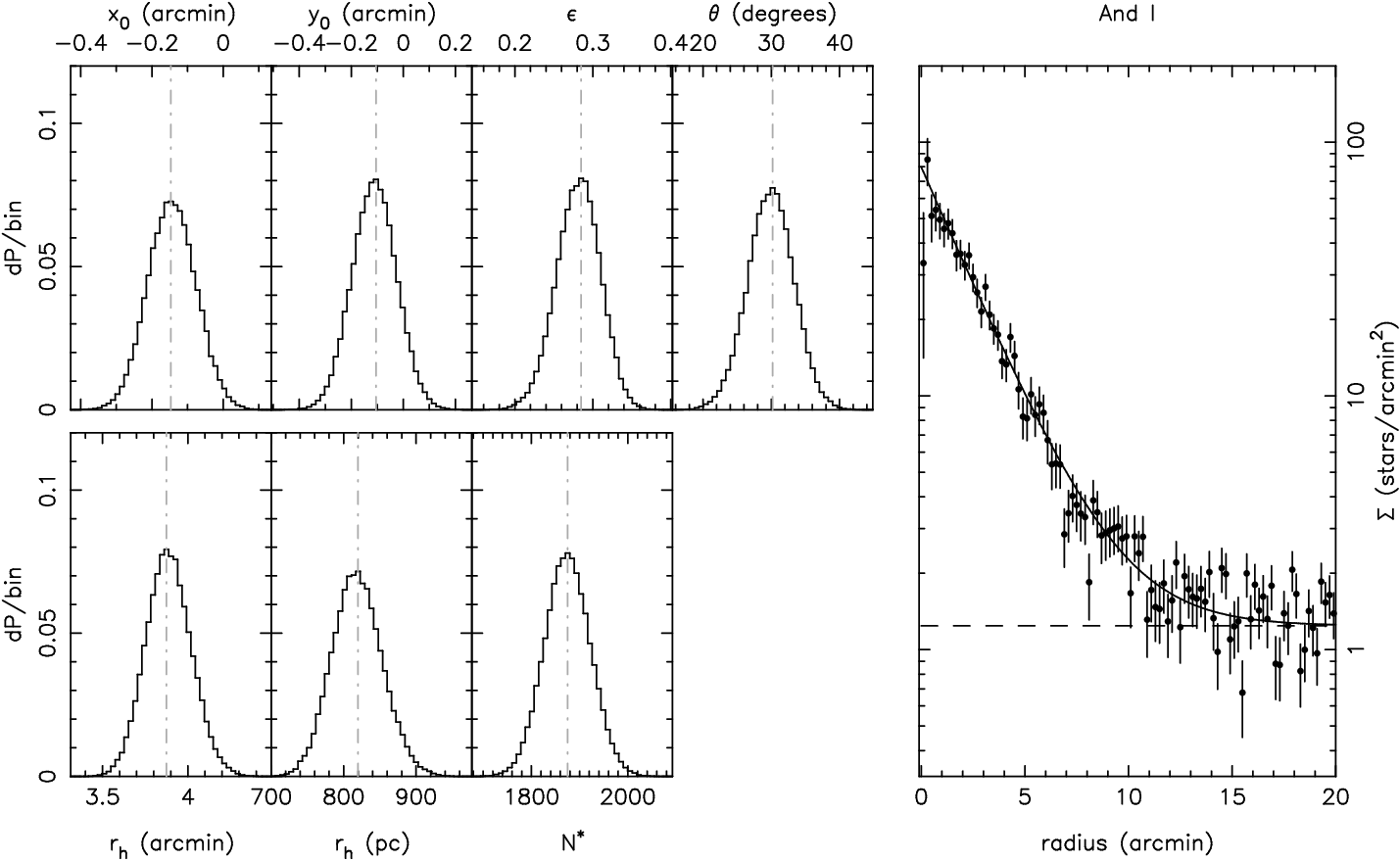}\hspace{1cm}
\includegraphics[width=0.4\hsize,angle=270]{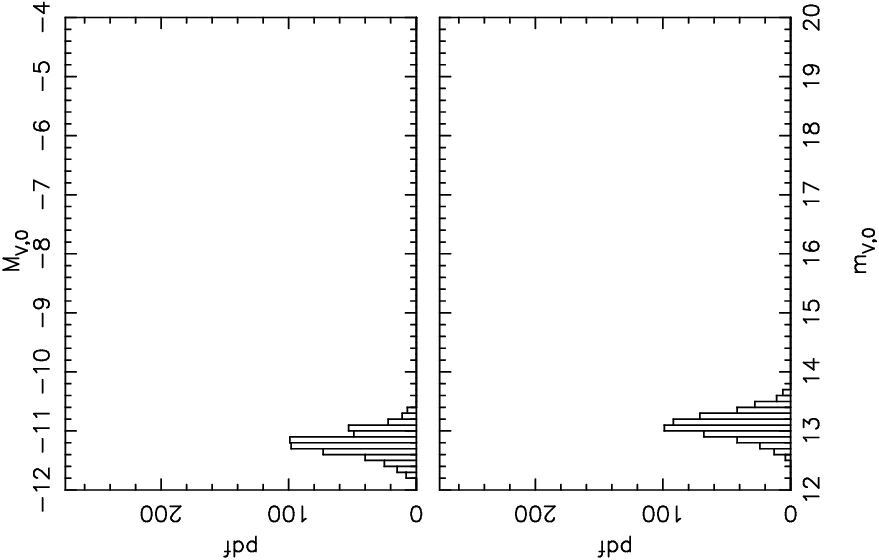}
\caption{\label{posterior_pdf_AndI}\emph{Left:} Marginalized one-dimensional posterior PDFs for the structural parameter inference of And~I. From top-left to bottom-right, the panels show the PDFs for the centroid ($x_0,y_0$), the ellipticity ($\epsilon$), the position angle ($\theta$), the angular and the physical half-light radii ($r_h$), and the mask-corrected number of stars that populate the dwarf galaxy within the chosen CMD selection box ($N^*$). The dashed vertical gray lines highlight the modes of the PDFs. \emph{Middle panel:} Comparison of the favored exponential radial density profile (full line) with the PAndAS data binned in elliptical annuli following the favored structural centroid, ellipticity, position angle, and number of stars, corrected for regions of the survey that are masked out (dots, with the error bars representing Poisson uncertainties). The dashed line indicates the favored estimate of the flat background density ($\Sigma_b$). \emph{Right:} PDF for the apparent and absolute magnitudes of And~I (top and bottom, respectively).}
\end{center}
\end{figure*}

And~I is a bright dwarf galaxy that was initially discovered on photographic plates \citep{vandenbergh72}. The PDFs of the structural parameters we derive are all well behaved (left panels of Figure~\ref{posterior_pdf_AndI}; $r_h=3.9\pm0.1'$, $\epsilon=0.28\pm0.03$, $m_{V,0}=13.1\pm0.2$), the favored exponential model is entirely consistent with the data binned following the favored model (central panel), and the magnitudes of this bright dwarf galaxy are very well constrained due to the large number of stars on its RGB, which limits the impact of CMD `shot-noise' (right panels).

The latest study of the structure of And~I was performed by \citet{mcconnachie06b} from INT/WFC data reaching $\sim1.5$~magnitudes below the dwarf galaxy's TRGB and yielded a fairly similar picture to the one presented here ($r_h=2.9\pm0.1'$, $\epsilon = 0.22\pm0.04$, $m_{V,0}=12.7\pm0.1$), albeit with some differences. However, the updated size of the system is significantly larger than before. This could be linked to the S-shaped outskirts of And~I reported by \citet{mcconnachie06b} and the deeper PAndAS data being more sensitive to tidally affected low-surface brightness regions extending further out. We also find that And~I is slightly but significantly fainter than previously estimated, likely indicative of small systematics in either our or the INT/WFC study for cases such as this one where the RGB is very well populated and the CMD `shot-noise' minimal.

\subsection{And II}
\begin{figure*}
\begin{center}
\includegraphics[width=0.4\hsize,angle=270]{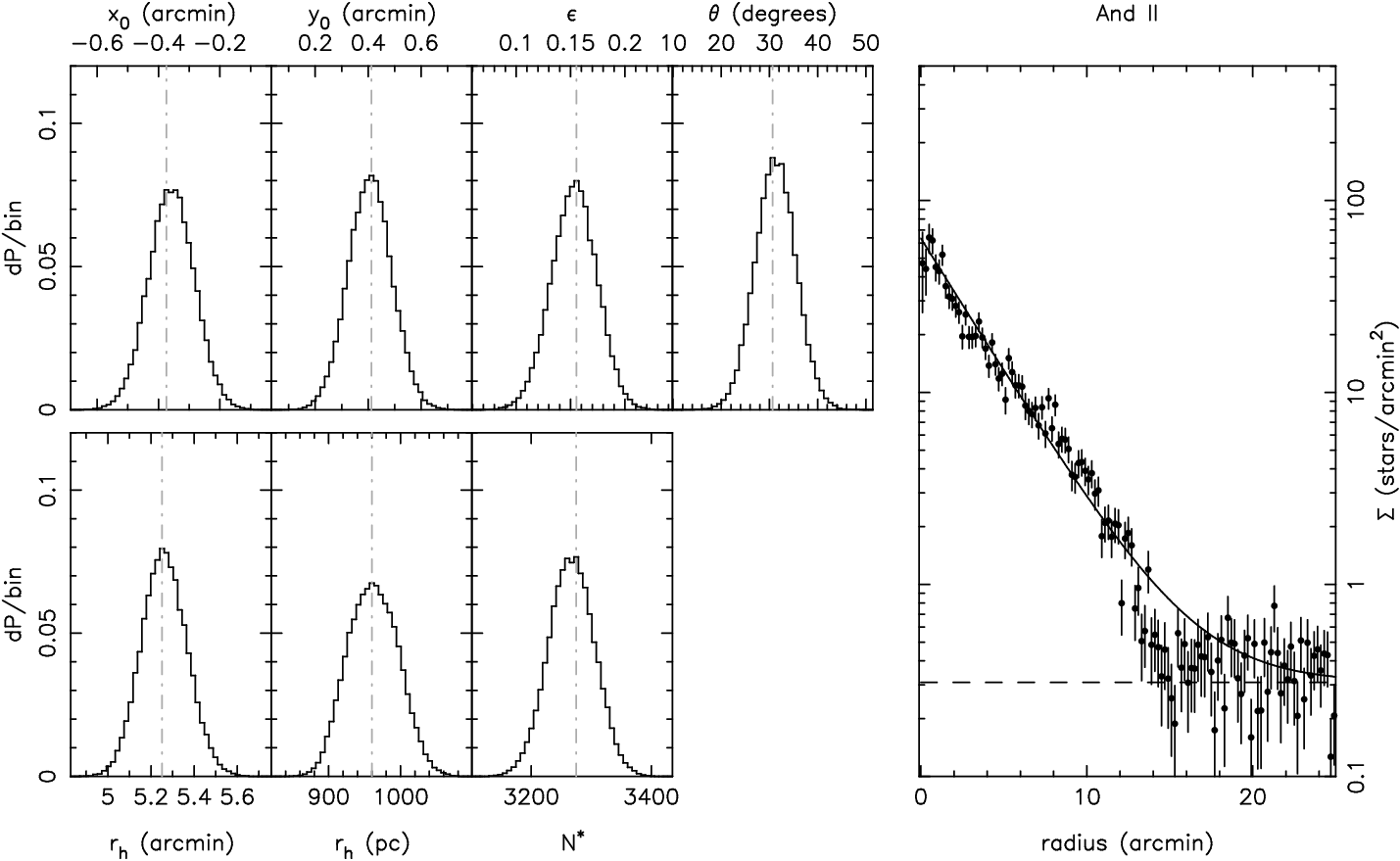}\hspace{1cm}
\includegraphics[width=0.4\hsize,angle=270]{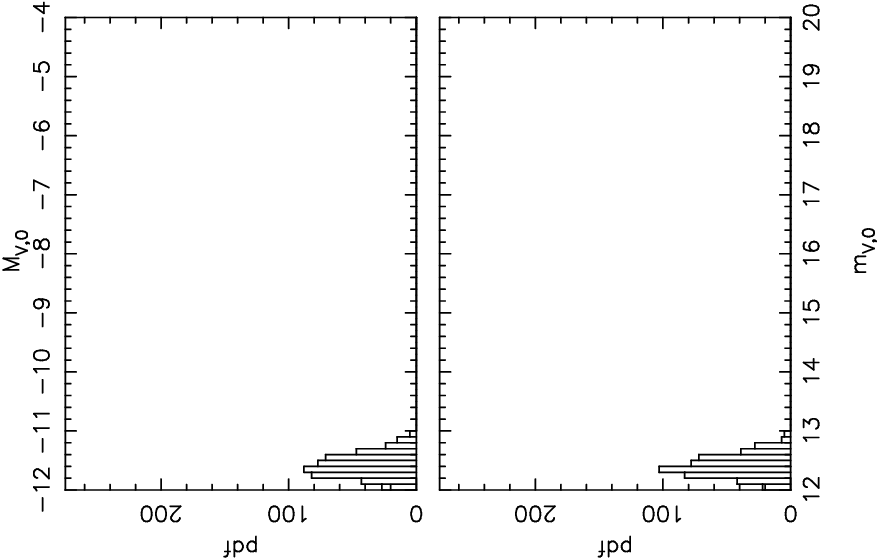}
\caption{\label{posterior_pdf_AndII}Same as Figure~\ref{posterior_pdf_AndI} for And~II.}
\end{center}
\end{figure*}

And~II was also discovered by \citet{vandenbergh72} from photographic plate studies. Its diagnostic plots are presented in Figure~\ref{posterior_pdf_AndII}. As for And~I, the PDFs are well behaved ($r_h=5.3\pm0.1'$, $\epsilon = 0.16\pm0.02$, $m_{V,0}=12.4\pm0.2$). However, the favored exponential profile shows discrepancies with the data binned following the favored model. This is not surprising as And~II is known to host at least two distinct components with different chemistry and, more importantly, different radial density profiles \citep{mcconnachie06b,mcconnachie07a,weisz14b}. A recent study of RGB velocities further points to complex dynamics \citep{amorisco14}.

Our simple exponential radial density fit is evidently not an exact representation of this dwarf galaxy's properties but it nevertheless provides an easy way to compare And~II with other, fainter dwarf galaxies that may also contain distinct stellar populations, even though they would likely escape detection. Our results are similar to the ones presented by \citet{mcconnachie06b} from their shallower INT/WFC data ($r_h = 5.9\pm0.1'$, $\epsilon = 0.20\pm0.08$, $m_{V,0} = 11.7\pm0.2$).

\subsection{And III}
\begin{figure*}
\begin{center}
\includegraphics[width=0.4\hsize,angle=270]{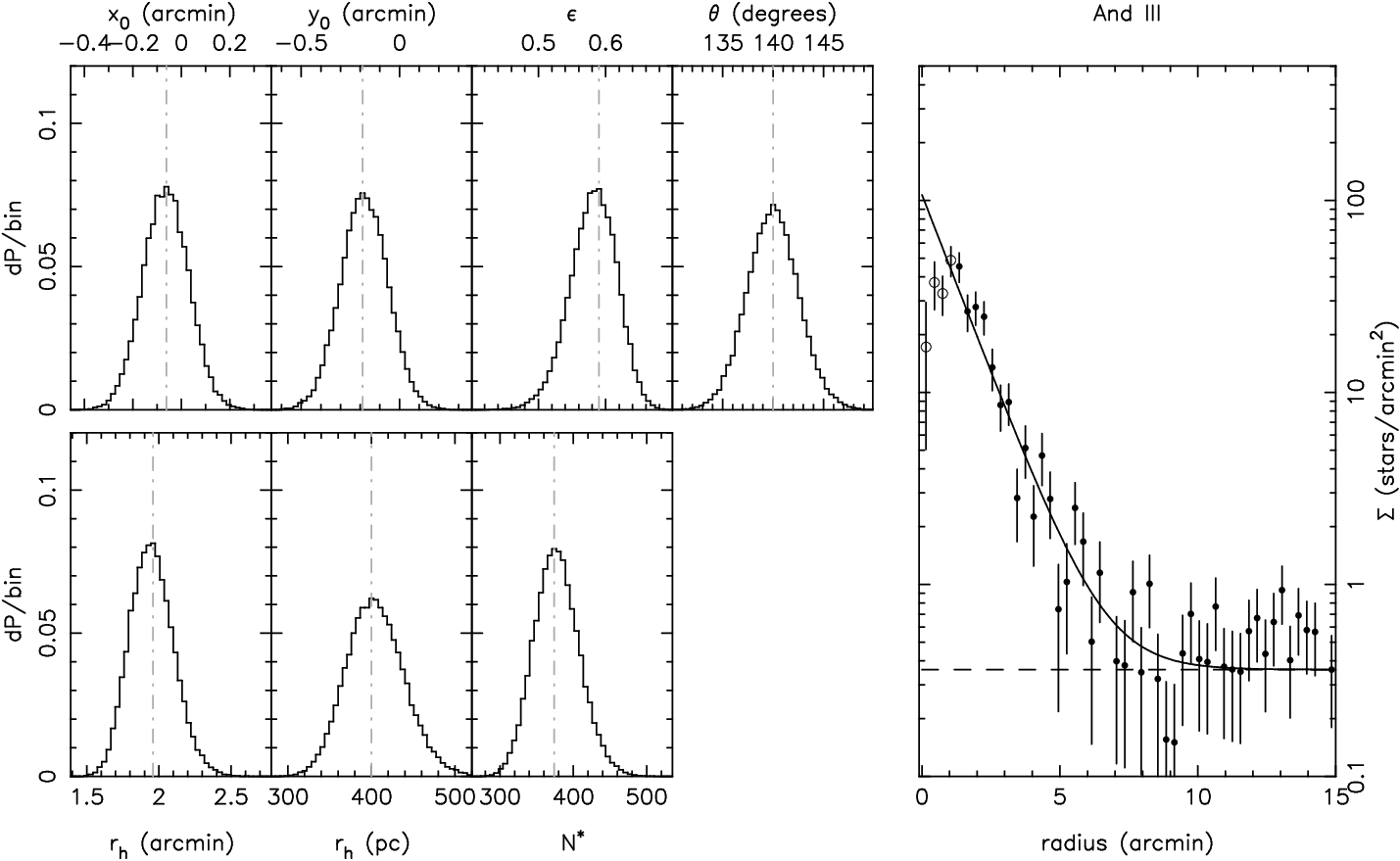}\hspace{1cm}
\includegraphics[width=0.4\hsize,angle=270]{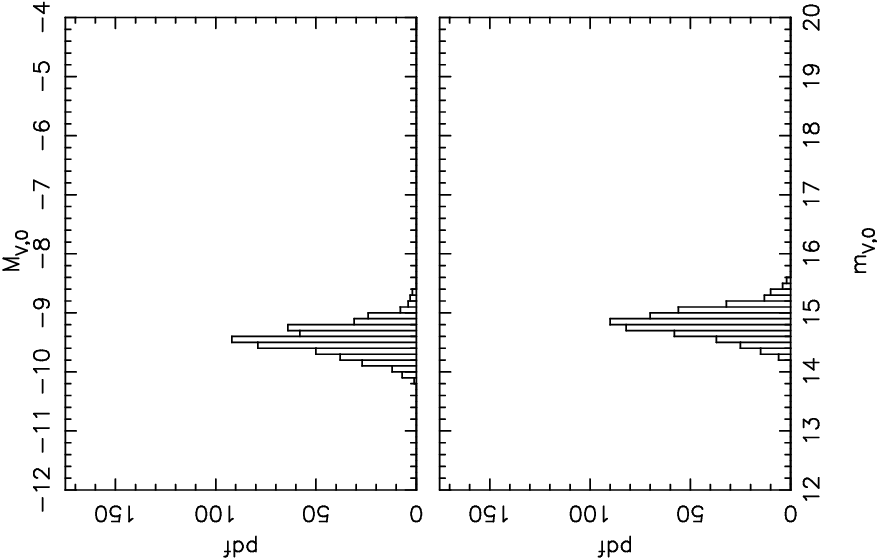}
\caption{\label{posterior_pdf_AndIII}Same as Figure~\ref{posterior_pdf_AndI} for And~III. In the central panel comparing the radial density profile and the favored model, hollow circles represent regions affected by crowding that was not taken into account for the structural parameter inference.}
\end{center}
\end{figure*}

And~III is much fainter than the previous two dwarf galaxies but was nevertheless found during the same analysis \citep{vandenbergh72}. The high central density of this fairly compact and reasonably bright dwarf galaxy leads to some crowding near its center in PAndAS. We consequently mask out an elliptical region of $1.2'$ when we infer the structural properties. The density estimates from this region are shown as hollow circles in the profile panel of Figure~\ref{posterior_pdf_AndIII} ($r_h=2.0\pm0.2'$, $\epsilon = 0.59\pm0.04$, $m_{V,0}=14.8\pm0.2$). A chip gap also masks out a relatively large region in the densest parts of the dwarf galaxy but the other panels of the figure nevertheless show well-behaved PDFs and our results are consistent with those of \citet[][$r_h=1.7\pm0.1'$, $\epsilon = 0.52\pm0.02$, $m_{V,0} = 14.4\pm0.3$]{mcconnachie06b}, once again from shallower INT/WFC imaging.

\subsection{And V}
\begin{figure*}
\begin{center}
\includegraphics[width=0.4\hsize,angle=270]{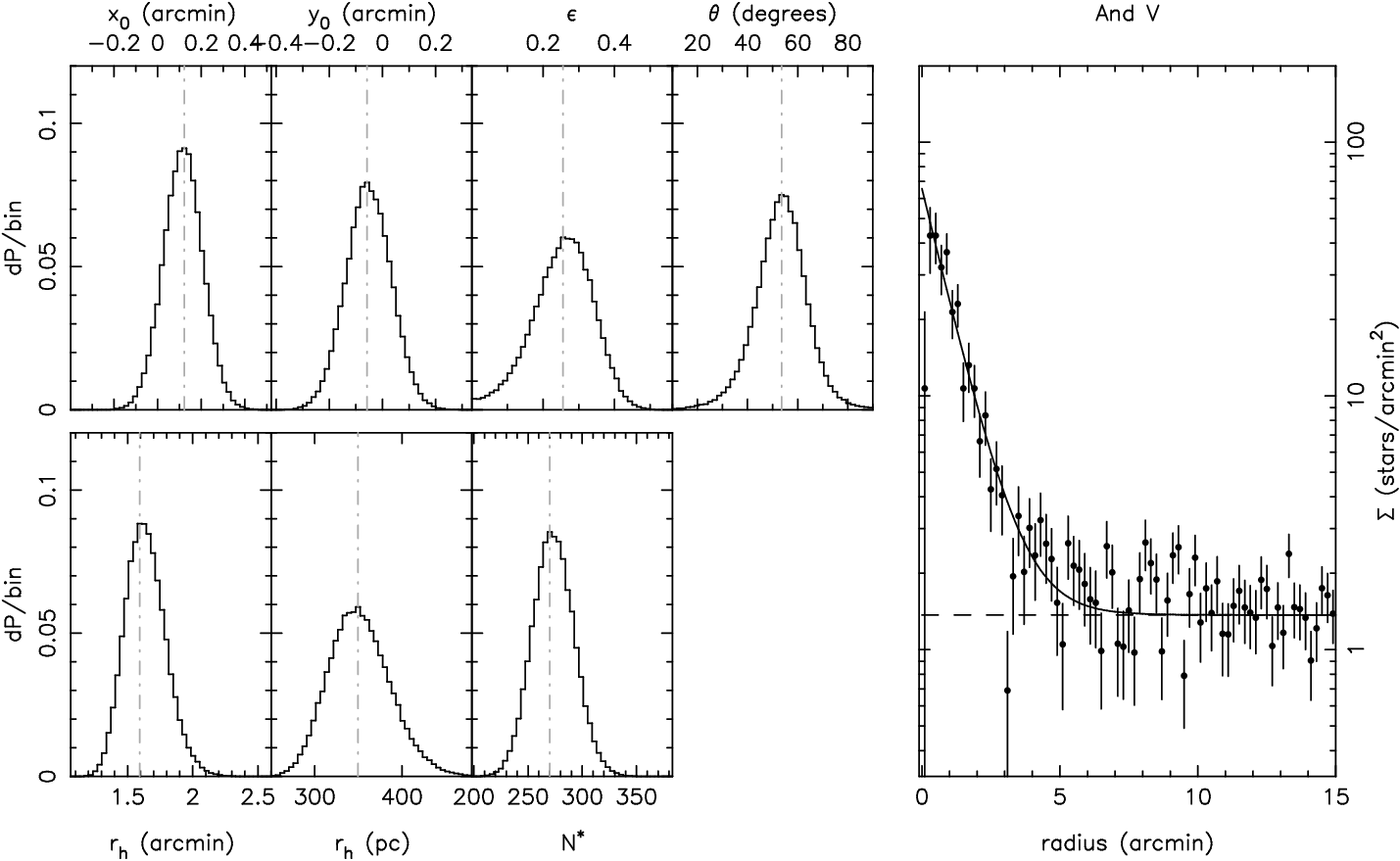}\hspace{1cm}
\includegraphics[width=0.4\hsize,angle=270]{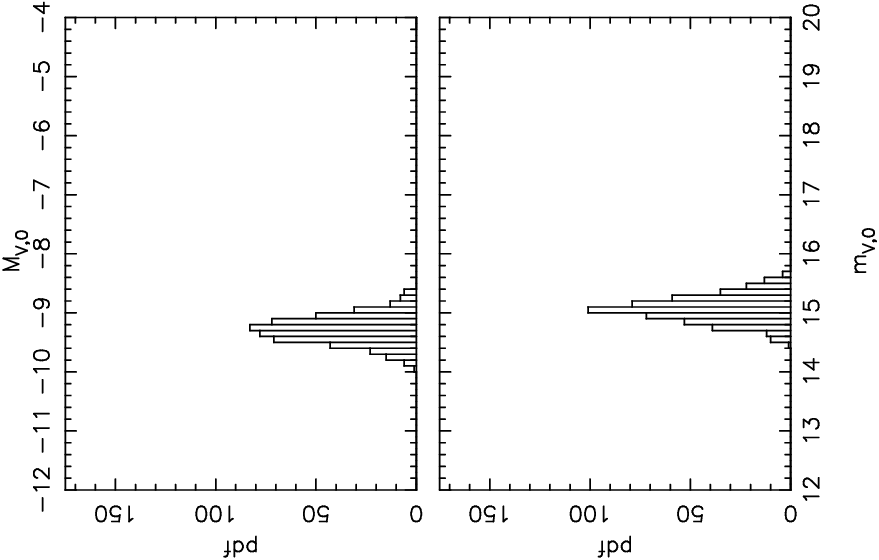}
\caption{\label{posterior_pdf_AndV}Same as Figure~\ref{posterior_pdf_AndI} for And~V.}
\end{center}
\end{figure*}

And~V was found in digitized photographic plates by \citet{armandroff98}. Our results for this dwarf galaxy are presented in Figure~\ref{posterior_pdf_AndV} and paint a very similar picture to the one obtained for And~III ($r_h=1.6^{+0.2}_{-0.1}$$'$, $\epsilon = 0.26^{+0.09}_{-0.07}$, $m_{V,0}=15.1\pm0.2$). Here as well, our results are consistent with those of \citet[][$r_h=1.4\pm0.1'$, $\epsilon=0.18\pm0.05$, $m_{V,0}=15.3\pm0.2$]{mcconnachie06b}.

Overall, the comparison of our inference of the properties of the four bright Andromeda satellites And~I, And~II, And~III, and And~V with those from the INT/WFC analysis of \citet{mcconnachie06b} show good agreement, despite the two analyses being performed with completely different techniques and different data sets, thereby giving confidence in the current analysis before turning to fainter and less dense stellar systems.

Our results for And~V are also very consistent with those of \citet{collins11b} from Subaru/SuprimeCam data of similar depth to the PAndAS data ($r_h = 1.3\pm0.1'$, $\epsilon = 0.17\pm0.02$).

\subsection{And IX}
\begin{figure*}
\begin{center}
\includegraphics[width=0.4\hsize,angle=270]{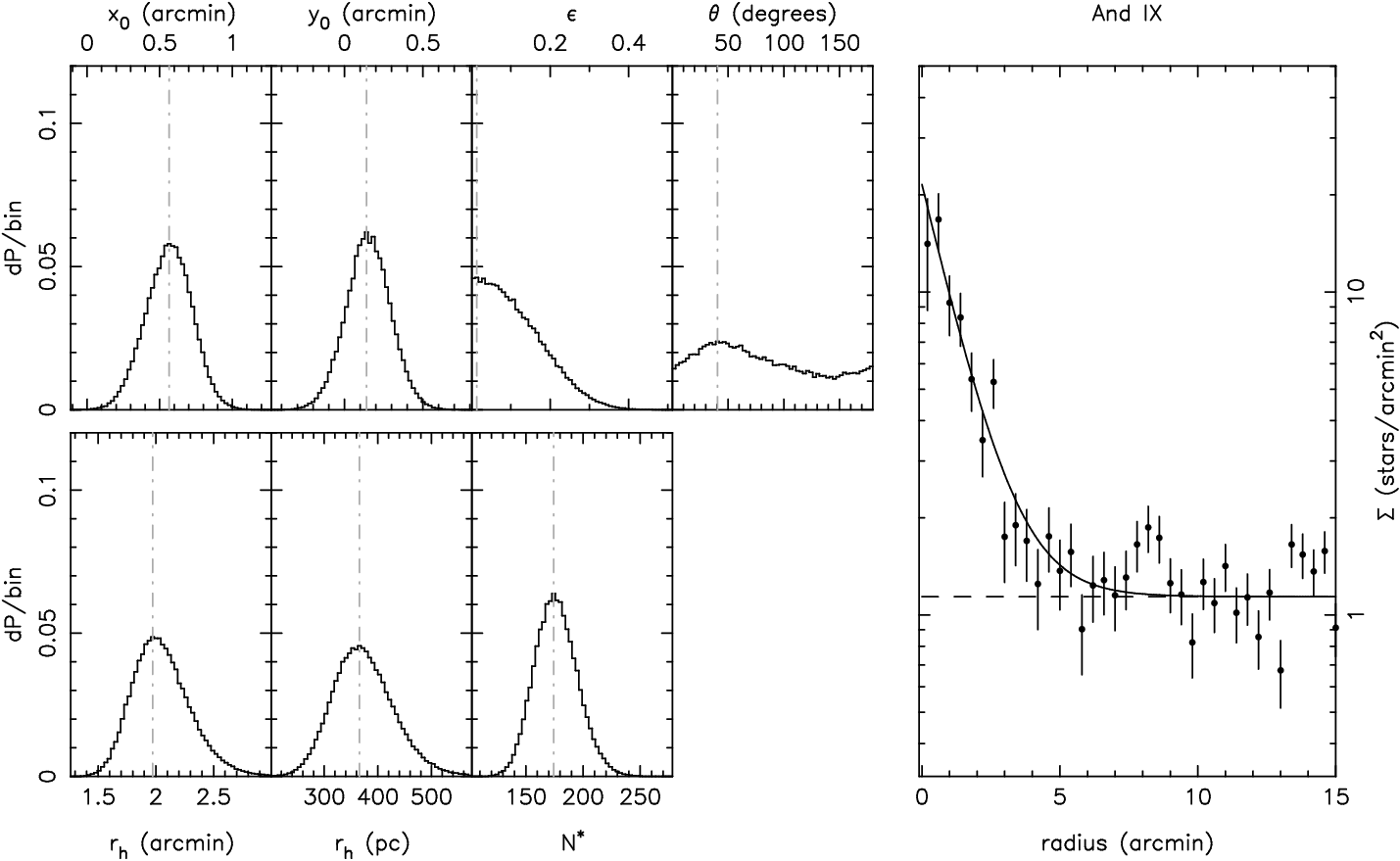}\hspace{1cm}
\includegraphics[width=0.4\hsize,angle=270]{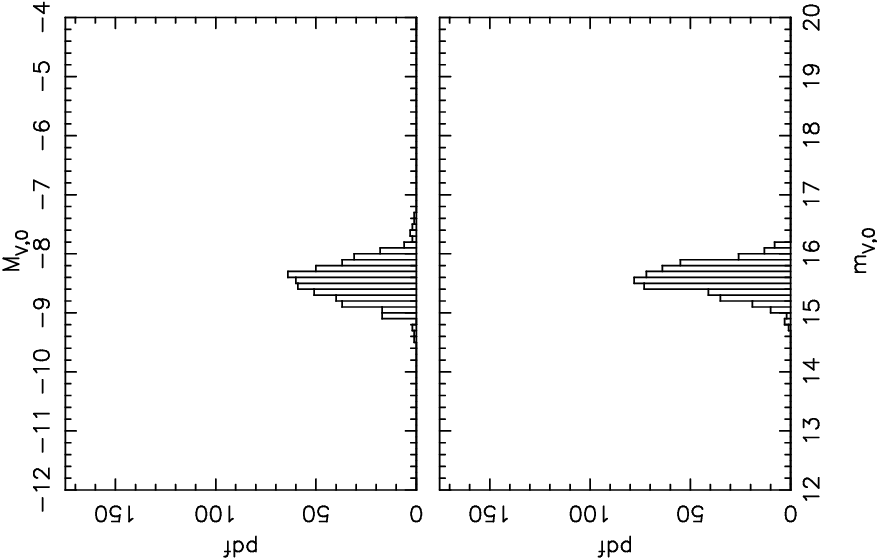}
\caption{\label{posterior_pdf_AndIX}Same as Figure~\ref{posterior_pdf_AndI} for And~IX.}
\end{center}
\end{figure*}

Figure~\ref{posterior_pdf_AndIX} summarizes our results for the fainter And~IX ($r_h=2.0^{+0.3}_{-0.2}$$'$, $\epsilon = 0.00^{+0.16}_{-0.00}$, $M_{V,0}=-8.5\pm0.3$), which was found in the SDSS by \citet{zucker04b}. We infer a round system (explaining the poorly constrained position angle), with a smaller size compared to the previous inference from Subaru/SuprimeCam data of similar depth to the PAndAS data by \citet[][assumption of circular symmetry, $r_h=2.5\pm0.1'$, $M_{V,0} = -8.1^{+0.4}_{-0.1}$]{collins10}. It should however be noted that this previous analysis binned the data and relied on a smaller field of view, both of which can impact size estimates (M08, \citealt{munoz12}). Our magnitude estimate is slightly brighter, although statistically consistent with that of the two previous analyses.

\subsection{And X}
\label{section_X}
\begin{figure*}
\begin{center}
\includegraphics[width=0.4\hsize,angle=270]{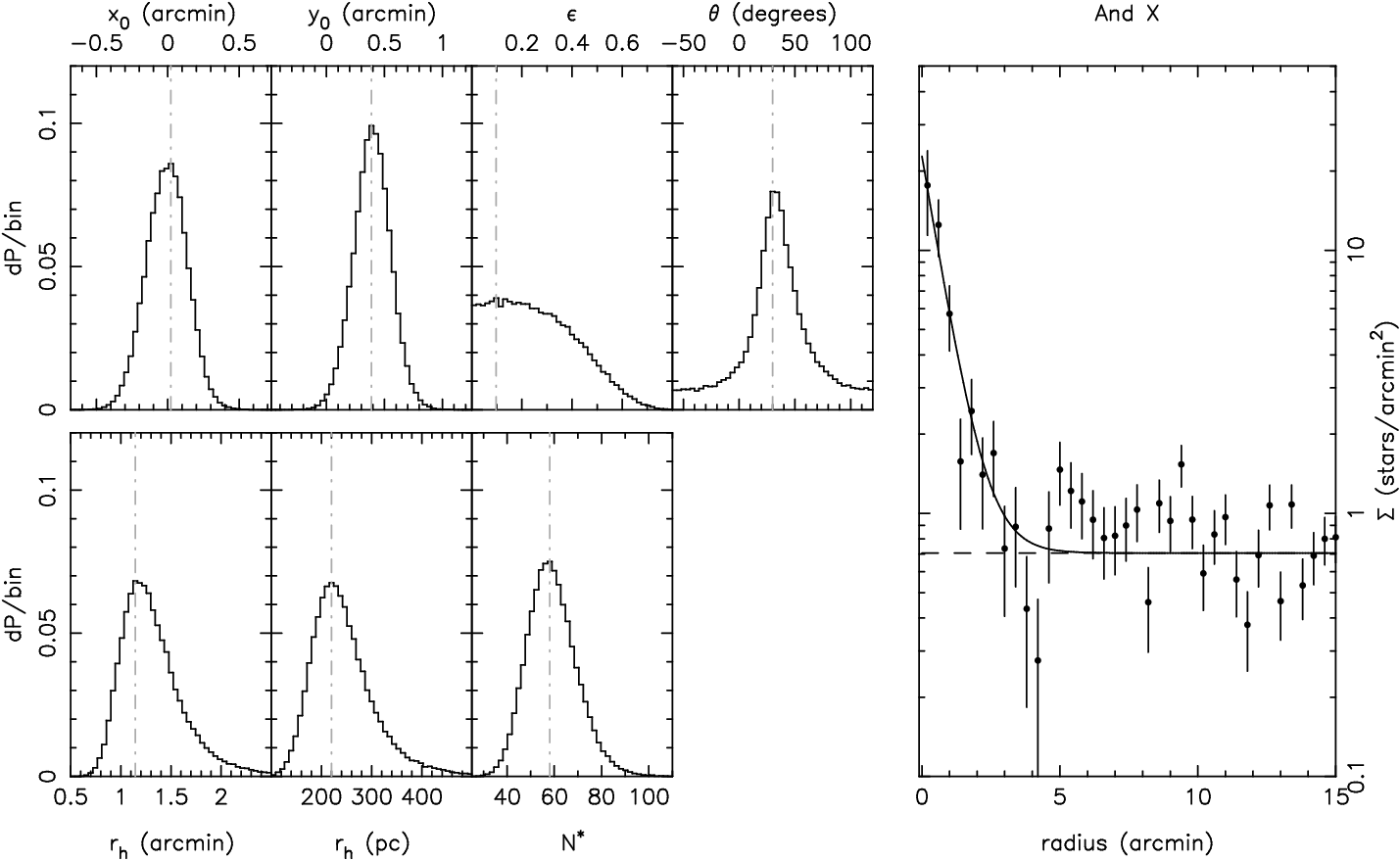}\hspace{1cm}
\includegraphics[width=0.4\hsize,angle=270]{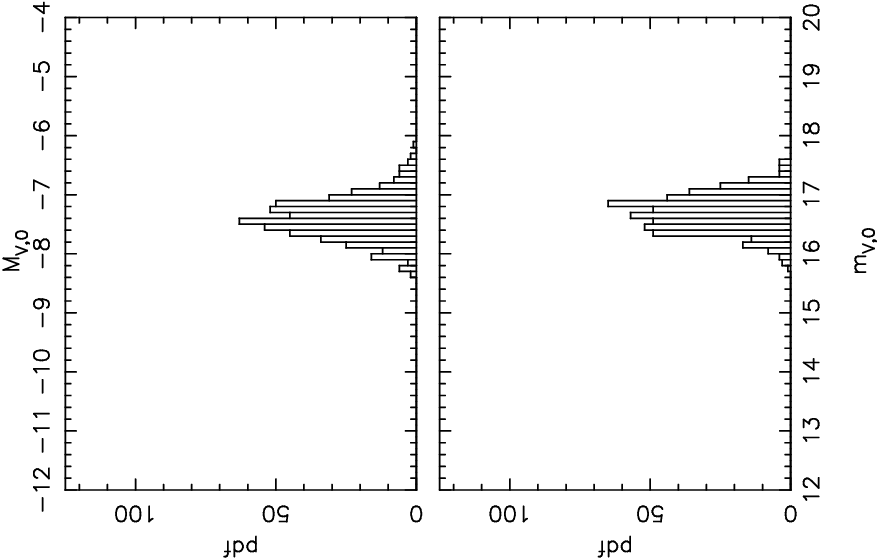}
\caption{\label{posterior_pdf_AndX}Same as Figure~\ref{posterior_pdf_AndI} for And~X.}
\end{center}
\end{figure*}

The results for And~X, discovered in the SDSS by \citet{zucker07}, are summarized in Figure~\ref{posterior_pdf_AndX} ($r_h=1.1^{+0.4}_{-0.2}$$'$, $\epsilon = 0.10^{+0.34}_{-0.10}$, $M_{V,0}=-7.4\pm0.3$) and are consistent with the structural parameters and magnitudes presented in \citet{brasseur11a} from LBC/LBT photometry that is deeper than the PAndAS data ($r_h = 1.3\pm0.1'$, $\epsilon = 0.44\pm0.06$, $M_{V,0} = -7.4\pm0.9$). We find that And~X is fairly round, but the uncertainties on our ellipticity inference fully include the larger and more accurate ellipticity measurement obtained from the LBC/LBT data with a method similar to the one described in this paper.

It should be noted that we enforced a stricter prior on the half-light radius, with $r_\mathrm{max}=6'$, to prevent a small fraction of the MCMC iterations from wandering off toward models with large $r_h$ that clearly do not represent the structure of And~X but likely track the structure in the field. And~X lies in a region with some M31 halo stellar substructure, leading to a contamination level that is not perfectly flat.

\subsection{And XI}
\begin{figure*}
\begin{center}
\includegraphics[width=0.4\hsize,angle=270]{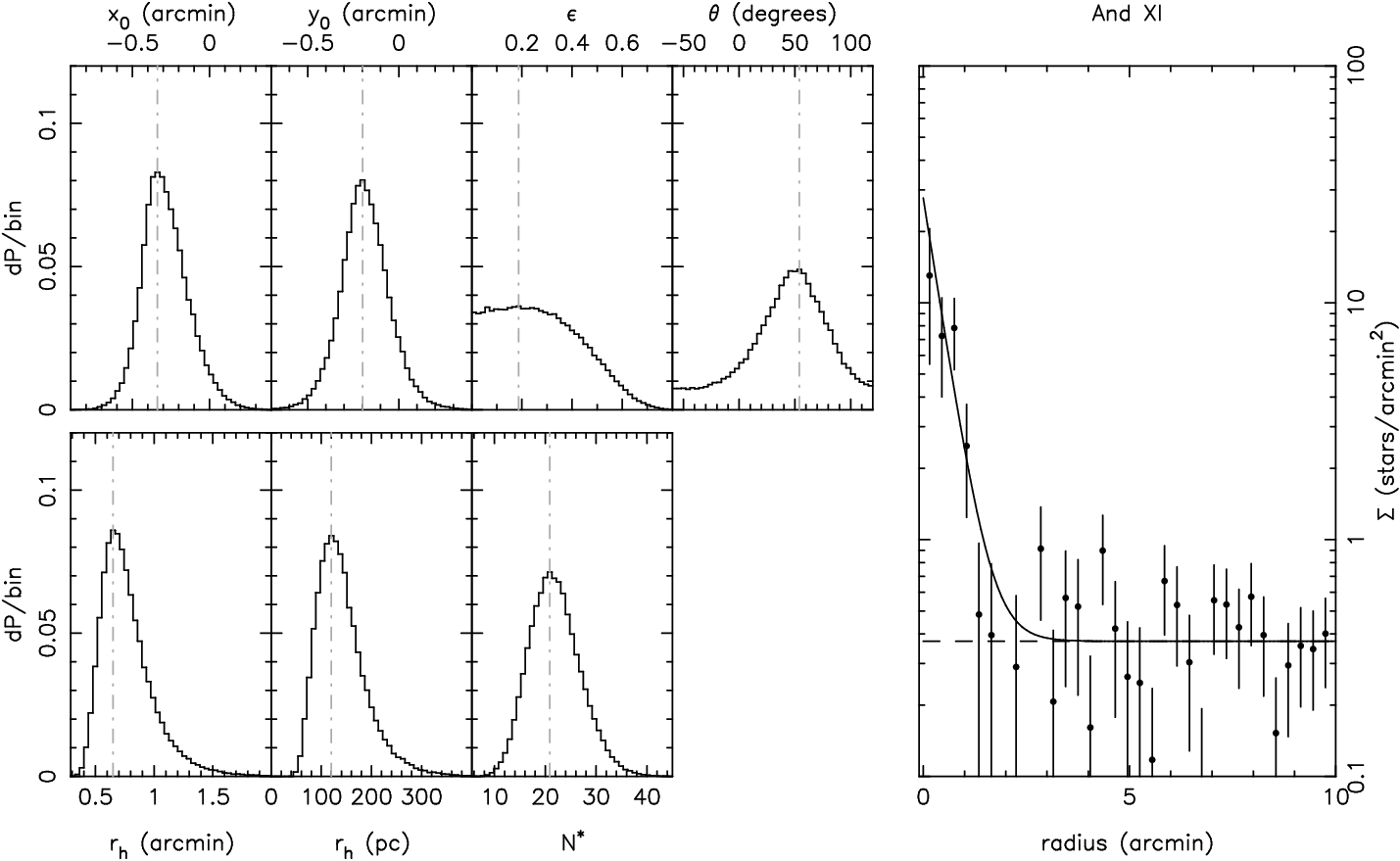}\hspace{1cm}
\includegraphics[width=0.4\hsize,angle=270]{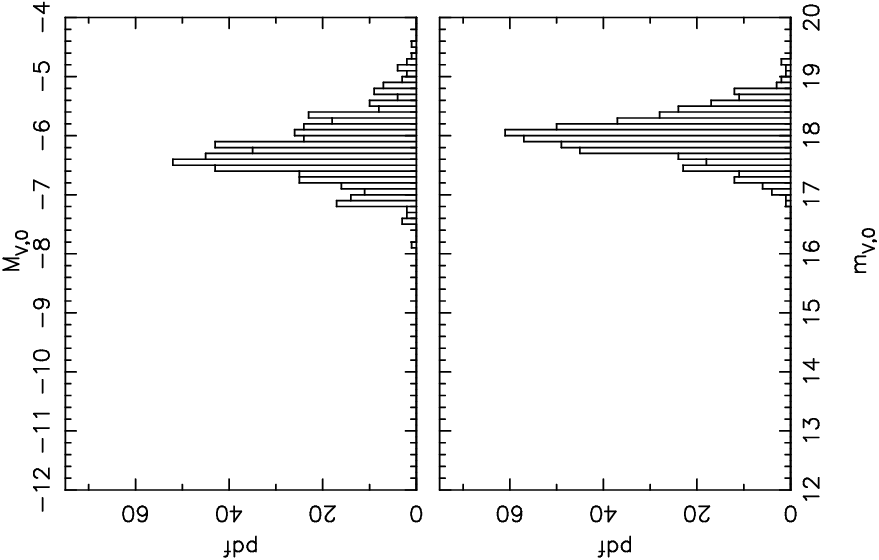}
\caption{\label{posterior_pdf_AndXI}Same as Figure~\ref{posterior_pdf_AndI} for And~XI.}
\end{center}
\end{figure*}

And~XI was discovered among the first set of pre-PAndAS MegaCam/CFHT observations \citep{martin06b} and later characterized in more depth by \citet{collins10}. Our results, summarized in Figure~\ref{posterior_pdf_AndXI} ($r_h=0.6\pm0.2'$, $\epsilon = 0.19^{+0.28}_{-0.19}$, $M_{V,0}=-6.3^{+0.6}_{-0.4}$) are consistent with those of this latter paper, based on Subaru/SuprimeCam data deeper than the PAndAS data (assumption of circular symmetry, $r_h = 0.7\pm0.1'$, $M_{V,0} = -6.9^{+0.5}_{-0.1}$). The width of the PDF for $m_V$ highlights that CMD `shot-noise' is important for this dwarf galaxy and must be taken into account.

\subsection{And XII}
\begin{figure*}
\begin{center}
\includegraphics[width=0.4\hsize,angle=270]{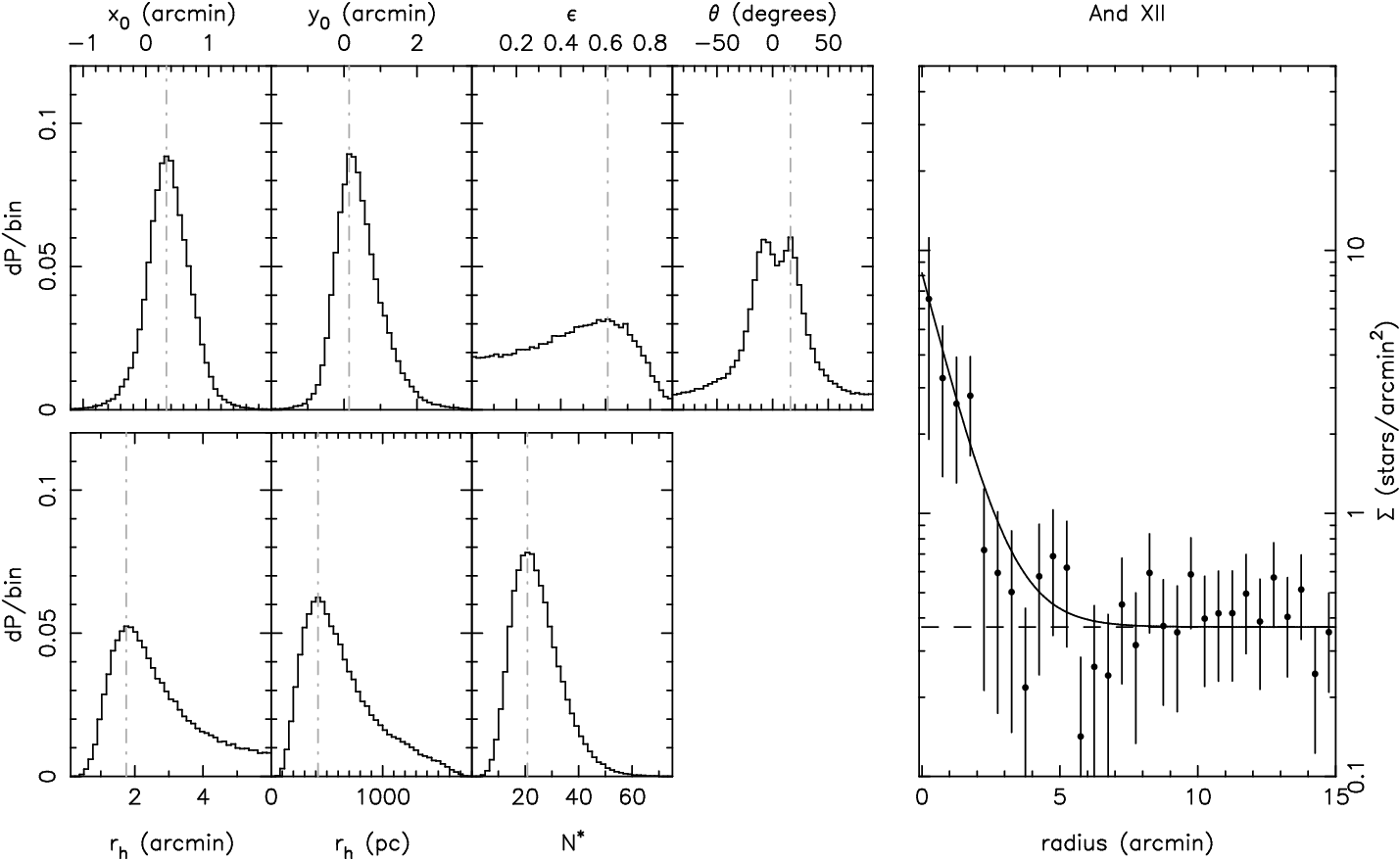}\hspace{1cm}
\includegraphics[width=0.4\hsize,angle=270]{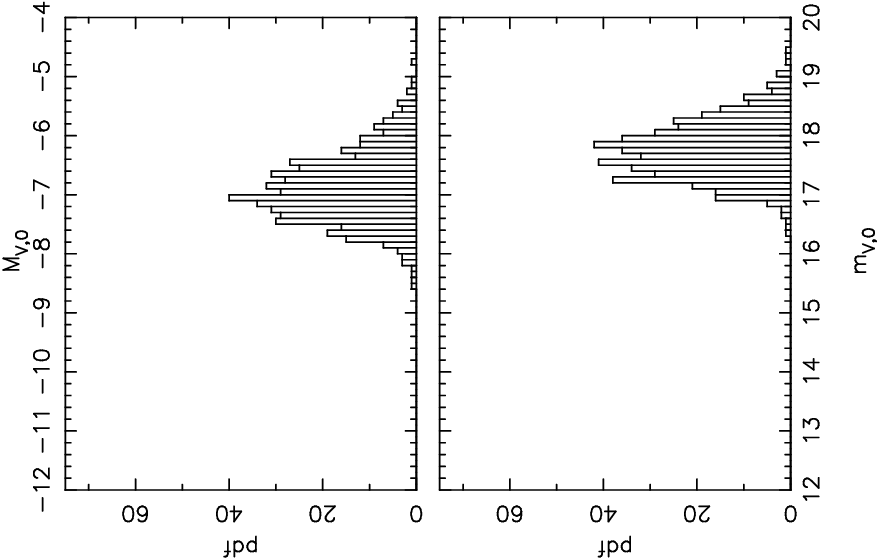}
\caption{\label{posterior_pdf_AndXII}Same as Figure~\ref{posterior_pdf_AndI} for And~XII.}
\end{center}
\end{figure*}

Like And~XI, And~XII was discovered by \citet{martin06b} and characterized more thoroughly from deeper Subaru/SuprimeCam data by \citet[][assumption of circular symmetry, $r_h = 1.1\pm0.2'$, $M_{V,0} = -6.4^{+0.1}_{-0.5}$]{collins10}. It is one of the faintest dwarf galaxies found around M31 and, as such, proves difficult to analyse with PAndAS data alone. This is reflected in the wide PDFs of Figure~\ref{posterior_pdf_AndXII}, in particular for the half-light radius ($r_h=1.8^{+1.2}_{-0.7}$$'$, $\epsilon = 0.61^{+0.16}_{-0.48}$, $M_{V,0}=-7.0^{+0.7}_{-0.5}$). In addition, we were forced to use a stricter prior on this parameter ($r_\mathrm{max}=6.0'$) to prevent the chain from diverging. Despite these limitations, the PAndAS results are fully consistent with the results of \citet{collins10}. CMD `shot-noise' is once again significant for And~XII.

\subsection{And XIII}
\begin{figure*}
\begin{center}
\includegraphics[width=0.4\hsize,angle=270]{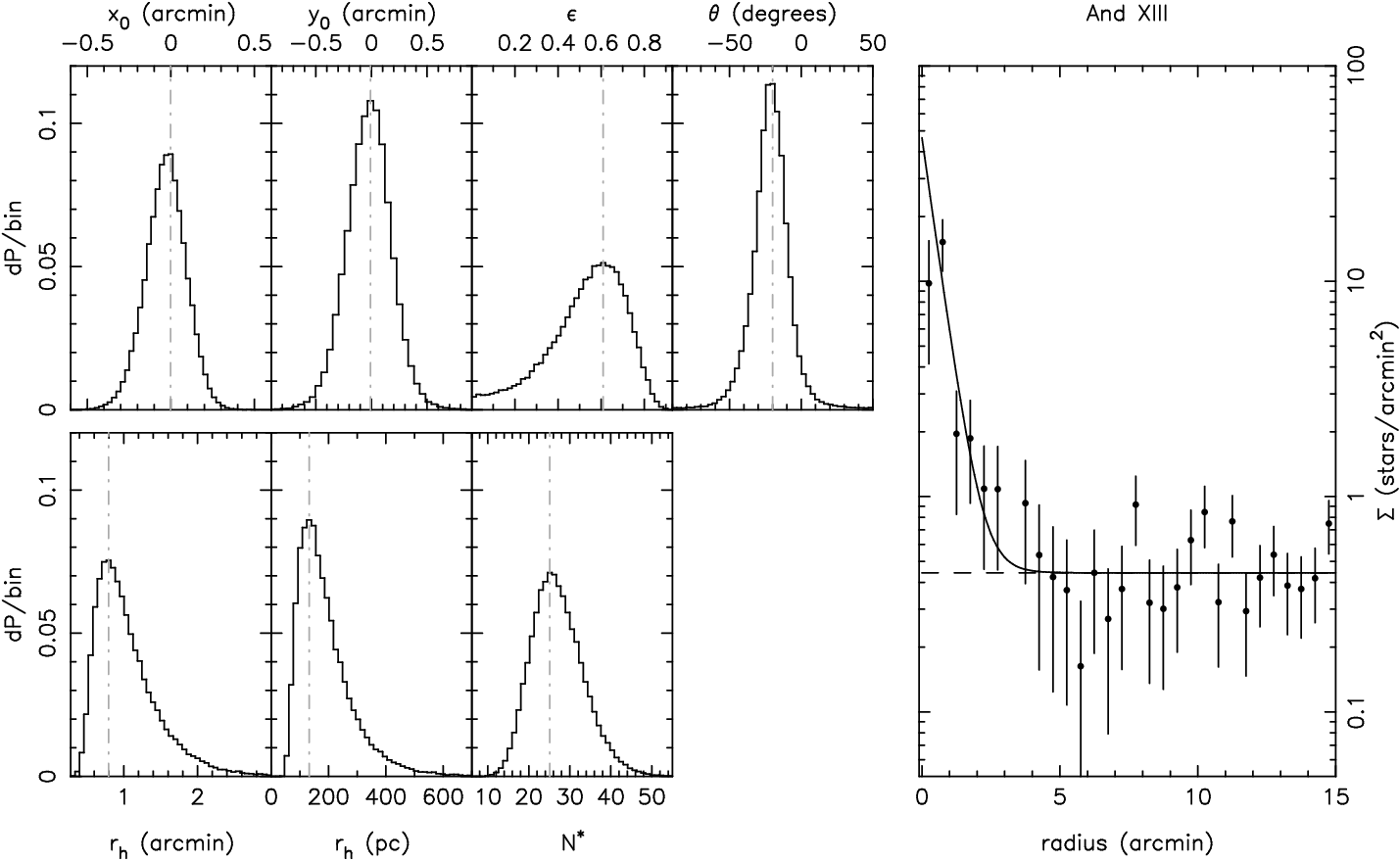}\hspace{1cm}
\includegraphics[width=0.4\hsize,angle=270]{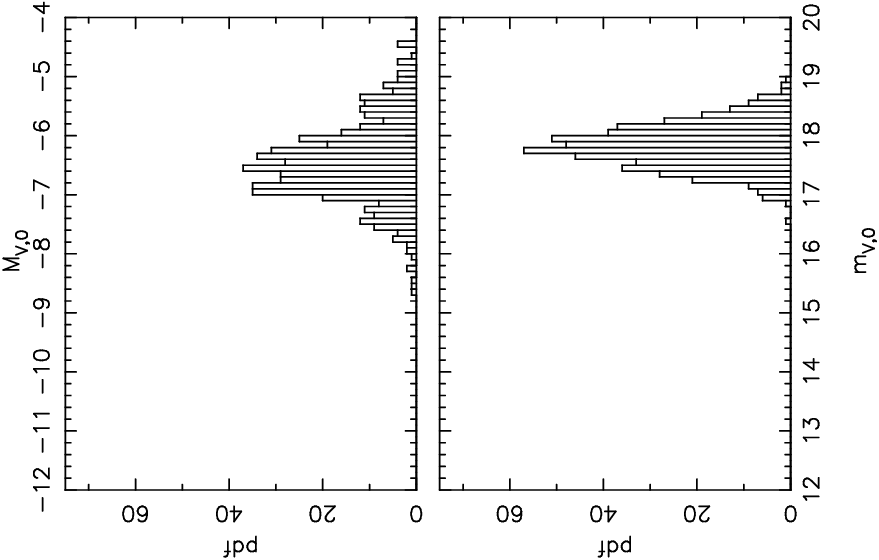}
\caption{\label{posterior_pdf_AndXIII}Same as Figure~\ref{posterior_pdf_AndI} for And~XIII.}
\end{center}
\end{figure*}

Like the previous two satellites, And~XIII was discovered by \citet{martin06b} and studied by \citet{collins10} from deeper SuprimeCam/Subaru photometry. Our conclusions, illustrated in Figure~\ref{posterior_pdf_AndXIII} ($r_h=0.8^{+0.4}_{-0.3}$$'$, $\epsilon = 0.61^{+0.14}_{-0.20}$, $M_{V,0}=-6.5^{+0.7}_{-0.5}$), are fully consistent with those of the SuprimeCam analysis (assumption of circular symmetry, $r_h =0.7\pm0.1'$, $M_V = -6.7^{+0.4}_{-0.1}$). CMD `shot-noise' is significant for And~XIII and at the same level as for And~XI and And~XII. This is not surprising since these three dwarf galaxies share very similar properties, with $-7.0 \simlt M_{V,0} \simlt -6.3$.

\subsection{And XIV}
\begin{figure*}
\begin{center}
\includegraphics[width=0.4\hsize,angle=270]{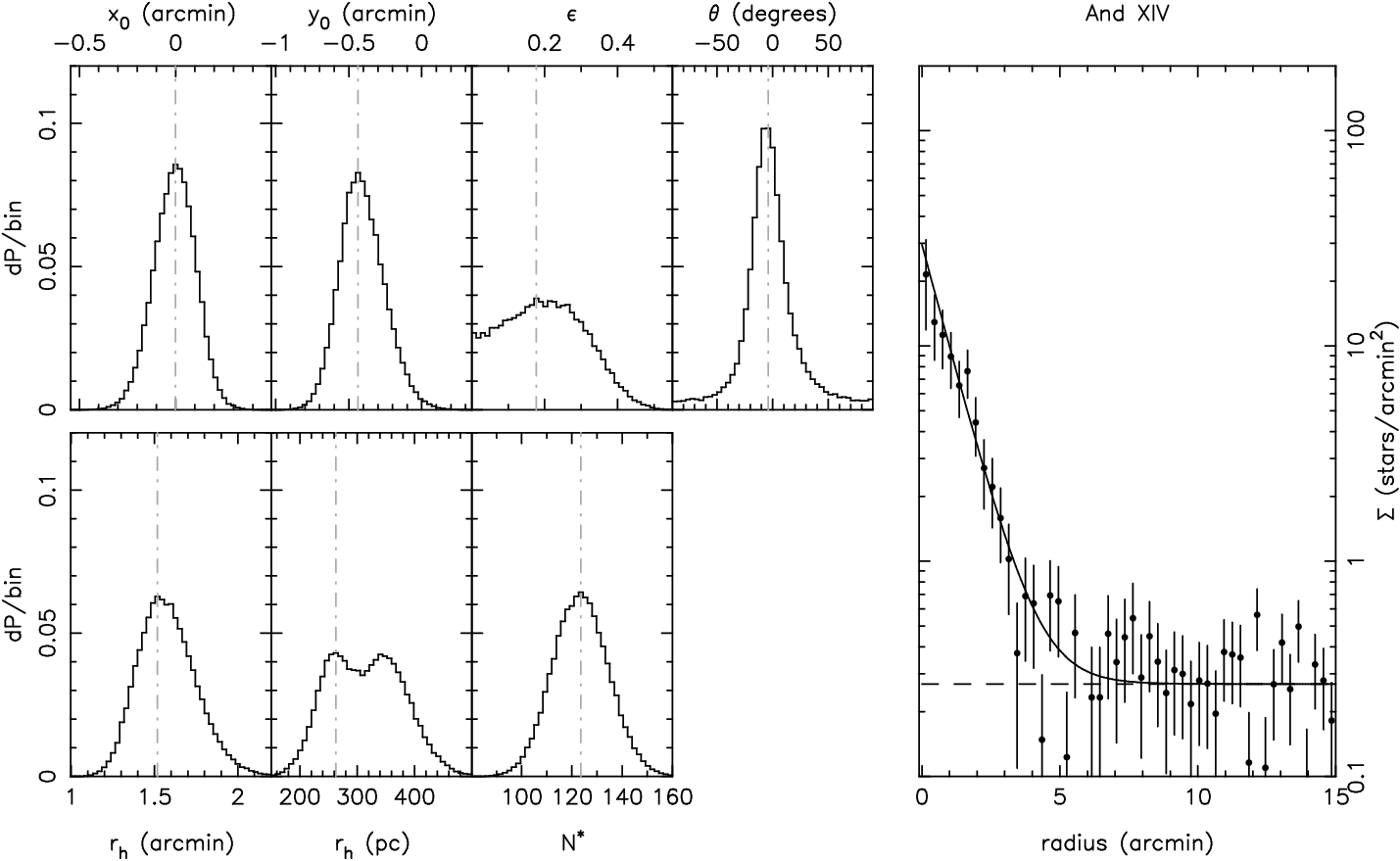}\hspace{1cm}
\includegraphics[width=0.4\hsize,angle=270]{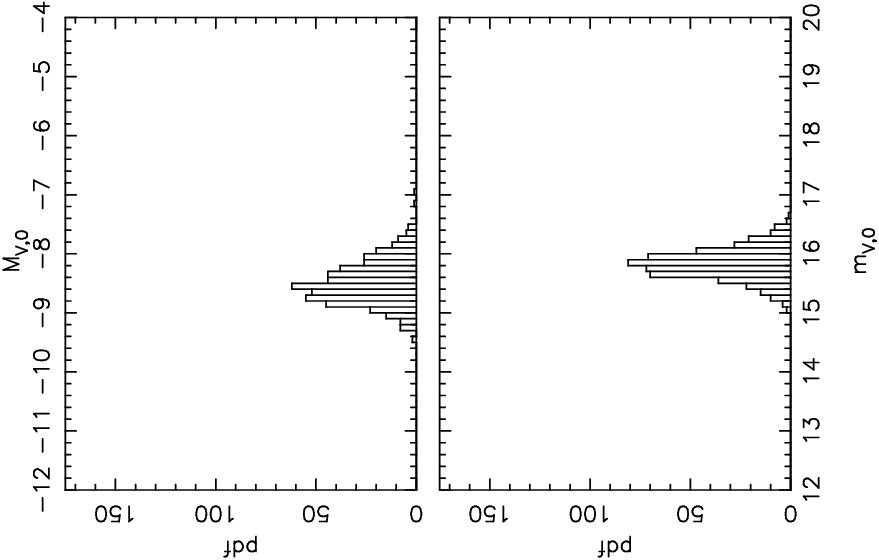}
\caption{\label{posterior_pdf_AndXIV}Same as Figure~\ref{posterior_pdf_AndI} for And~XIV.}
\end{center}
\end{figure*}

And~XIV was serendipitously discovered by \citet{majewski07} before it was observed in PAndAS. For the analysis of this dwarf galaxy, we need to mask a large background galaxy and a bright foreground star, which, combined, represent a sizable fraction of the region within $2r_h$ of the centroid. And~XIV is however dense enough that we derive good constraints on its structural parameters and magnitudes, as shown in Figure~\ref{posterior_pdf_AndXIV} ($r_h=1.5\pm0.2'$, $\epsilon = 0.17^{+0.16}_{-0.17}$, $m_{V,0}=15.8\pm0.3$). We are in very good agreement with the values of \citet{majewski07}, recalculated by \citet{mcconnachie12} for the parameters we use in our analysis ($r_h = 1.7\pm0.8$, $\epsilon = 0.31\pm0.09$, $m_{V,0} \sim 15.9$). It should be noted that the complex distance modulus PDF \citet{aconn12} obtained for And~XIV translates into complex physical $r_h$ and $M_V$ PDFs.

\subsection{And XV}
\begin{figure*}
\begin{center}
\includegraphics[width=0.4\hsize,angle=270]{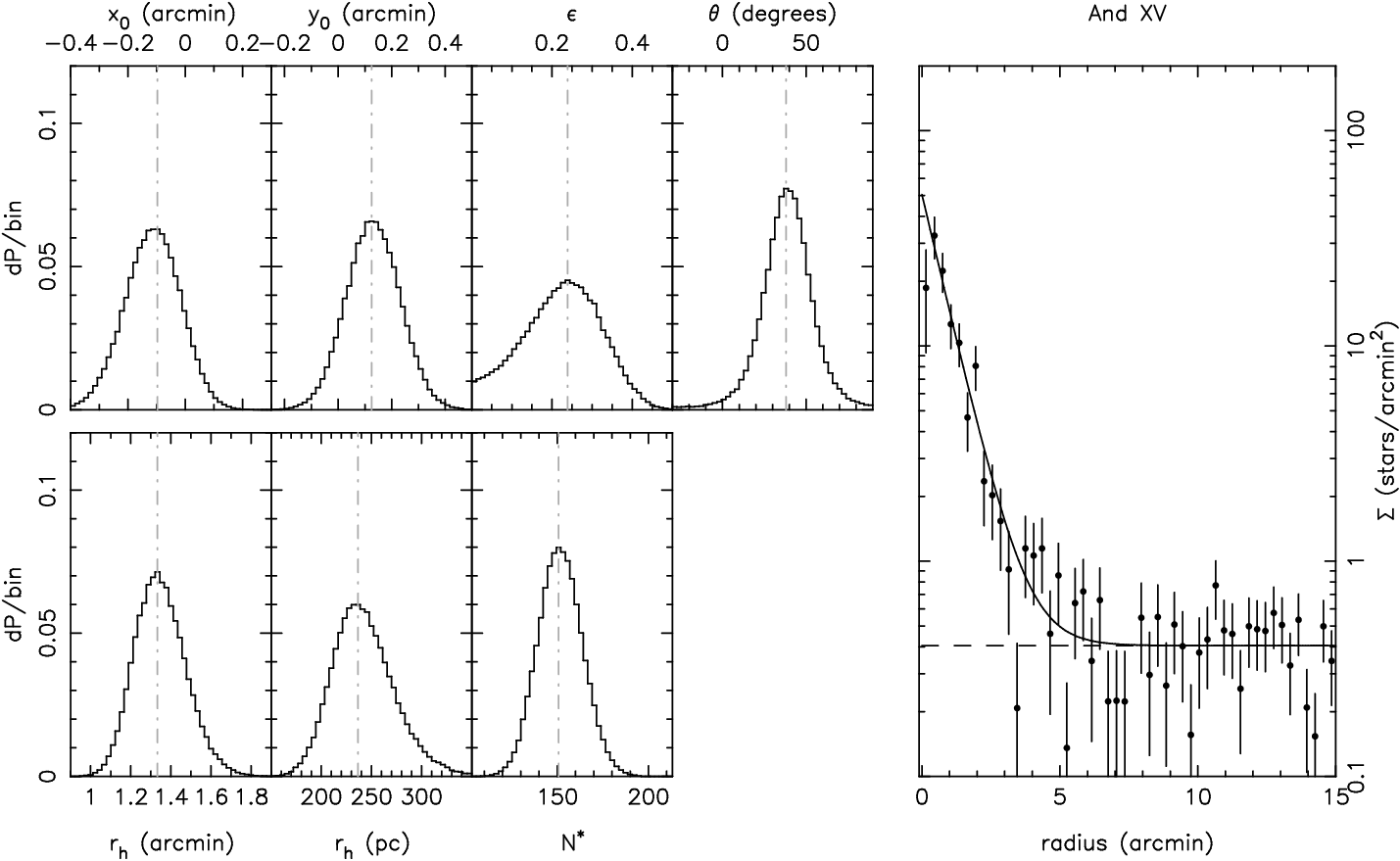}\hspace{1cm}
\includegraphics[width=0.4\hsize,angle=270]{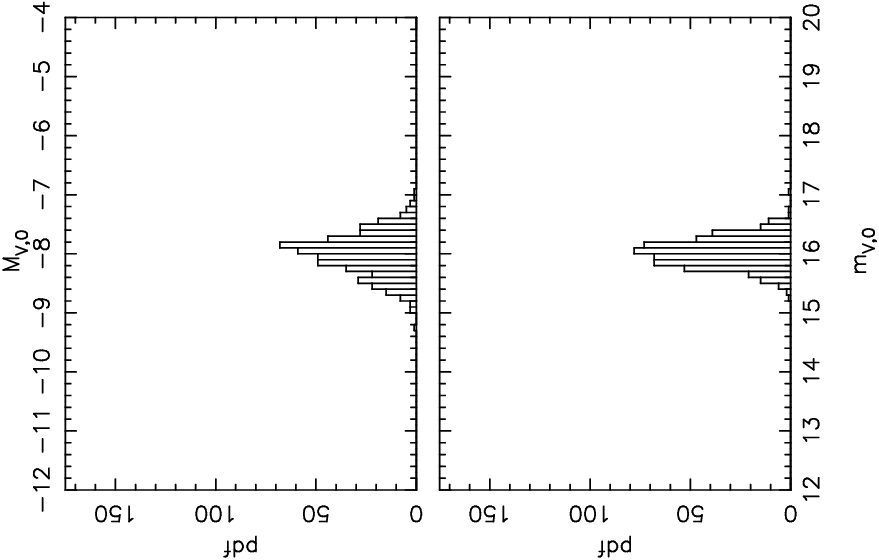}
\caption{\label{posterior_pdf_AndXV}Same as Figure~\ref{posterior_pdf_AndI} for And~XV.}
\end{center}
\end{figure*}

And~XV was found within the pre-PAndAS MegaCam/CFHT data \citep{ibata07}, whose analysis assumed circular symmetry and yielded $r_h = 1.2\pm0.1'$ and $m_{V,0} = 14.3\pm0.3$. Our results (Figure~\ref{posterior_pdf_AndXV}) include the first inference of the ellipticity for this dwarf galaxy, which we find to be somewhat elliptical ($r_h=1.3\pm0.1'$, $\epsilon = 0.24\pm0.10$, $m_{V,0}=16.0\pm0.2$). The apparent magnitude we derive is significantly fainter than that of \citet{ibata07} and is not related to their assumption of spherical symmetry since values of $N^*$ only increase by $\sim10$ percent if we enforce a circular model. A comparison of And~XV's CMD with those of And~IX and And~X (Figure~\ref{CMDs}), all three located at similar distances ($24.0<m-M<24.4$; \citealt{aconn12}), shows that the RGB of And~XV has a density bracketed by those of the two other dwarf galaxies. Therefore, their magnitudes should also bracket that of And~XV; this is indeed what we find. The value from \citet{ibata07} would make And~XV brighter than And~III or And~V, which is clearly not supported by Figure~\ref{CMDs}. This implies a flaw in our previous analysis, the origin of which we discuss next.

\subsection{And XVI}
\begin{figure*}
\begin{center}
\includegraphics[width=0.4\hsize,angle=270]{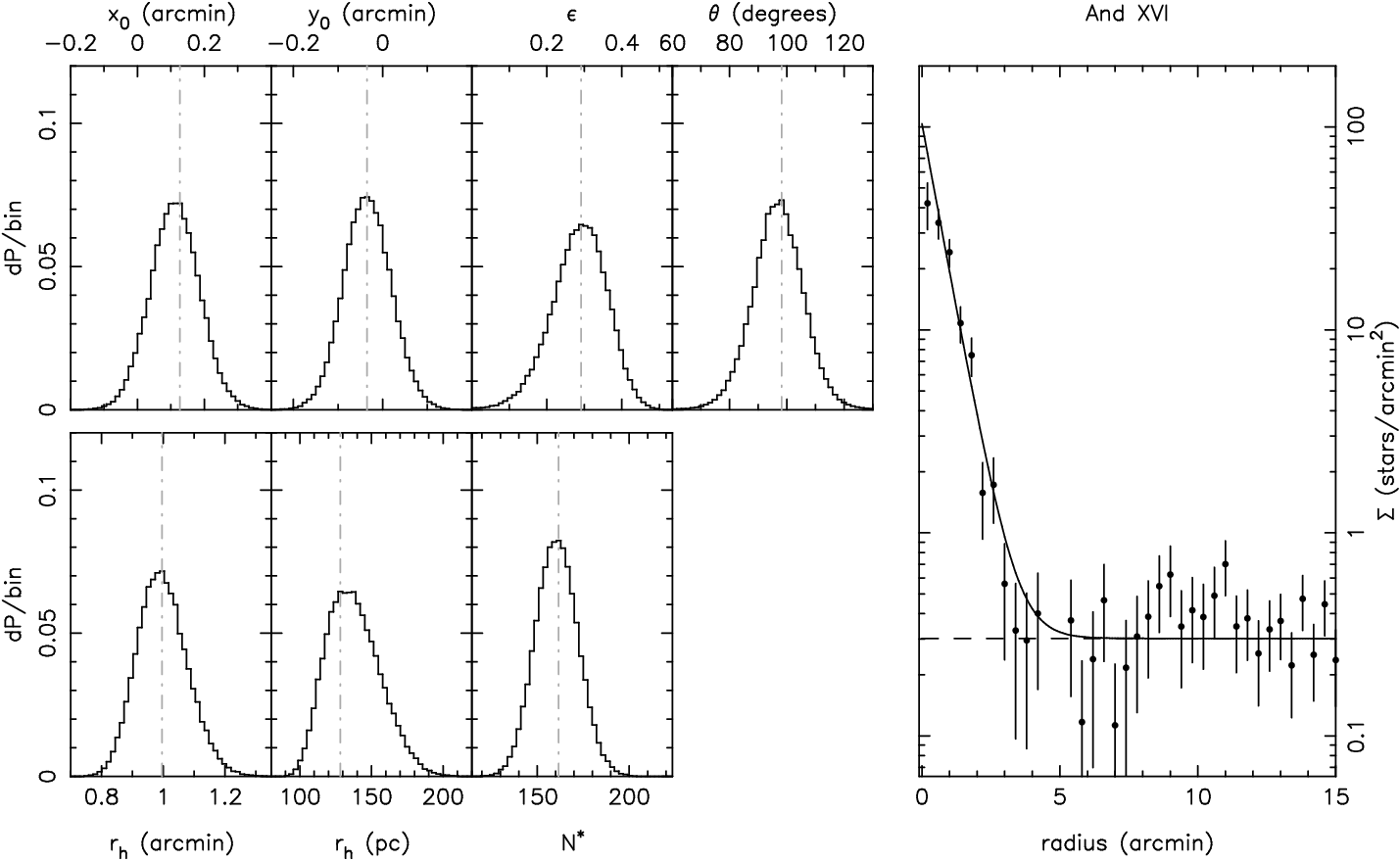}\hspace{1cm}
\includegraphics[width=0.4\hsize,angle=270]{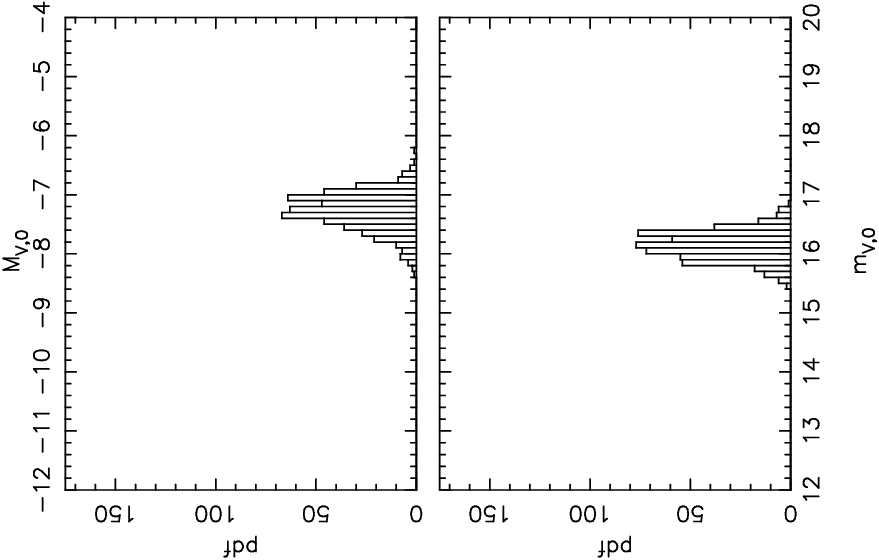}
\caption{\label{posterior_pdf_AndXVI}Same as Figure~\ref{posterior_pdf_AndI} for And~XVI.}
\end{center}
\end{figure*}

Like And~XV, And~XVI was discovered in the pre-PAndAS MegaCam/CFHT study by \citet{ibata07} and, as before, the structural parameters we infer ($r_h=1.0\pm0.1'$, $\epsilon = 0.29\pm0.08$) are compatible with the one they derive, despite their assumption of circular symmetry ($r_h = 0.9\pm0.1'$). However, our estimate of the apparent magnitude of the dwarf galaxy is, again, significantly fainter ($m_{V,0} = 16.1\pm0.3$ vs. $14.4\pm0.3$). Because And~XVI is much closer to us ($m-M = 23.6\pm0.2$), the comparison of its CMD with those of other dwarf galaxies is made harder by the presence of And~XVI's horizontal branch in Figure~\ref{CMDs}. With this in mind and focusing on the brighter regions of the RGB, And~XVI appears to have a slightly lower number of RGB stars than And~XV, closer to what is seen for And~X. This visual comparison is in full agreement with our magnitude estimates for the three dwarf galaxies once we account for the differences in distance ($M_{V,0}^\mathrm{XV}=-8.0^{+0.3}_{-0.4}<M_{V,0}^\mathrm{XVI}=-7.3\pm0.3\simeq M_{V,0}^\mathrm{X}=-7.4\pm0.3$).

\citet{ibata07} derived the magnitudes of And~XV and And~XVI via their stellar counts within the region limited by $r_h$ and by comparison with the same measurement applied to And~III. An investigation of the details of the analysis however reveals that we did not account for the crowding at the center of And~III or the significant PAndAS chip gap near the center of the dwarf galaxy (Figure~\ref{map_III_V}), which likely led us to significantly underestimate the stellar counts within $r_h$ for And~III in the original analysis. This erroneously led us to conclude that And~XV and And~XVI had similar magnitudes to And~III when they are in fact both much fainter.

This effect is now corrected in the new results, which give the picture of a fairly elliptical and faint dwarf galaxy (Figure~\ref{posterior_pdf_AndXVI}).

\subsection{And XVII}
\begin{figure*}
\begin{center}
\includegraphics[width=0.4\hsize,angle=270]{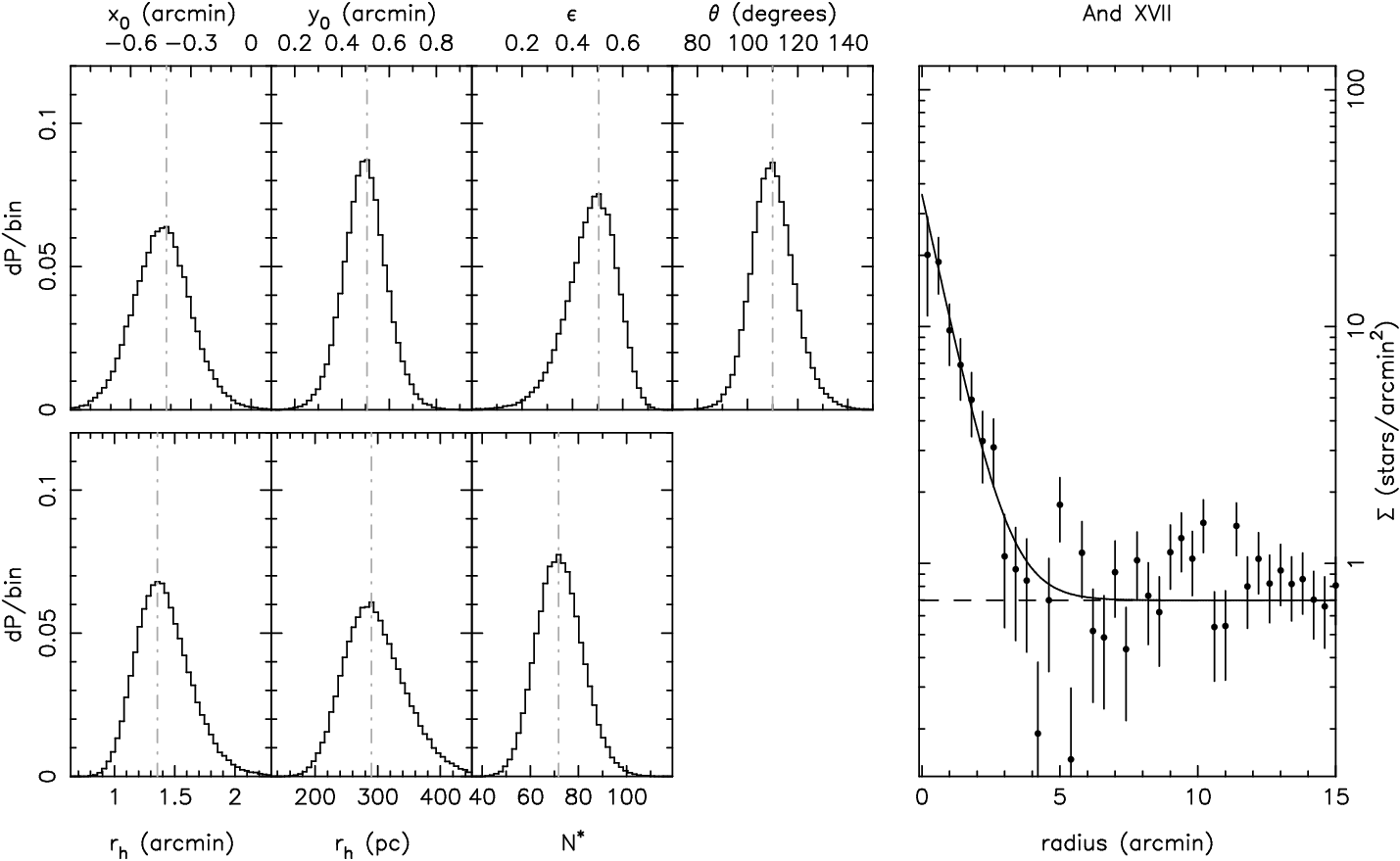}\hspace{1cm}
\includegraphics[width=0.4\hsize,angle=270]{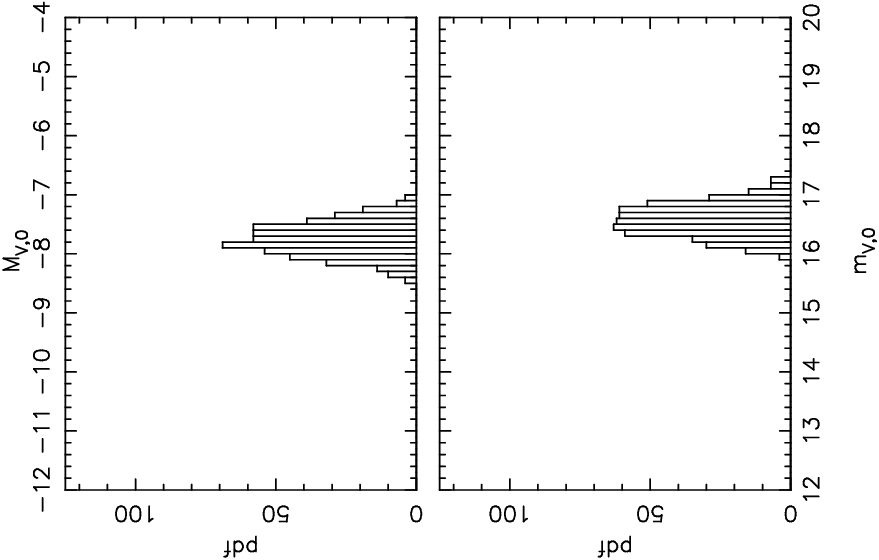}
\caption{\label{posterior_pdf_AndXVII}Same as Figure~\ref{posterior_pdf_AndI} for And~XVII.}
\end{center}
\end{figure*}

Discovered in an INT/WFC survey of the inner regions of the M31 halo \citep{irwin08}, And~XVII was later studied in more detail by \citet[][$r_h=1.2\pm0.1'$, $\epsilon=0.27\pm0.06$, $m_{V,0} = 15.8\pm0.4$]{brasseur11a} with LBC/LBT imaging that is deeper than the PAndAS data. Our results are shown in Figure~\ref{posterior_pdf_AndXVII} ($r_h=1.4\pm0.3'$, $\epsilon = 0.50\pm0.10$, $m_{V,0}=16.6\pm0.3$). Although our size estimate agrees with the one based on the LBC data measured with a similar technique, our ellipticity is larger than before and we derive a fainter apparent magnitude. A comparison of the CMD of And~XVII with that of And~XV seems to favor And~XVII being at most as bright as And~XV and therefore supports our magnitude measurement. It however remains unclear why our ellipticity measurement is $\sim2\sigma$ discrepant with the literature value. It could be related to \citet{brasseur11a} only using a much smaller spatial region for their analysis than we do here, which could potentially bias their results \citep{munoz12}.

\subsection{And XVIII}
\begin{figure*}
\begin{center}
\includegraphics[width=0.4\hsize,angle=270]{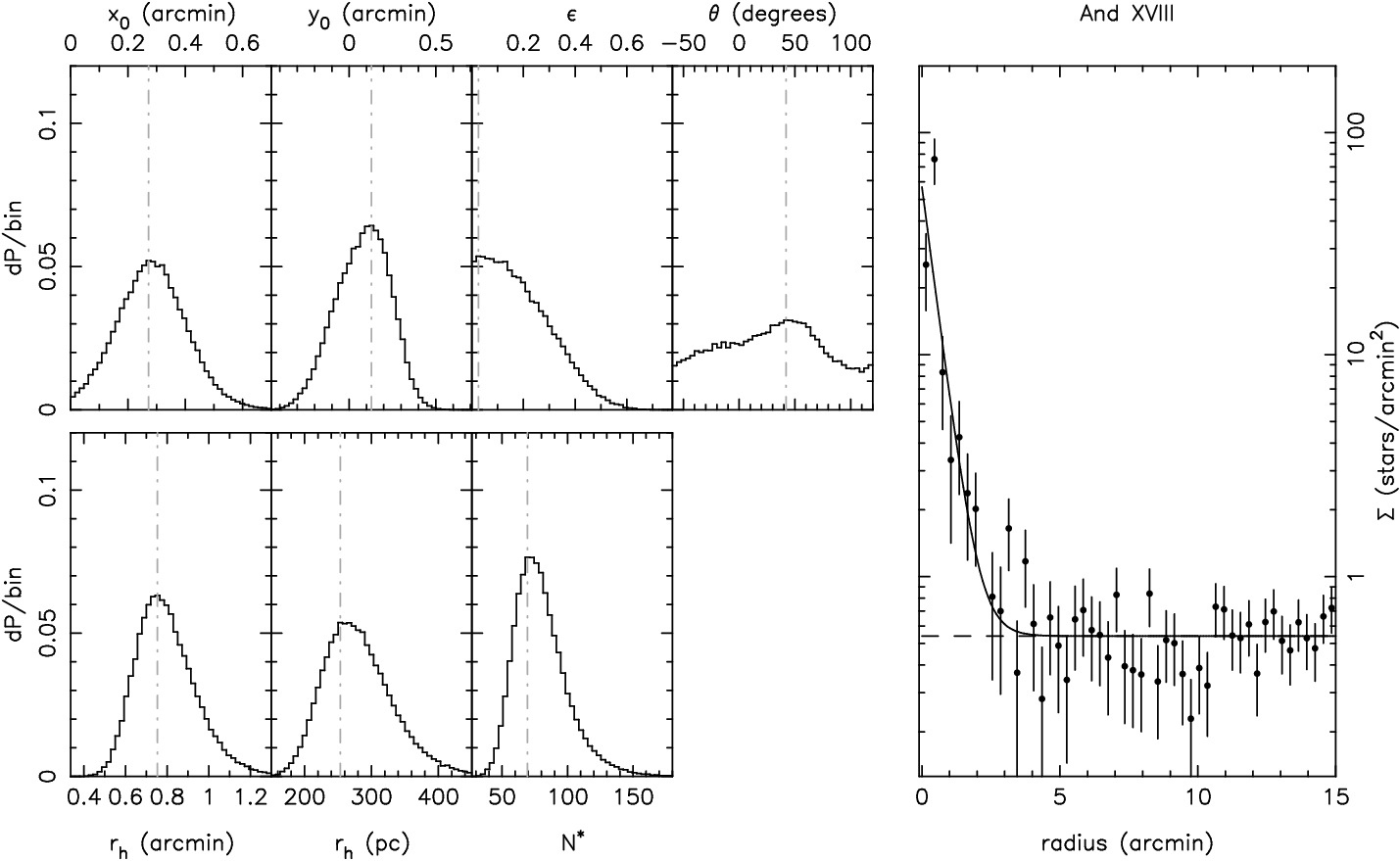}\hspace{1cm}
\includegraphics[width=0.4\hsize,angle=270]{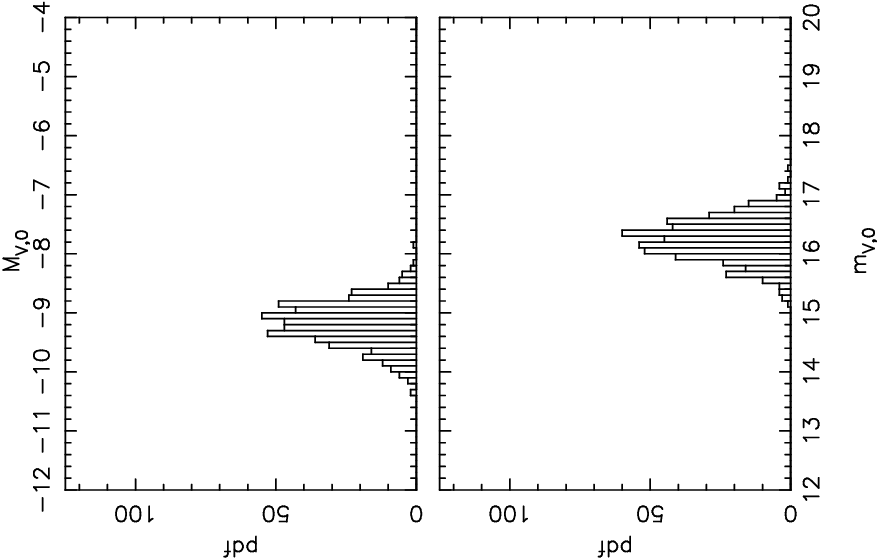}
\caption{\label{posterior_pdf_AndXVIII}Same as Figure~\ref{posterior_pdf_AndI} for And~XVIII.}
\end{center}
\end{figure*}

And~XVIII was found in another batch of pre-PAndAS MegaCam/CFHT data by \citet{mcconnachie08}. This previous analysis differs from the current one in its assumption of circular symmetry in order to cope with half of the dwarf galaxy overlapping a large chip gap (Figure~\ref{map_XVII_XVIII}) and it only provided a faint limit to the system's magnitude for the same reason ($r_h = 0.9\pm0.1'$, $m_{V,0} \leq 16.0$). We do without this assumption here and rely on the masking of this region to derive the global structural properties of And~XVIII ($r_h=0.8\pm0.1'$, $\epsilon = 0.03^{+0.28}_{-0.03}$, $m_{V,0}=16.2\pm0.4$). Although the ellipticity is not well constrained, our size estimate is fully compatible with that found by \citet{mcconnachie08} and we can finally derive the dwarf galaxy's apparent magnitude, which, combined with its large distance behind M31 ($m-M=25.42\pm0.08$; \citealt{aconn12}), implies that And~XVIII is a fairly bright M31 satellite ($M_{V,0}=-9.2^{+0.3}_{-0.4}$). Our results are summarized in Figure~\ref{posterior_pdf_AndXVII}.

\subsection{And XIX}
\label{section_XIX}
\begin{figure*}
\begin{center}
\includegraphics[width=0.4\hsize,angle=270]{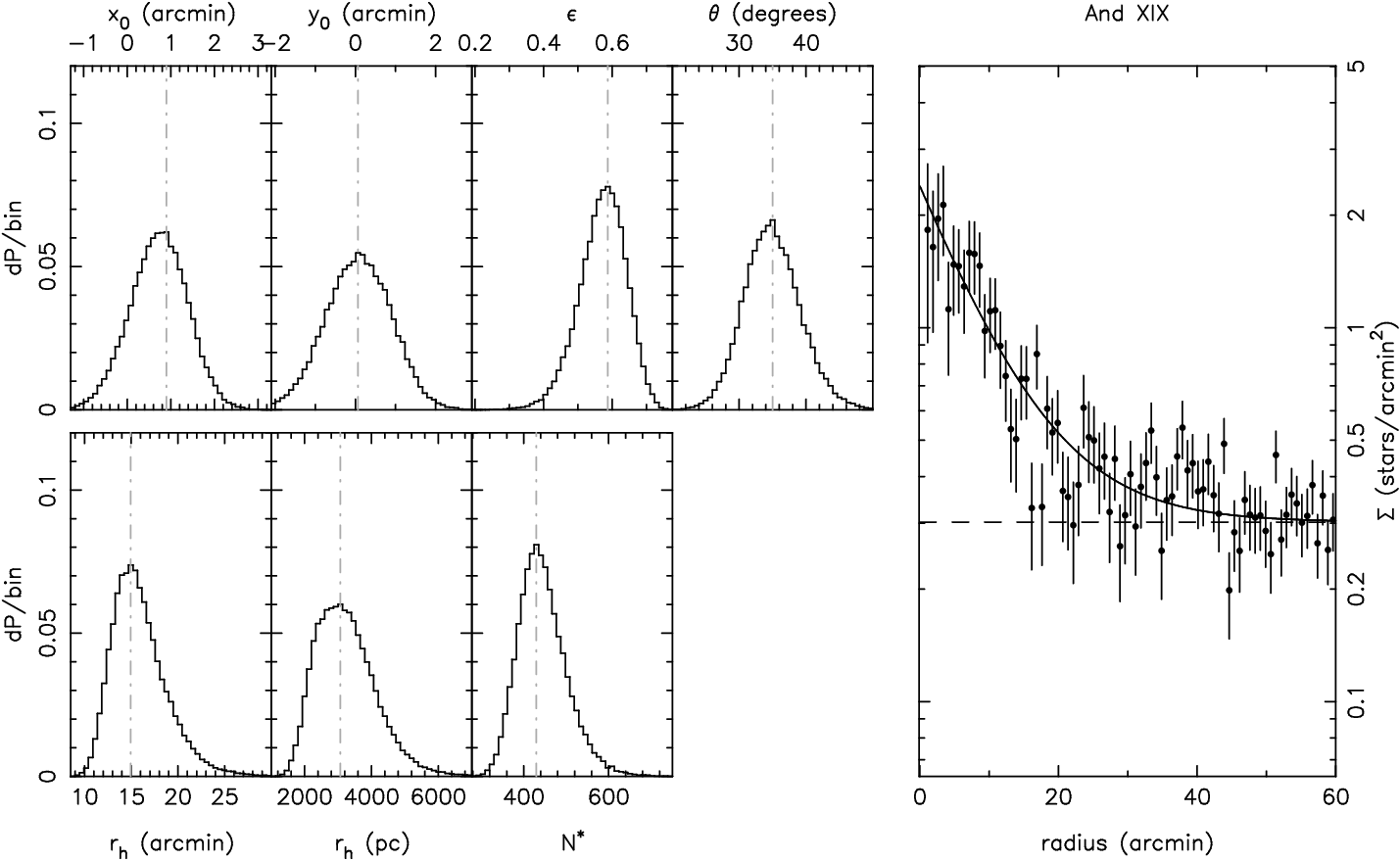}\hspace{1cm}
\includegraphics[width=0.4\hsize,angle=270]{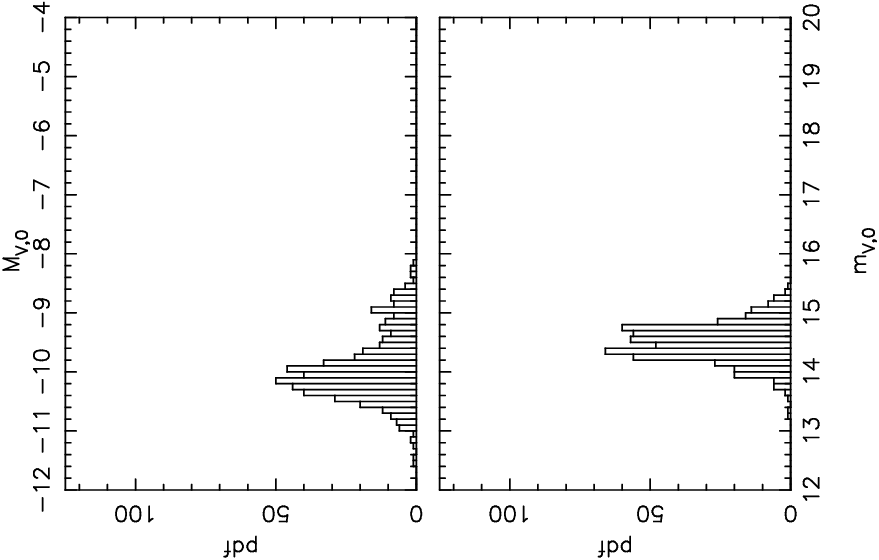}
\caption{\label{posterior_pdf_AndXIX}Same as Figure~\ref{posterior_pdf_AndI} for And~XIX.}
\end{center}
\end{figure*}

And~XIX was also found by \citet{mcconnachie08} in pre-PAndAS MegaCam/CFHT data. Our results, presented in Figure~\ref{posterior_pdf_AndXIX} ($r_h=14.2^{+3.4}_{-1.9}$$'$, $\epsilon = 0.58^{+0.05}_{-0.10}$, $m_{V,0}=14.5\pm0.3$), differ drastically from the literature values ($r_h = 6.2\pm0.1'$, $\epsilon=0.17\pm0.02$, $m_{V,0}=15.6\pm0.6$), which we derived from what now clearly appears as a spatial region of the survey that was much too small for the analysis of this dwarf galaxy. This led to the much smaller half-light radius (see also \citealt{munoz12} for the impact of using a small field of view to derive the size of a dwarf galaxy). In the current analysis, we use a sample that extends $\pm70'$ away from And~XIX's centroid, leading to the inference of a much larger system. The fact that we also find And~XIX to be very elliptical implies that the stellar stream it is embedded in \citep[][see also Figure~2 of \citealt{bate14}]{mcconnachie08} further impacts our results and may be responsible for the positive counts above the favored radial density profile at $\sim40'$ (central panel of Figure~\ref{posterior_pdf_AndXIX}). In addition, the large extent of the region considered means that the assumption of a flat background is likely an approximation to the true stellar density in the chosen CMD selection box.

With a half-light radius as large as $14.2^{+3.4}_{-1.9}$$'$ or $3065^{+935}_{-1065}\pc$, it is uncertain whether And~XIX can still be classified as a dwarf galaxy or whether it is so disrupted that its properties cannot be directly compared to other pristine dwarf galaxies. The only Local Group dwarf spheroidal galaxy with a similar size is Sagittarius \citep{majewski03}, which is clearly being pulled apart by gravitational tides. Alternatively, And~XIX could be a fainter and local counterpart of the `extremely diffuse' galaxies recently found in the Coma cluster \citep{vandokkum15}. Ultimately, a detailed spectroscopic analysis of the system's kinematics, developing the initial analysis of \citet{collins13} over the large body of the system, is mandatory to understand the state of And~XIX and whether it remains a bound system. 

\subsection{And XX}
\begin{figure*}
\begin{center}
\includegraphics[width=0.4\hsize,angle=270]{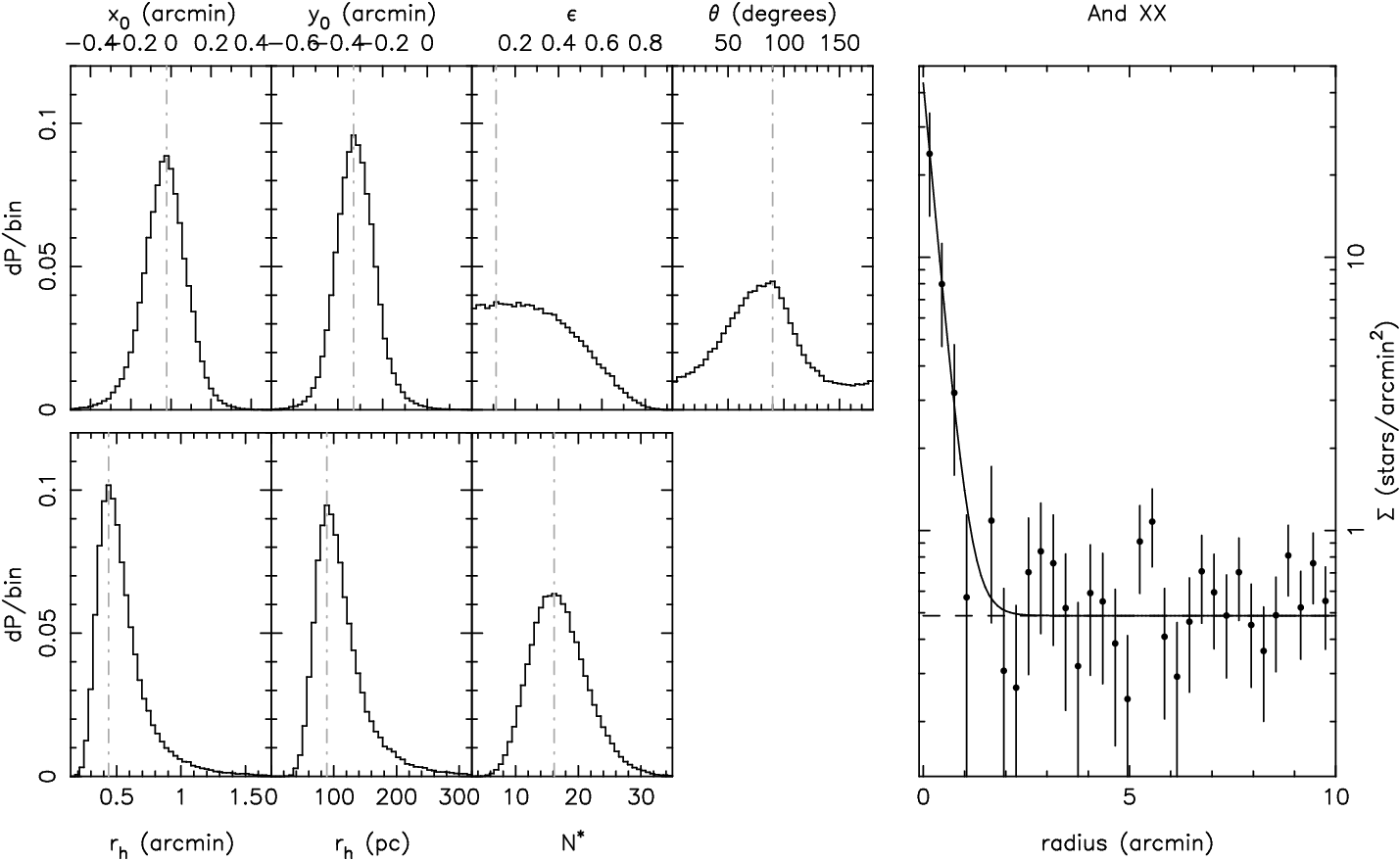}\hspace{1cm}
\includegraphics[width=0.4\hsize,angle=270]{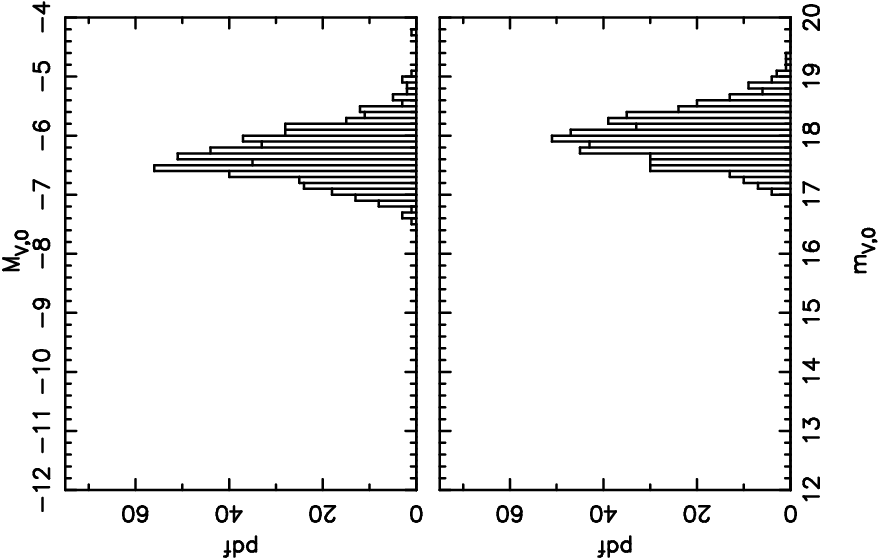}
\caption{\label{posterior_pdf_AndXX}Same as Figure~\ref{posterior_pdf_AndI} for And~XX.}
\end{center}
\end{figure*}

And~XX is the last of the three dwarf galaxies discovered by \citet{mcconnachie08}, who found it to be extremely small and faint ($r_h = 0.5\pm0.1'$, $\epsilon=0.30\pm0.15$, $m_{V,0} = 18.2\pm0.8$). We concur with this assessment, as can be seen in Figure~\ref{posterior_pdf_AndXX} ($r_h=0.4^{+0.2}_{-0.1}$$'$, $\epsilon = 0.11^{+0.41}_{-0.11}$, $m_{V,0}=18.0\pm0.4$). CMD `shot-noise' is significant for this faint galaxy but our technique that relies on sampling artificial CMDs provides better constraints on $m_{V,0}$ than what was previously achieved by summing up the flux of the few observed stars.

\subsection{And XXI}
\begin{figure*}
\begin{center}
\includegraphics[width=0.4\hsize,angle=270]{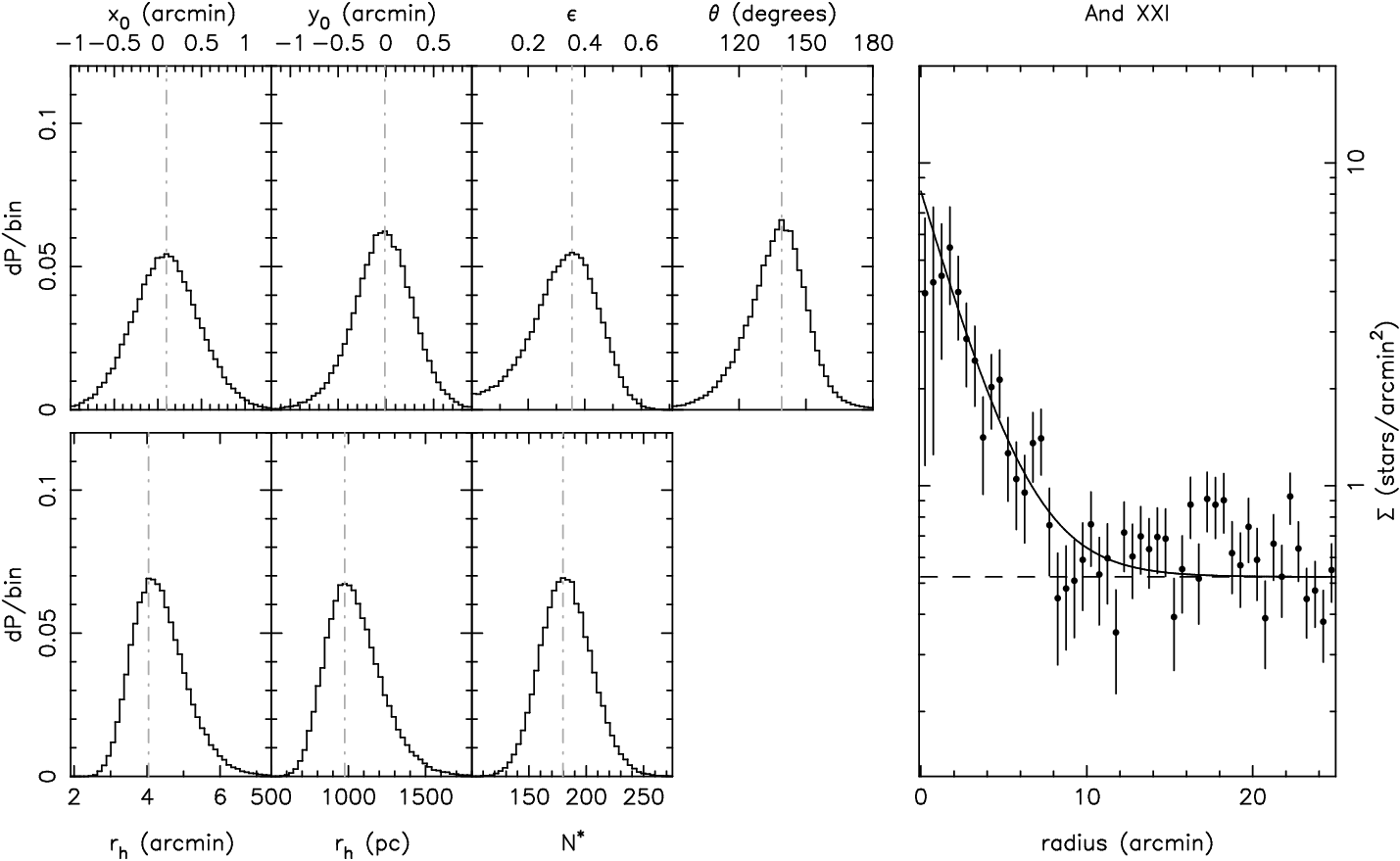}\hspace{1cm}
\includegraphics[width=0.4\hsize,angle=270]{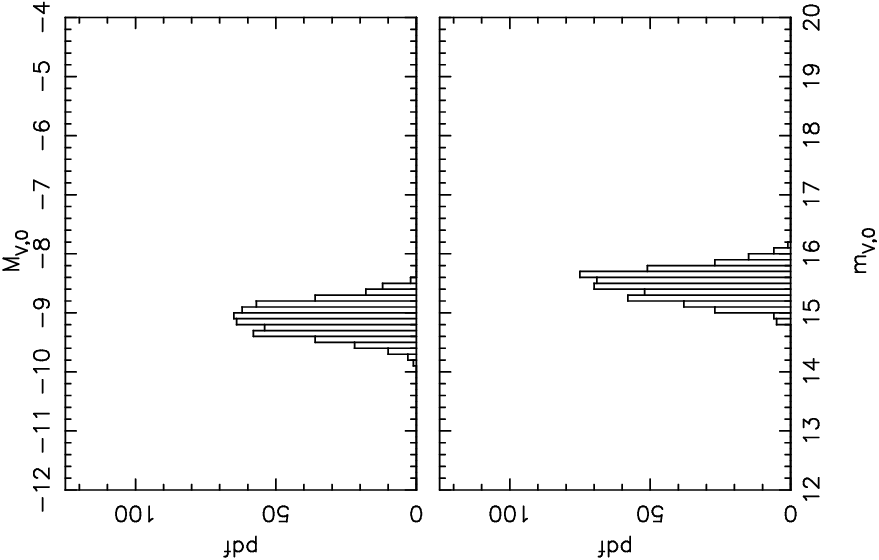}
\caption{\label{posterior_pdf_AndXXI}Same as Figure~\ref{posterior_pdf_AndI} for And~XXI.}
\end{center}
\end{figure*}

And~XXI was found in PAndAS and its first analysis was presented in \citet[][$r_h=3.5\pm0.3'$, $\epsilon=0.20\pm0.07$, $m_{V,0}=14.8\pm0.6$]{martin09}.  Our updated analysis yields results that are consistent with the literature values (Figure~\ref{posterior_pdf_AndXXI}; $r_h=4.1^{+0.8}_{-0.4}$$'$, $\epsilon = 0.36^{+0.10}_{-0.13}$, $m_{V,0}=15.5\pm0.3$). The new magnitude measurement seems more consistent than the previous one if we are to compare the CMD of And~XXI with that of other dwarf galaxies of similar apparent magnitude in Fig~\ref{CMDs} (e.g., And~V, And~XXIII).

It should be noted that part of the central region of And~XXI is masked out by a chip gap, which is naturally taken into account in the analysis.

\subsection{And XXII}
\begin{figure*}
\includegraphics[width=0.4\hsize,angle=270]{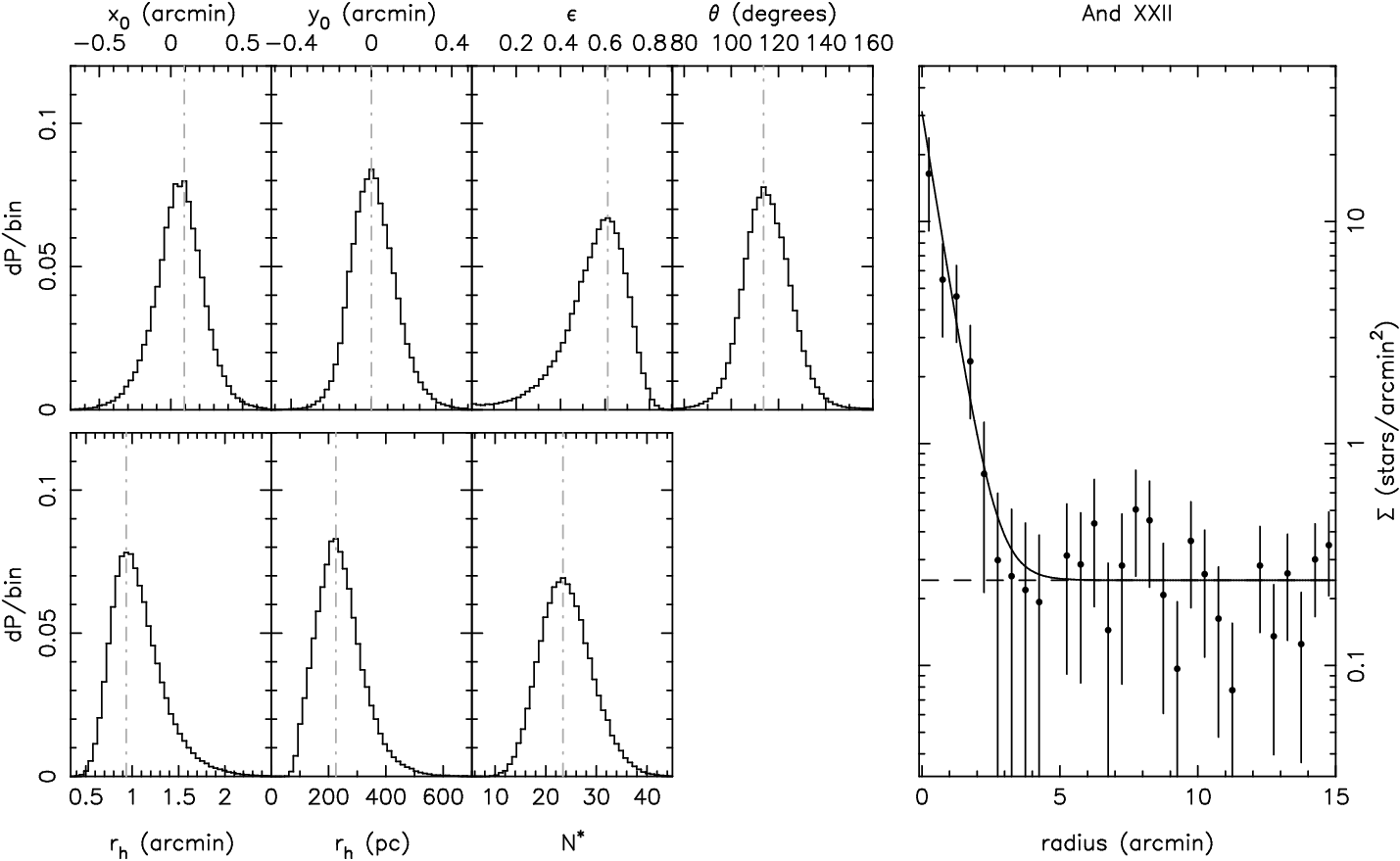}\hspace{1cm}
\includegraphics[width=0.4\hsize,angle=270]{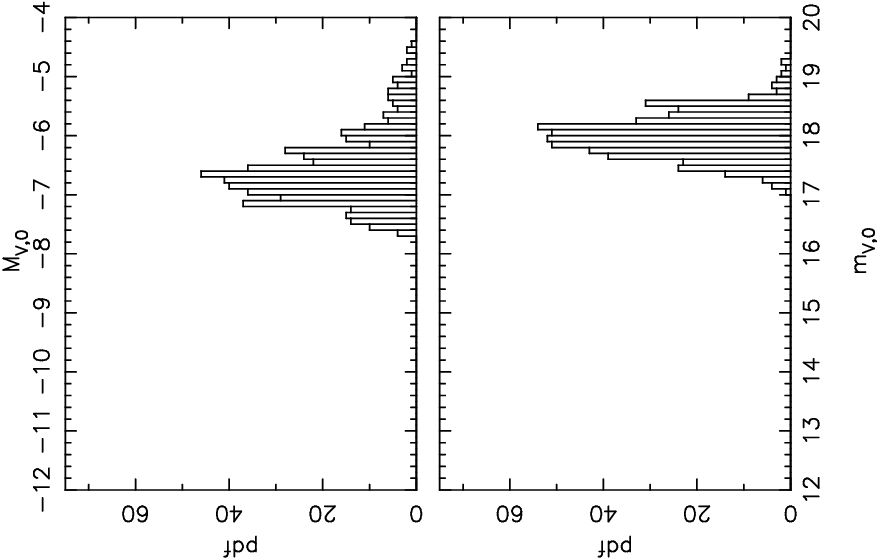}
\caption{\label{posterior_pdf_AndXXII}Same as Figure~\ref{posterior_pdf_AndI} for And~XXII.}
\end{figure*}

The discovery of And~XXII is also presented in \citet[][$r_h=0.9\pm0.1'$, $\epsilon=0.56\pm0.11$, $m_{V,0}=18.0\pm0.8$]{martin09}, based on PAndAS data. The new analysis yields results in perfect agreement with this previous analysis of the same data (Figure~\ref{posterior_pdf_AndXXII}; $r_h=0.9^{+0.3}_{-0.2}$$'$, $\epsilon = 0.61^{+0.10}_{-0.14}$, $m_{V,0}=18.0\pm0.4$), but with a better constrained magnitude measurement, despite a significant amount of CMD `shot-noise'. The new measurements also supersede the structural analysis of \citet{chapman13} based on the same data.

\subsection{And XXIII}
\begin{figure*}
\begin{center}
\includegraphics[width=0.4\hsize,angle=270]{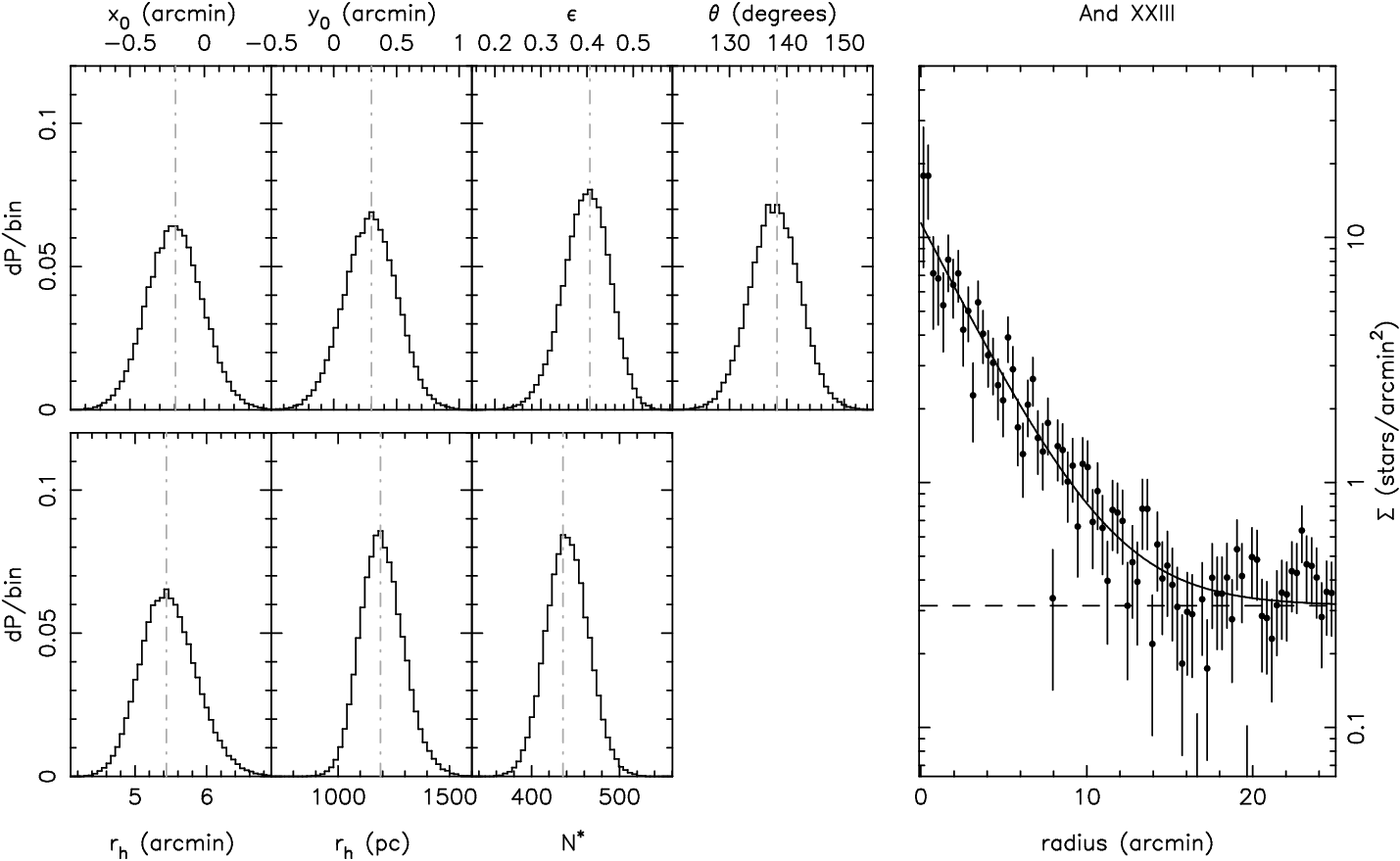}\hspace{1cm}
\includegraphics[width=0.4\hsize,angle=270]{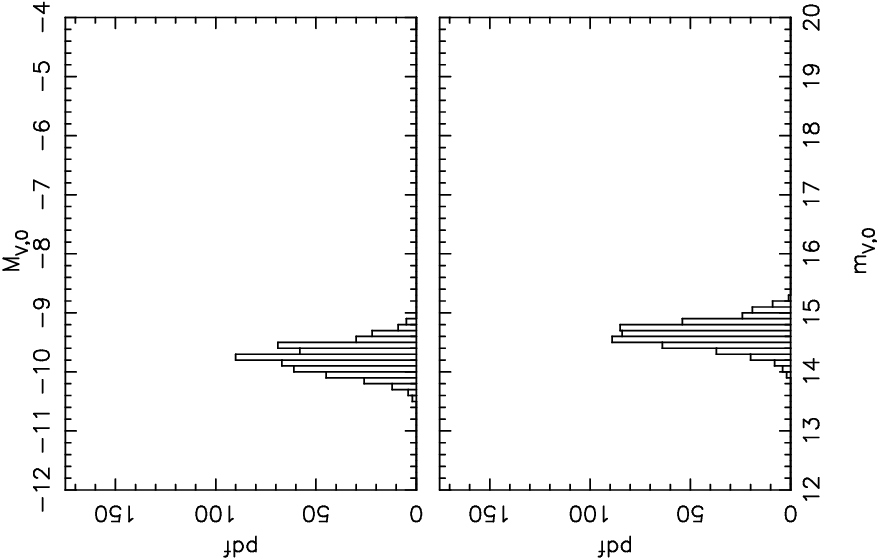}
\caption{\label{posterior_pdf_AndXXIII}Same as Figure~\ref{posterior_pdf_AndI} for And~XXIII.}
\end{center}
\end{figure*}

The discovery of And~XXIII from PAndAS data is presented in \citet[][$r_h = 4.6\pm0.2'$, $\epsilon=0.40\pm0.05$, $m_{V,0} = 14.2\pm0.5$]{richardson11} and our results (Figure~\ref{posterior_pdf_AndXXIII}; $r_h=5.4\pm0.4'$, $\epsilon = 0.41^{+0.05}_{-0.06}$, $m_{V,0}=14.6\pm0.2$) are in good agreement with this initial analysis, even though we now find the size to be slightly larger than originally estimated. And~XXIII is among the most luminous M31 companions with an absolute magnitude of $M_V = -9.8\pm0.1$, brighter even than And~III and And~V, but owes its late discovery to its large size and, therefore, its low surface brightness ($\mu_0 = 27.5\pm0.1$ mag/arcsec$^2$).

\subsection{And XXIV}
\begin{figure*}
\begin{center}
\includegraphics[width=0.4\hsize,angle=270]{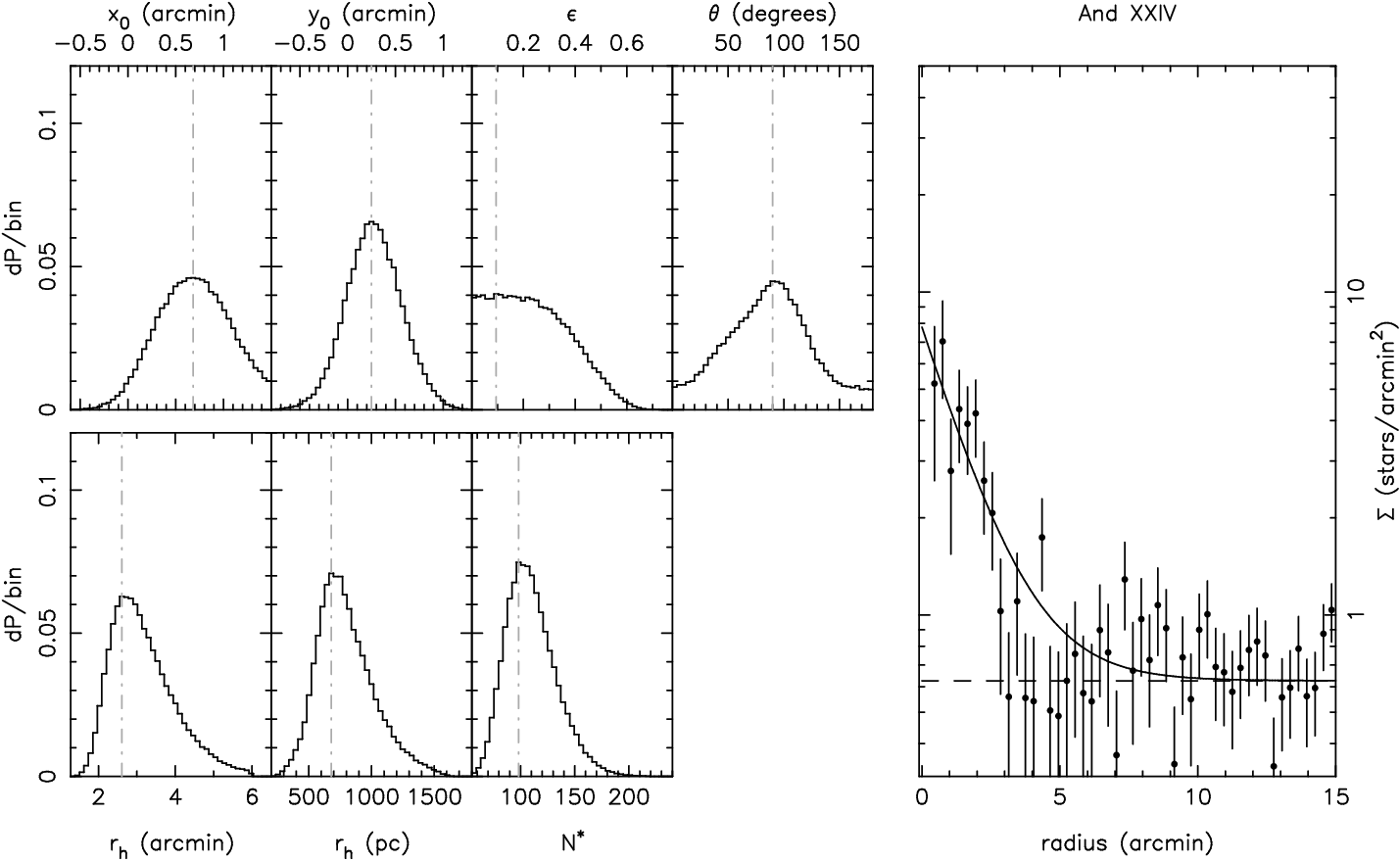}\hspace{1cm}
\includegraphics[width=0.4\hsize,angle=270]{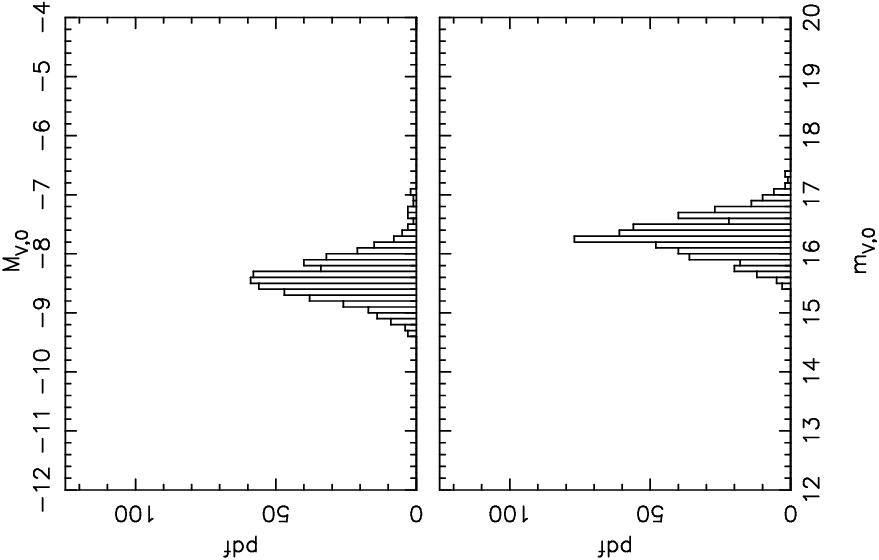}
\caption{\label{posterior_pdf_AndXXIV}Same as Figure~\ref{posterior_pdf_AndI} for And~XXIV.}
\end{center}
\end{figure*}

Like And~XXIII, And~XXIV was discovered in PAndAS by \citet{richardson11}. The agreement is good between the initial estimate of this dwarf galaxy's parameters ($r_h=2.1\pm0.1'$, $\epsilon=0.25\pm0.05$, $m_{V,0} = 16.3\pm0.5$) and our results (Figure~\ref{posterior_pdf_AndXXIV}; $r_h=2.6^{+1.0}_{-0.5}$$'$, $\epsilon = 0.10^{+0.31}_{-0.10}$, $m_{V,0}=16.3\pm0.3$). It should be noted that we had to mask a fairly large region a few arcminutes east of And~XXIV because of bad-quality data at the edge of a PAndAS field (Figure~\ref{map_XXIII_XXIV}).

\subsection{And XXV}
\begin{figure*}
\begin{center}
\includegraphics[width=0.4\hsize,angle=270]{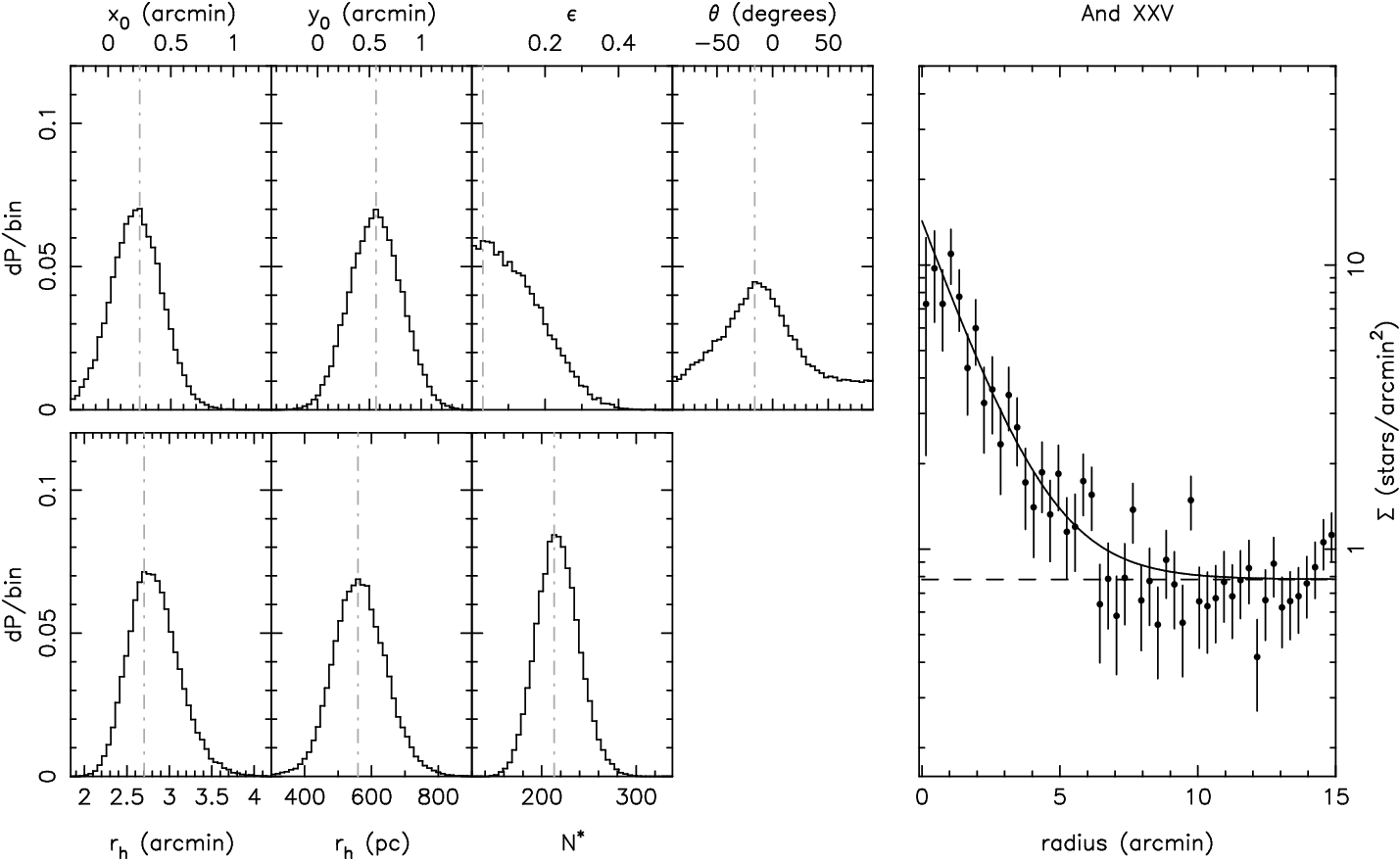}\hspace{1cm}
\includegraphics[width=0.4\hsize,angle=270]{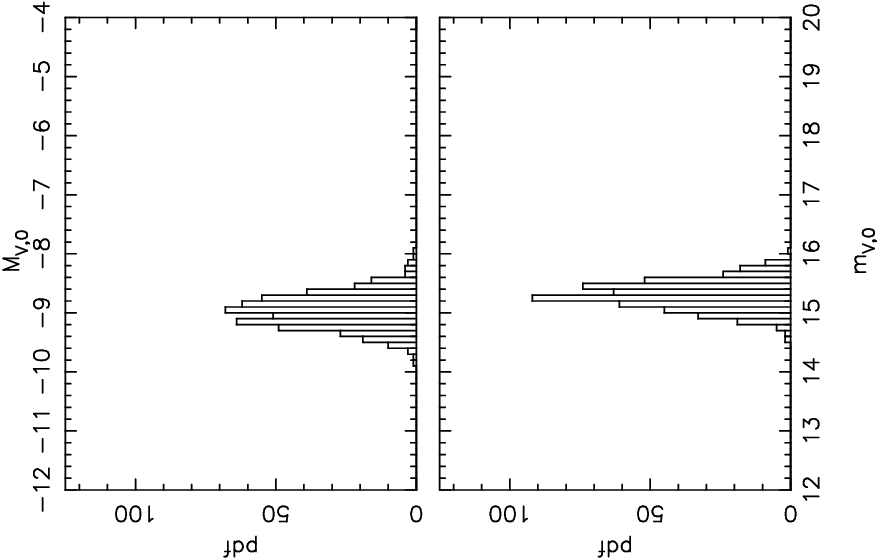}
\caption{\label{posterior_pdf_AndXXV}Same as Figure~\ref{posterior_pdf_AndI} for And~XXV.}
\end{center}
\end{figure*}

And~XXV was also found in PAndAS \citep[$r_h = 3.0\pm0.3'$, $\epsilon=0.25\pm0.05$, $m_{V,0} = 14.8\pm0.5$]{richardson11}. It is a fairly bright dwarf galaxy, as confirmed by our analysis (Figure~\ref{posterior_pdf_AndXXV}; $r_h=2.7^{+0.4}_{-0.2}$$'$, $\epsilon = 0.03^{+0.16}_{-0.03}$, $m_{V,0}=15.3^{+0.3}_{-0.2}$). We find the system to be rounder than previously estimated, even though the two measurements are compatible. Part of the large uncertainties are the consequence of a large chip gap slicing through And~XXV (Figure~\ref{map_XXV_XXVI}).

\subsection{And XXVI}
\begin{figure*}
\begin{center}
\includegraphics[width=0.4\hsize,angle=270]{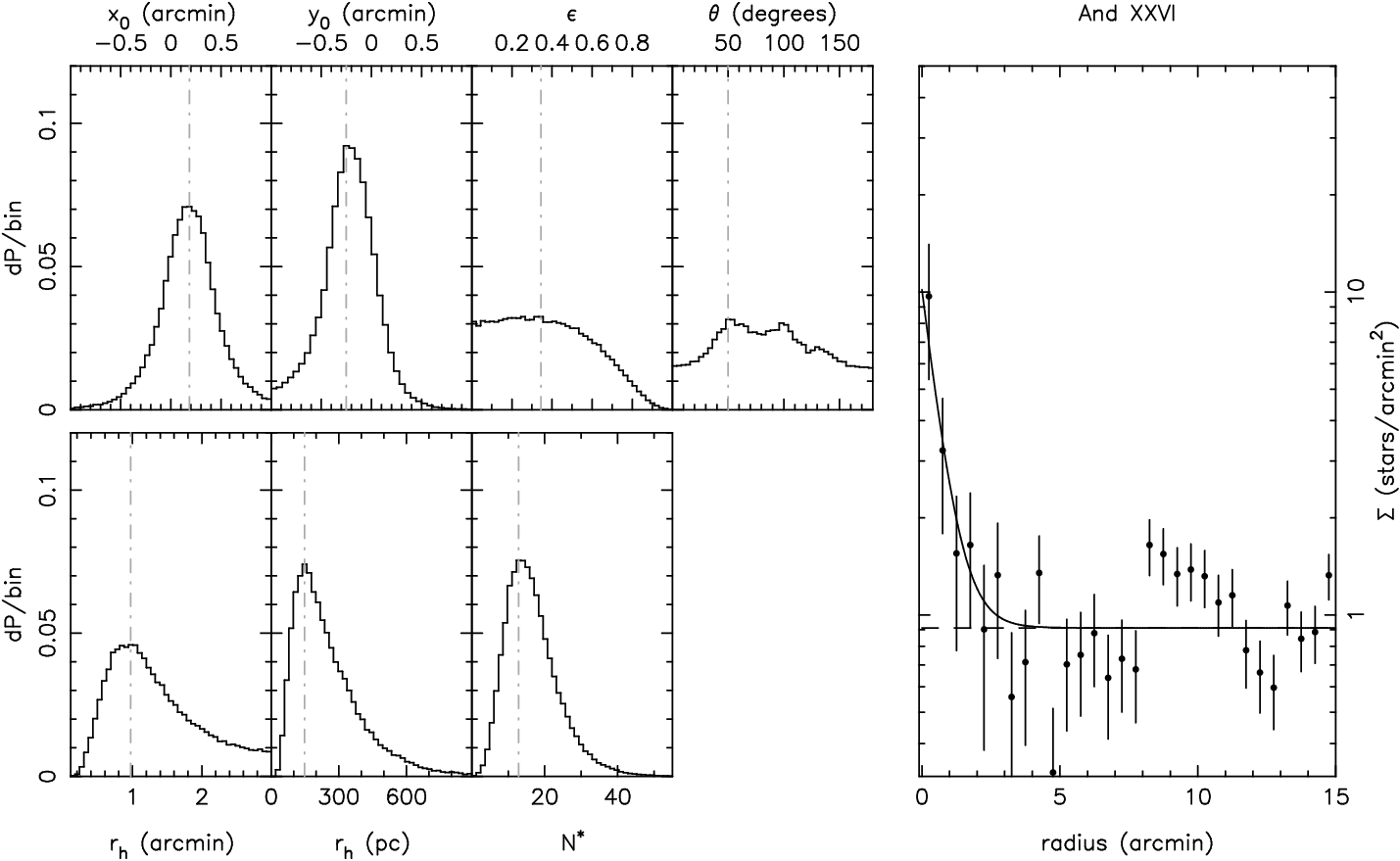}\hspace{1cm}
\includegraphics[width=0.4\hsize,angle=270]{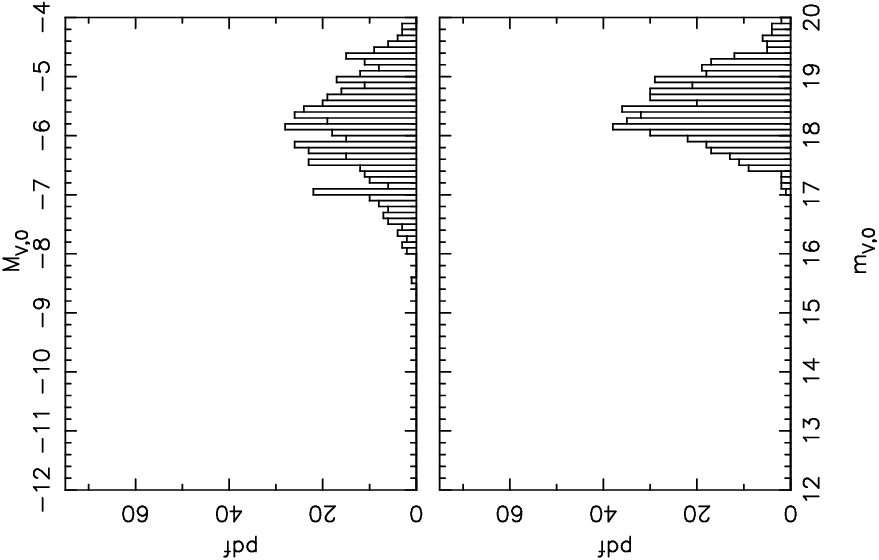}
\caption{\label{posterior_pdf_AndXXVI}Same as Figure~\ref{posterior_pdf_AndI} for And~XXVI.}
\end{center}
\end{figure*}

And~XXVI was discovered in PAndAS \citep[][$r_h=1.0\pm0.1'$, $\epsilon=0.25\pm0.05$, $m_{V,0}=17.3\pm0.5$]{richardson11}. The new analysis (Figure~\ref{posterior_pdf_AndI}; $r_h=1.0^{+0.6}_{-0.5}$$'$, $\epsilon = 0.35^{+0.33}_{-0.35}$, $m_{V,0}=18.5^{+0.7}_{-0.5}$) hints that this could be the faintest dwarf galaxy known around M31, although our results are rendered noisy by And~XXVI being located on one of the shallowest PAndAS fields and only a few arcminutes away from significant chip gaps (Figure~\ref{map_XXV_XXVI}). As a consequence, it was necessary to enforce an additional size prior ($r_\mathrm{max}=3'$) as well as centroid priors such that $x_0$ and $y_0$ do not deviate from the literature values by more than $1'$. Once these are in place, the analysis converges on parameters that are representative of the dwarf galaxy.

\subsection{Cas II/And XXX}
\begin{figure*}
\begin{center}
\includegraphics[width=0.4\hsize,angle=270]{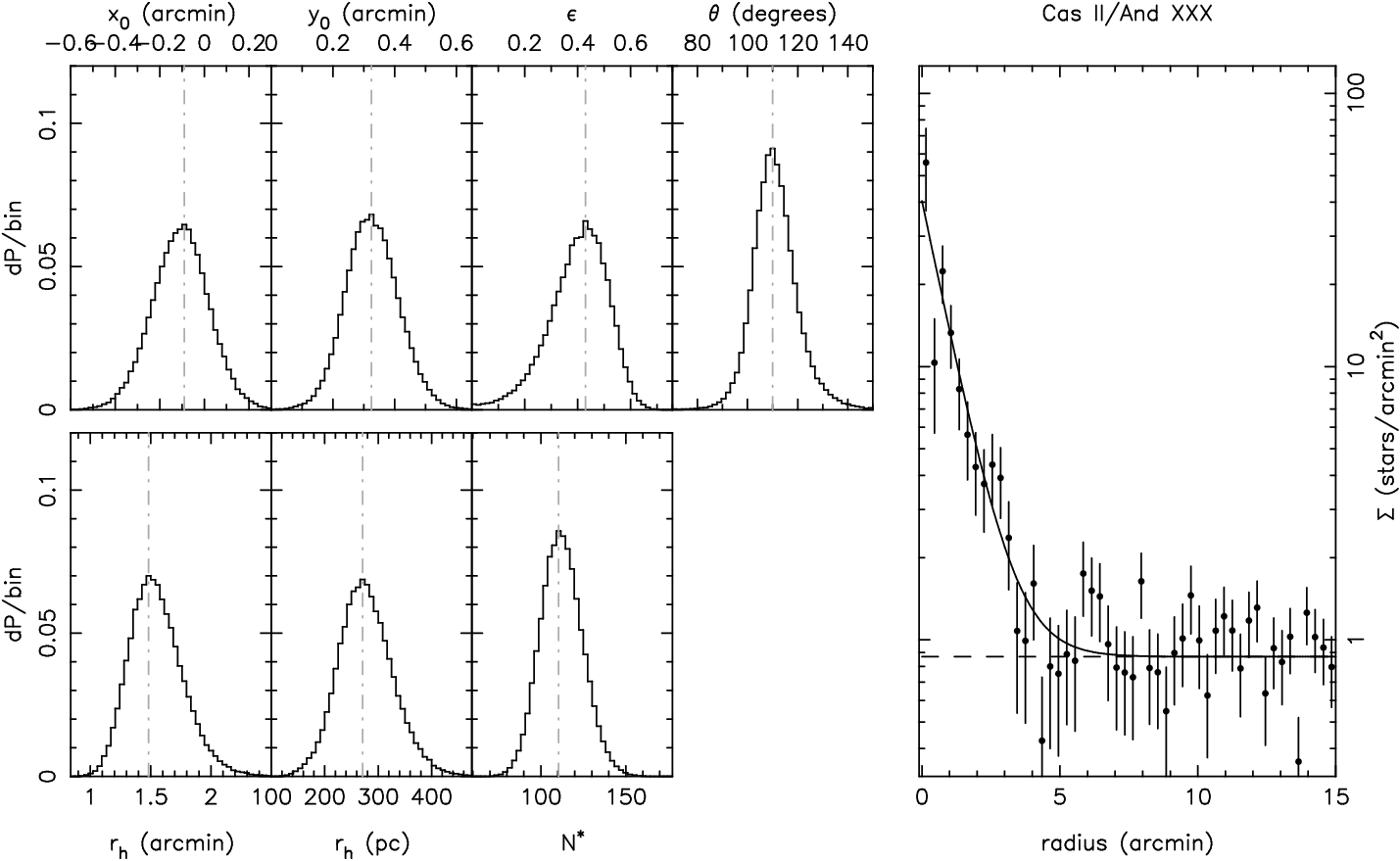}\hspace{1cm}
\includegraphics[width=0.4\hsize,angle=270]{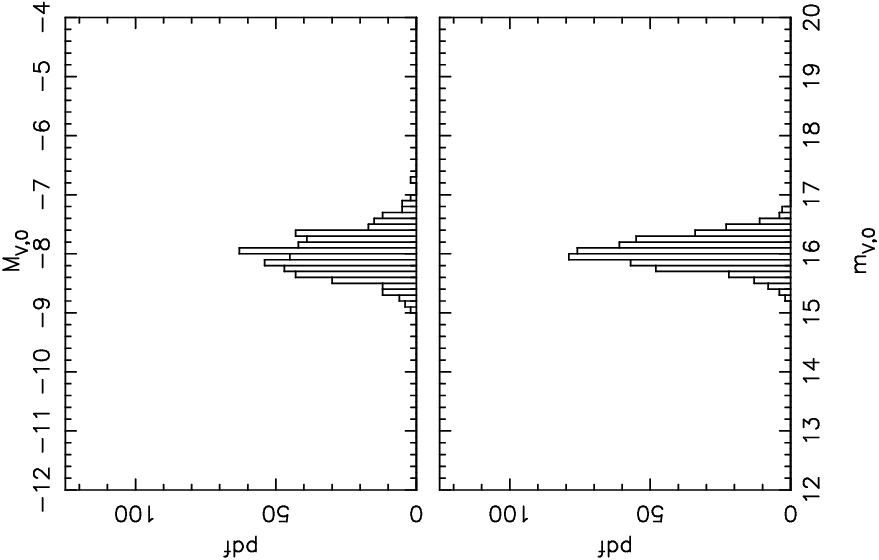}
\caption{\label{posterior_pdf_AndXXX}Same as Figure~\ref{posterior_pdf_AndI} for Cas~II/And~XXX.}
\end{center}
\end{figure*}

Although Cas~II/And~XXX was mentioned in the PAndAS dwarf galaxy distance and spectroscopic study papers \citep{aconn12,collins13}, this is the first detailed analysis of its structure and luminosity (Figure~\ref{posterior_pdf_AndXXX} and \ref{map_XXX}). Located very close to NGC~147 and NGC~185, Cas~II also shares distances \citep{conn12} and systemic velocities \citep{collins13} that are close to those of these two large elliptical galaxies. It is therefore likely that the three galaxies are part of the same group\footnote{Note, however, that Cas~III/And~XXXII \citep{martin13a}, located just outside the PAndAS footprint, north of NGC~147 and NGC~185, has a velocity that makes it incompatible with being part of this group of galaxies \citep{martin14b}.} \citep{arias16}.

Cas~II is reasonably faint ($m_{V,0} = 16.0^{+0.3}_{-0.2}$, $M_{V,0}=-8.0^{+0.4}_{-0.3}$) and has a typical size for galaxies of this luminosity ($r_h=1.5\pm0.2'$ or $270\pm50\pc$). It is also quite elliptical ($\epsilon=0.43^{+0.10}_{-0.12}$), although its major axis points neither toward NGC~147 nor toward NGC~185.

\subsection{The case of And XXVII}
\label{AndXXVII}
Despite there being a visual small overdensity of stars at the location of And~XXVII reported by \citet{richardson11}, it is impossible to get the algorithm to converge on a set of sensible parameters for this overdensity. The fit is entirely driven by the large stellar structure present in this region of the M31 stellar halo and favors an extremely large half-light radius ($r_h\sim30'$) with only a few hundred stars ($N^*\sim350$). Such a large half-light radius of $\sim6\kpc$ at the distance of M31 is implausible for a dwarf galaxy since, for the projected distance of And~XXVII from M31 ($\sim55\kpc$), this is close to the tidal radius of a stellar system with $10^9\msun$. Even if And~XXVII is located further from M31, the stellar body of the such a large system would likely still be significantly affected by tides. The structural parameters we infer more likely track the large stellar stream visible on the PAndAS stellar maps \citep[e.g.,][]{richardson11}. Already, the spectroscopic study of these stars by \citet{collins13} proved at odds with the expected properties of a dwarf galaxy as it revealed a large velocity dispersion of $\sim15\kms$ and structure in the velocity distribution. The combination of peculiar kinematics and structural parameters lead us to conclude that And~XXVII should not be classified as a dwarf galaxy and that we are likely observing a system that is in the final throes of its tidal disruption. A dedicated kinematic study shall hopefully shed more light onto this system (J. Preston et al., in preparation).

\section{Conclusions}
We have presented a homogeneous study of the structure and luminosity of 23 M31 dwarf spheroidal galaxies from the contiguous photometric survey PAndAS. The structural parameter analysis presented here builds on the work of M08 and explicitly constructs the likelihood function from the positions of individual stars, using an exponential radial profile that allows for flattening and a constant level of contamination. One of the outputs of the model is the number of member stars in a chosen CMD selection box, which we tie to the apparent magnitude of a dwarf galaxy through the sampling of an artificial, representative CMD. This step naturally accounts for CMD `shot-noise,' which becomes significant for faint dwarf galaxies whose RGB is poorly sampled. The 23 dwarf spheroidal galaxies in PAndAS range in absolute magnitude from the luminous And~II ($M_V=-11.6\pm0.2$), discovered more than 40 years ago, to the very dim And~XXVI ($M_V = -5.8^{+0.9}_{-1.0}$) discovered in PAndAS. Finally, we use the derived parameters to calculate the central surface brightnesses of the dwarf galaxies; these range from $\sim 25$ to $\sim29 \textrm{~mag/arcsec}^2$.

The main deviations from the literature values are:
\begin{itemize}
\item The luminosity of And~XV and And~XVI which were significantly overestimated in our previous analysis. We now find their total magnitudes are $M_V = -8.0^{+0.3}_{-0.4}$ and $-7.3\pm0.3$, respectively.
\item The size of And~XIX that was significantly underestimated in our previous analysis and now yields $r_h=3065^{+935}_{-1065}\pc$. It implies that And~XIX is significantly affected by tides and/or that it is a local counterpart to the `extremely diffuse' galaxies recently found in the Coma cluster \citep{vandokkum15}.
\item And~XXVI is found to be significantly fainter than in its discovery paper with $m_V=18.5^{+0.7}_{-0.5}$ and $M_V=-5.8^{+0.9}_{-1.0}$.
\item The impossibility to fit a sensible model to the stellar distribution around And~XXVII, leading us to conclude, in conjunction with its peculiar kinematics, that it is not a dwarf galaxy but likely an unbound stellar structure in the M31 halo.
\end{itemize}

The Bayesian framework we place ourselves in (sampling the prior PDFs on the distance and other relevant parameters) allows us to straightforwardly track the impact of uncertainties on the derived parameters. We further provide samplings from the Monte Carlo Markov Chains generated by our analysis in Appendix~\ref{chains}; these correspond to much more truthful representations of the analysis than the parameters listed in Table~\ref{table:param_list} \& \ref{table:luminosity_list}. Anyone who wishes to properly account for degeneracies between the parameters as well as for the peculiarity of the PDFs should strive to use these chains in their own analysis.

Combined with the \citet{aconn12} inference of the distance to these systems from the same data, the properties listed in Table~1 and their PDFs provide a characterization of a populated dwarf galaxy satellite system that should be free of most systematics that usually plague analyses based on data gathered in the literature, originating from different telescopes and instruments, and analyzed with different techniques\footnote{Although derived from another survey, it should also be noted that the structural parameters derived for Cas~III/And~XXXI, Lac~I/And~XXXII, and Per~I/And~XXXIII, three M31 dwarf galaxies discovered in Pan-STARRS1 data \citep{martin13a,martin13c}, were derived using the algorithm presented here.}. The benefits of this data set have started to be exploited in studies of the global properties of the M31 dwarf galaxies, such as a comparison of the properties of dwarf galaxies in and out of the vast and thin rotating disk of M31 satellites \citep{ibata13a}, as recently presented by \citet{collins15}; a recent study of the alignment of M31 satellites with their host \citep{barber15}; or of the intrinsic shape of the M31 satellite dwarf galaxies \citep{salomon15}.

Once we have determined the dwarf-galaxy-search completeness functions in the next paper in this series, we will endeavor to reliably characterize the global properties of a large fraction of the M31 dwarf galaxy system.

\acknowledgments
We are grateful to the CFHT observing team for gathering the PAndAS images, and for their continued support throughout the project. N.F.M. thanks Morgan Fouesneau for stimulating discussions. M.A.F. acknowledges support by NSF grant AST-1009652 to the University of Massachusetts. G.F.L. gratefully acknowledges financial support for his ARC Future Fellowship (FT100100268) and through the award of an ARC Discovery Project (DP110100678). This work is based on observations obtained with MegaPrime/MegaCam, a joint project of CFHT and CEA/DAPNIA, at the Canada-France-Hawaii Telescope, which is operated by the National Research Council (NRC) of Canada, the Institut National des Sciences de l'Univers of the Centre National de la Recherche Scientifique (CNRS) of France, and the University of Hawaii.

%\bibliography{/Users/martin/Work/Papers/Biblio}
%\bibliographystyle{apj}
% Bibtex will create a .bbs file in the directory and before sending to the editor, I should replace the bibliography call by this file.

\appendix
\section{Stellar distributions}
\begin{figure*}
\begin{center}
\includegraphics[width=0.47\hsize,angle=270]{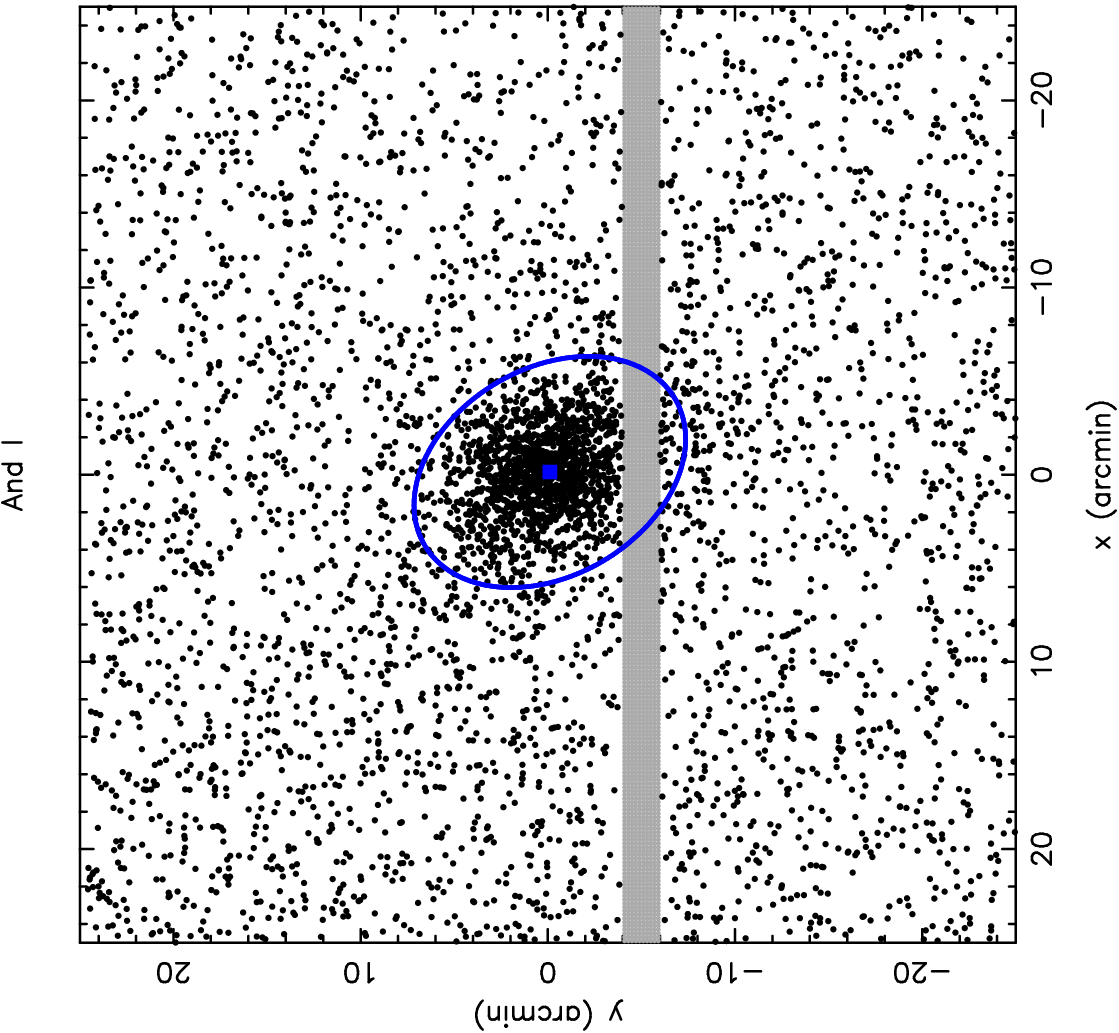}
\hspace{1cm}
\includegraphics[width=0.47\hsize,angle=270]{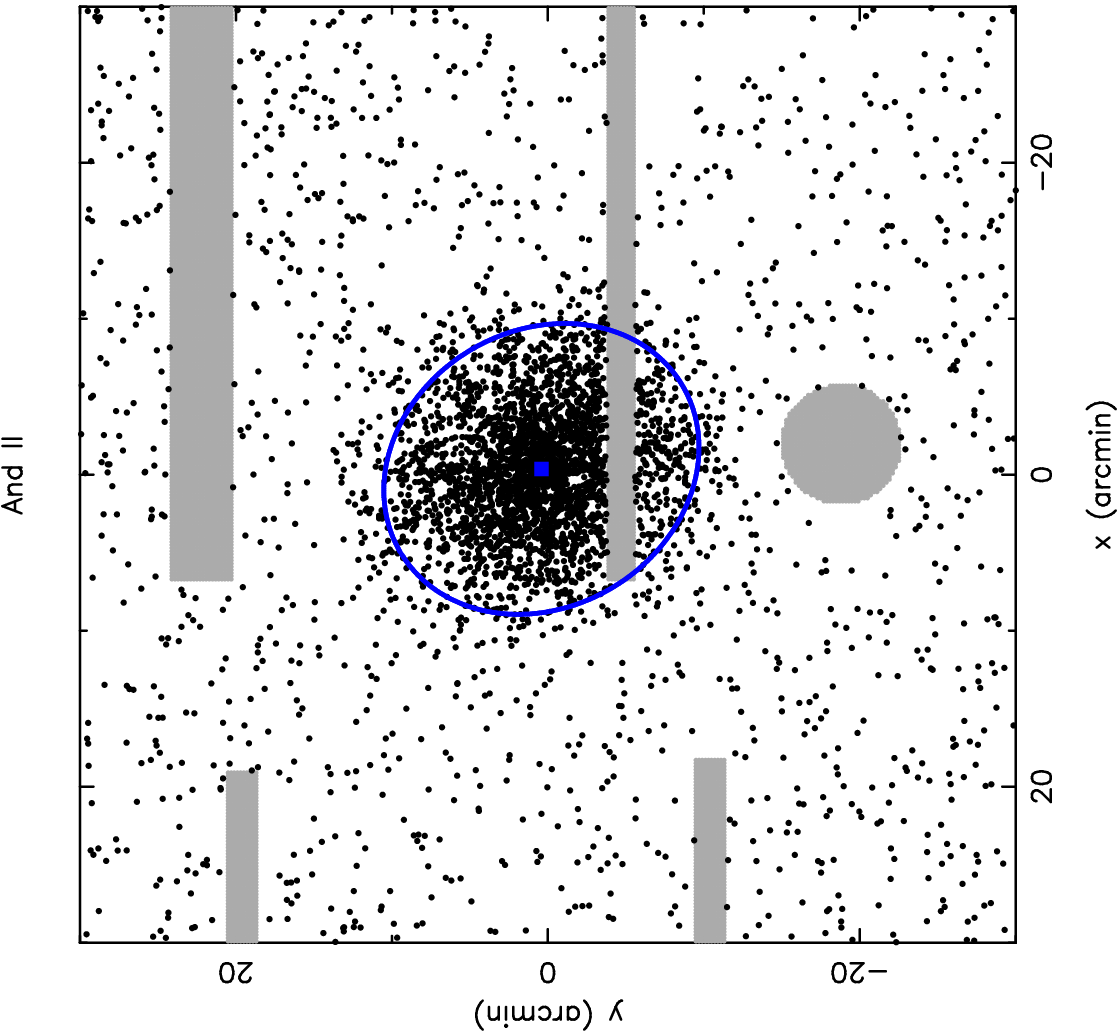}
\caption{\label{map_I_II}Distribution of CMD-selected stars in the region $\mathcal{A}$ around And~I and And~II. Regions in gray were masked because of chip gaps, bright stars, or bad data. The represented $(x,y)$ coordinates correspond to offsets from the chosen literature values of the centroid to the dwarf galaxies, as defined in section~\ref{structure}. As such the favored centroids inferred from the current analysis and represented by the blue squares can be offset from (0,0). The ellipses represents $2r_h$, as defined by the favored models.}
\end{center}
\end{figure*}

\begin{figure*}
\begin{center}
\includegraphics[width=0.47\hsize,angle=270]{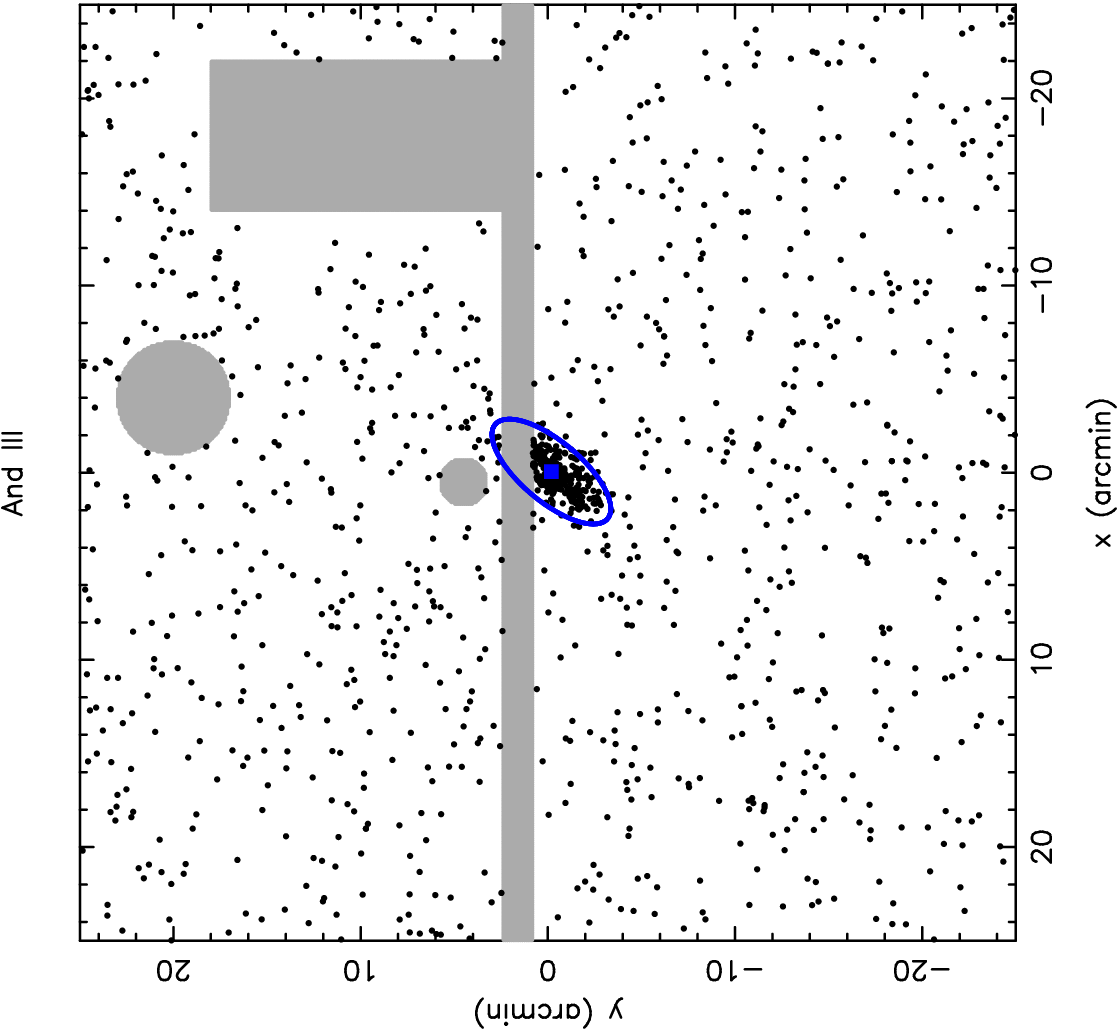}
\hspace{1cm}
\includegraphics[width=0.47\hsize,angle=270]{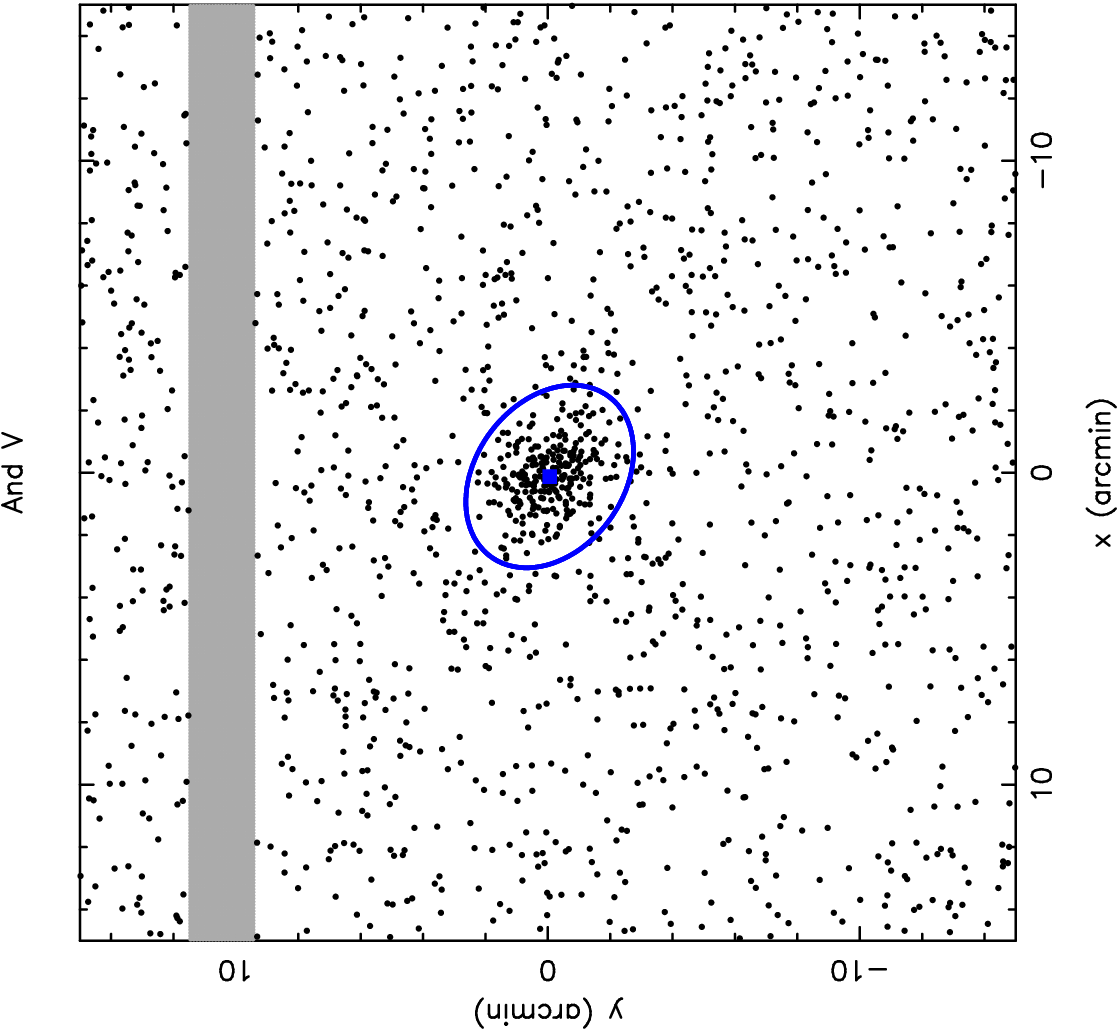}
\caption{\label{map_III_V}Same as Figure~\ref{map_I_II} but for And~III and And~V.}
\end{center}
\end{figure*}

\begin{figure*}
\begin{center}
\includegraphics[width=0.47\hsize,angle=270]{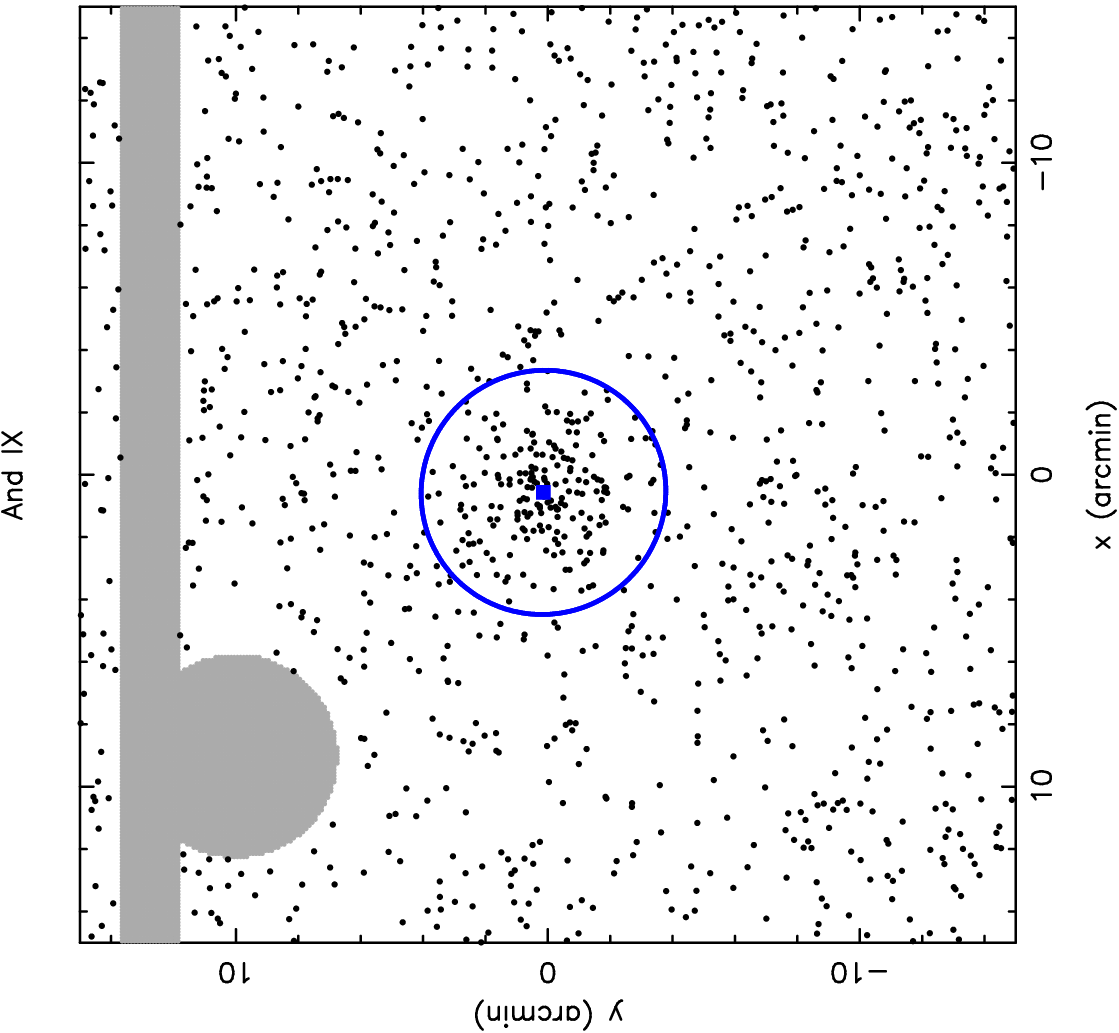}
\hspace{1cm}
\includegraphics[width=0.47\hsize,angle=270]{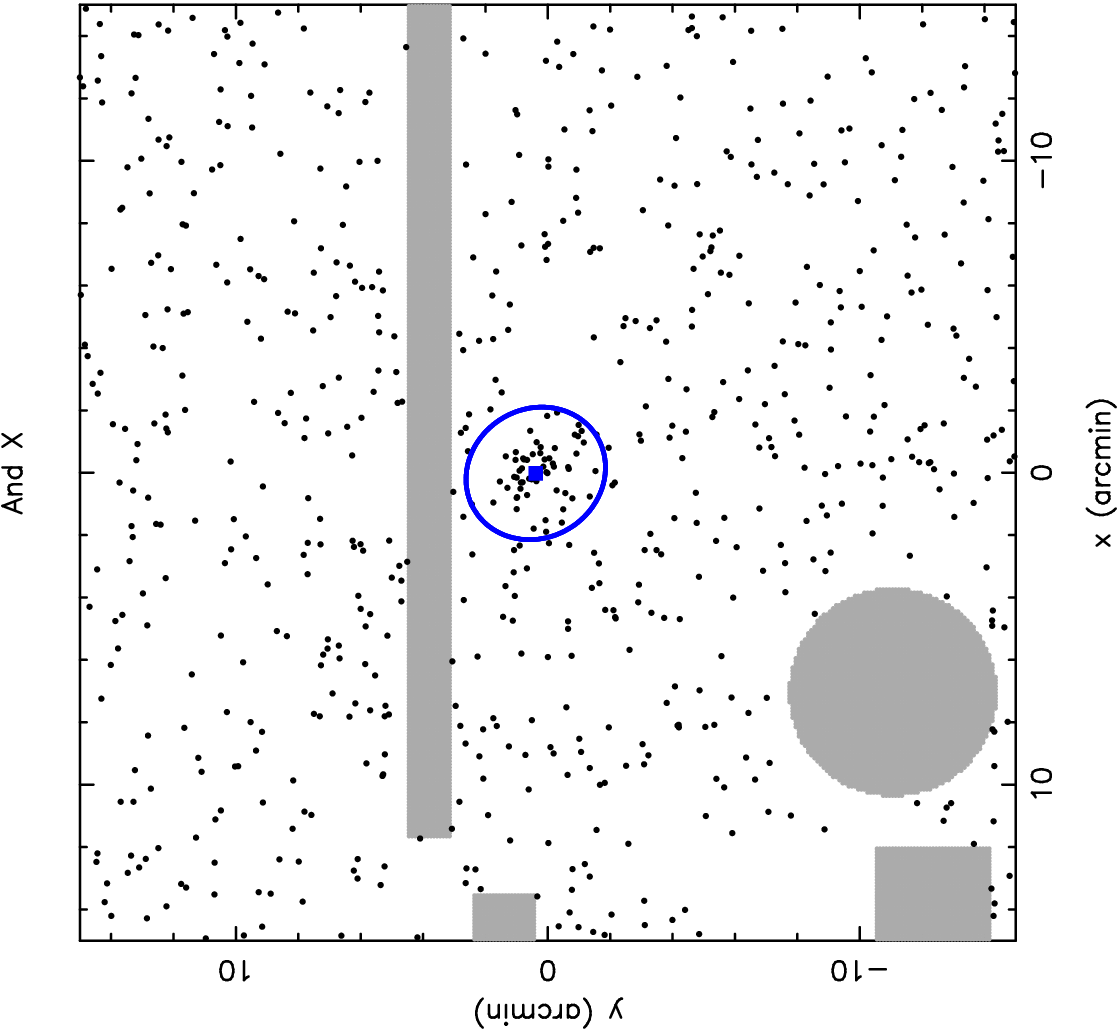}
\caption{\label{map_IX_X}Same as Figure~\ref{map_I_II} but for And~IX and And~X.}
\end{center}
\end{figure*}

\begin{figure*}
\begin{center}
\includegraphics[width=0.47\hsize,angle=270]{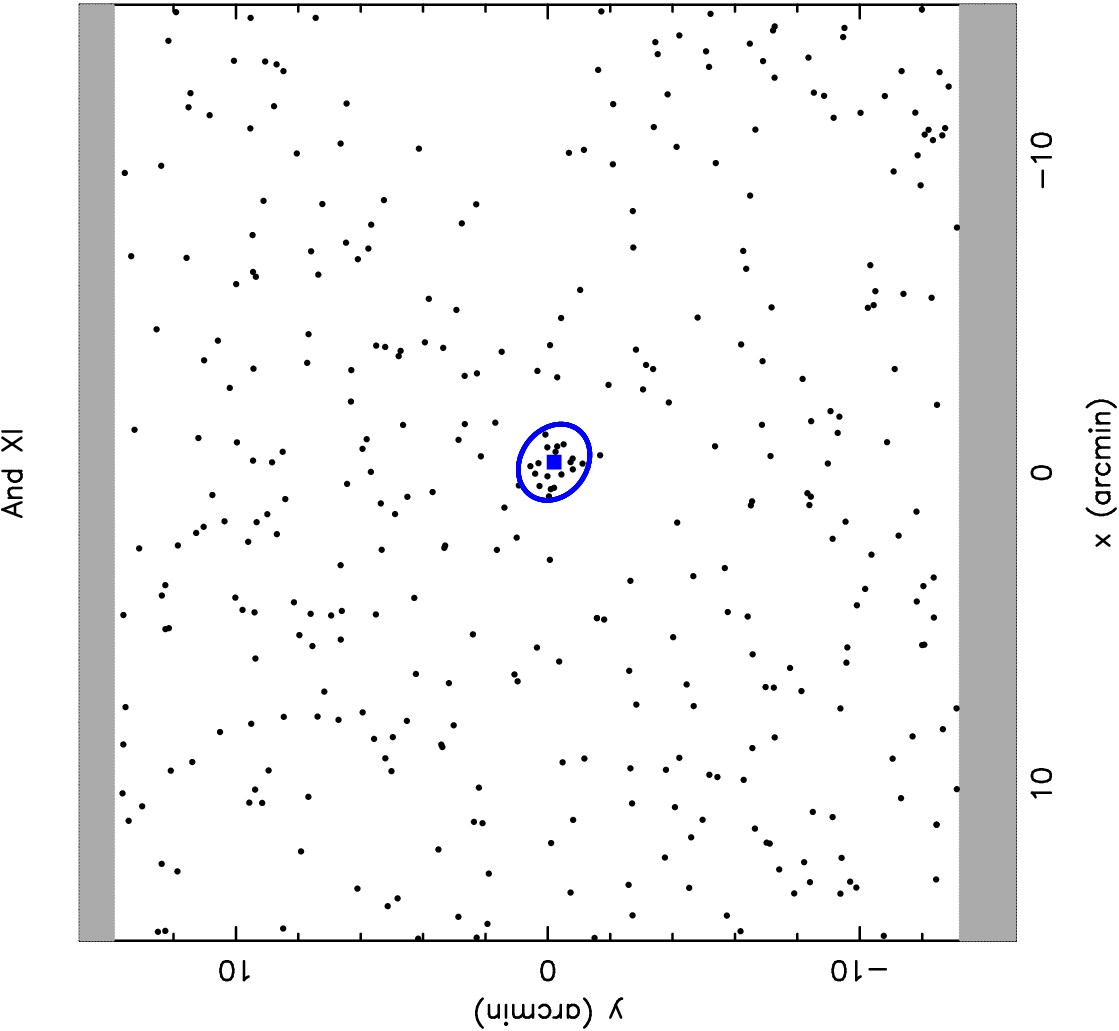}
\hspace{1cm}
\includegraphics[width=0.47\hsize,angle=270]{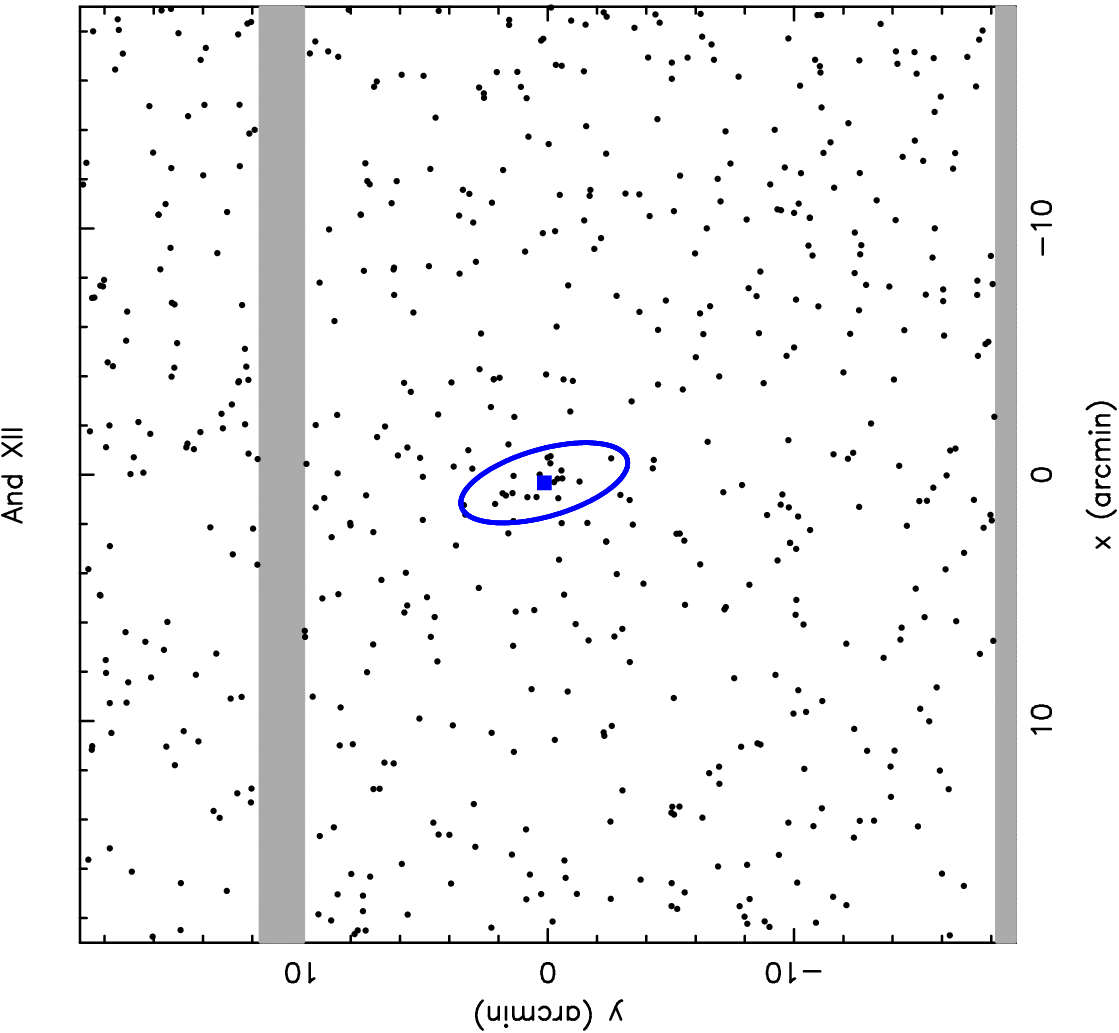}
\caption{\label{map_XI_XII}Same as Figure~\ref{map_I_II} but for And~XI and And~XII.}
\end{center}
\end{figure*}

\begin{figure*}
\begin{center}
\includegraphics[width=0.47\hsize,angle=270]{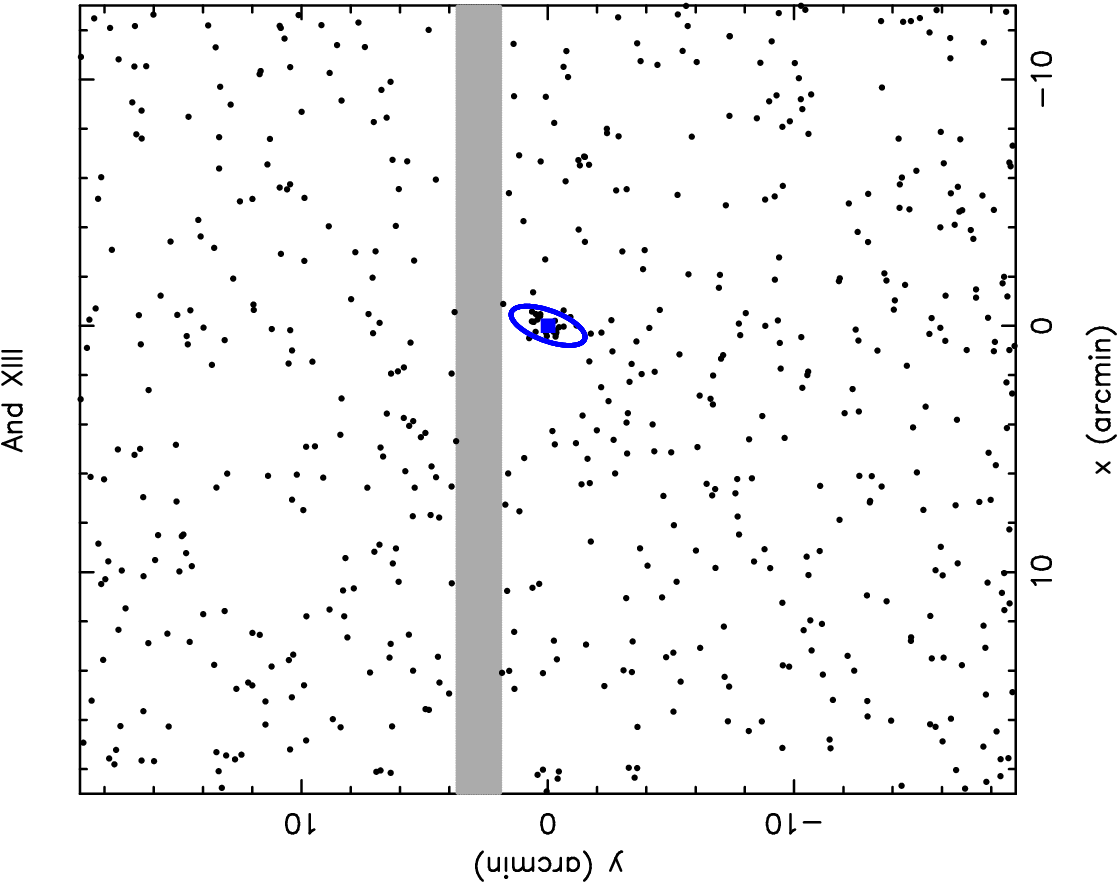}
\hspace{1cm}
\includegraphics[width=0.47\hsize,angle=270]{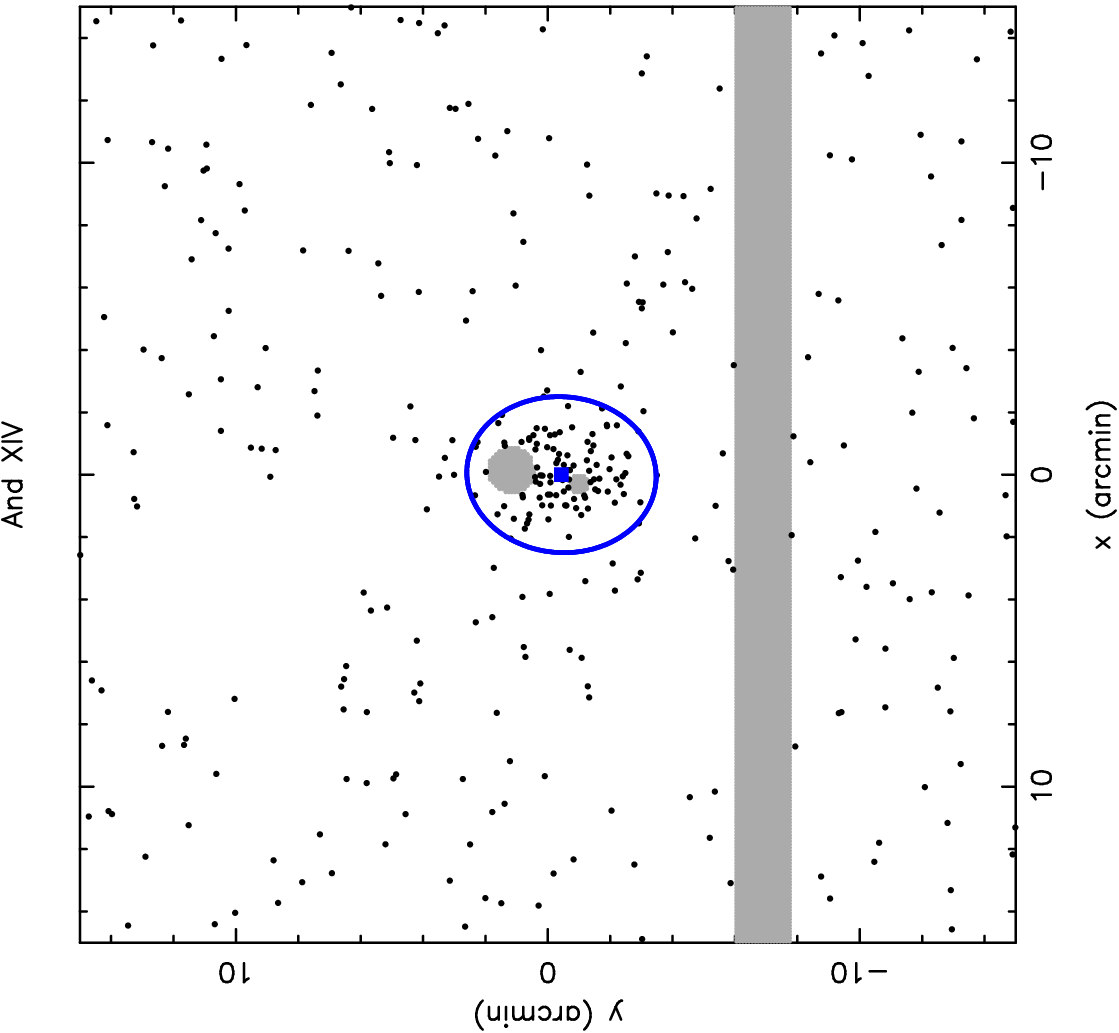}
\caption{\label{map_XIII_XIV}Same as Figure~\ref{map_I_II} but for And~XIII and And~XIV.}
\end{center}
\end{figure*}

\begin{figure*}
\begin{center}
\includegraphics[width=0.47\hsize,angle=270]{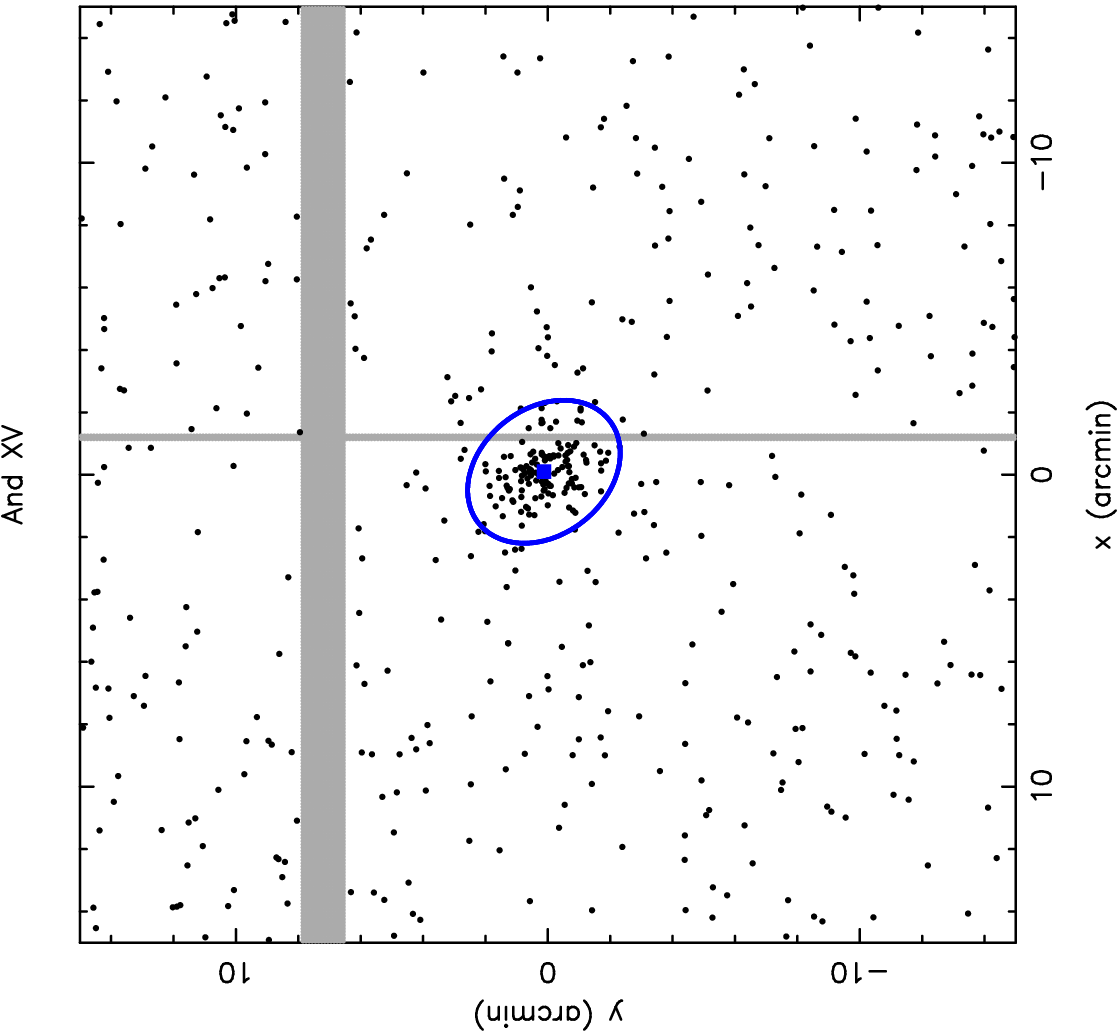}
\hspace{1cm}
\includegraphics[width=0.47\hsize,angle=270]{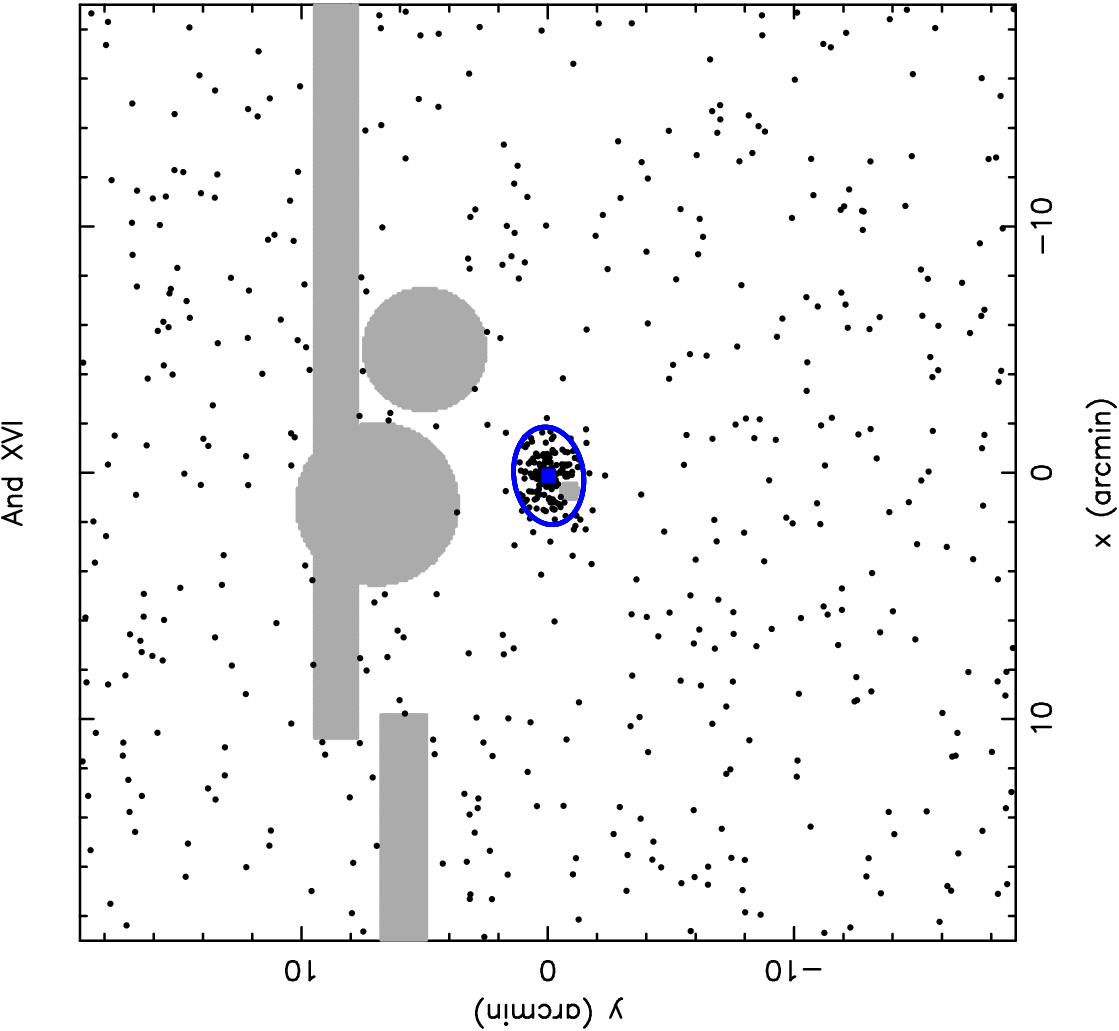}
\caption{\label{map_XV_XVI}Same as Figure~\ref{map_I_II} but for And~XV and And~XVI.}
\end{center}
\end{figure*}

\begin{figure*}
\begin{center}
\includegraphics[width=0.47\hsize,angle=270]{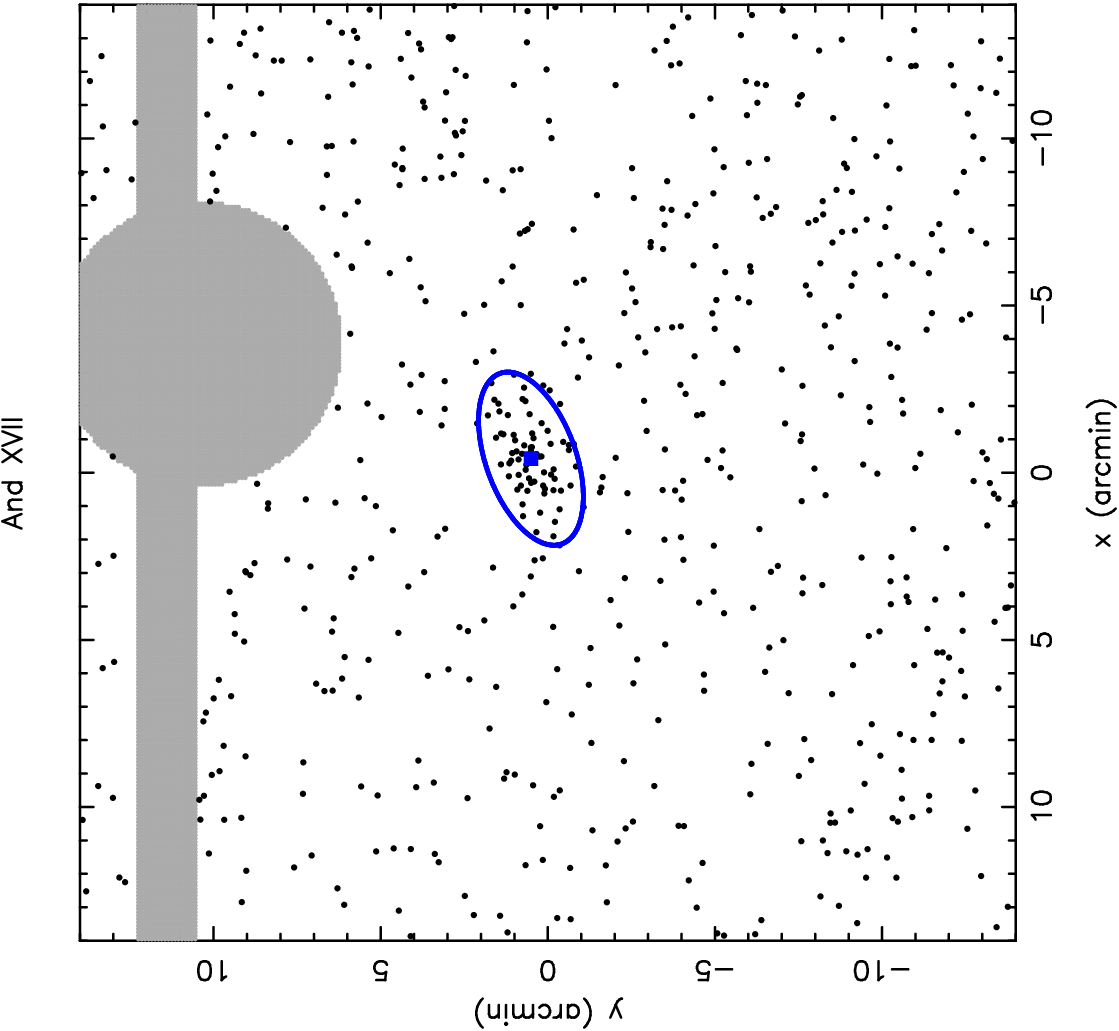}
\hspace{1cm}
\includegraphics[width=0.47\hsize,angle=270]{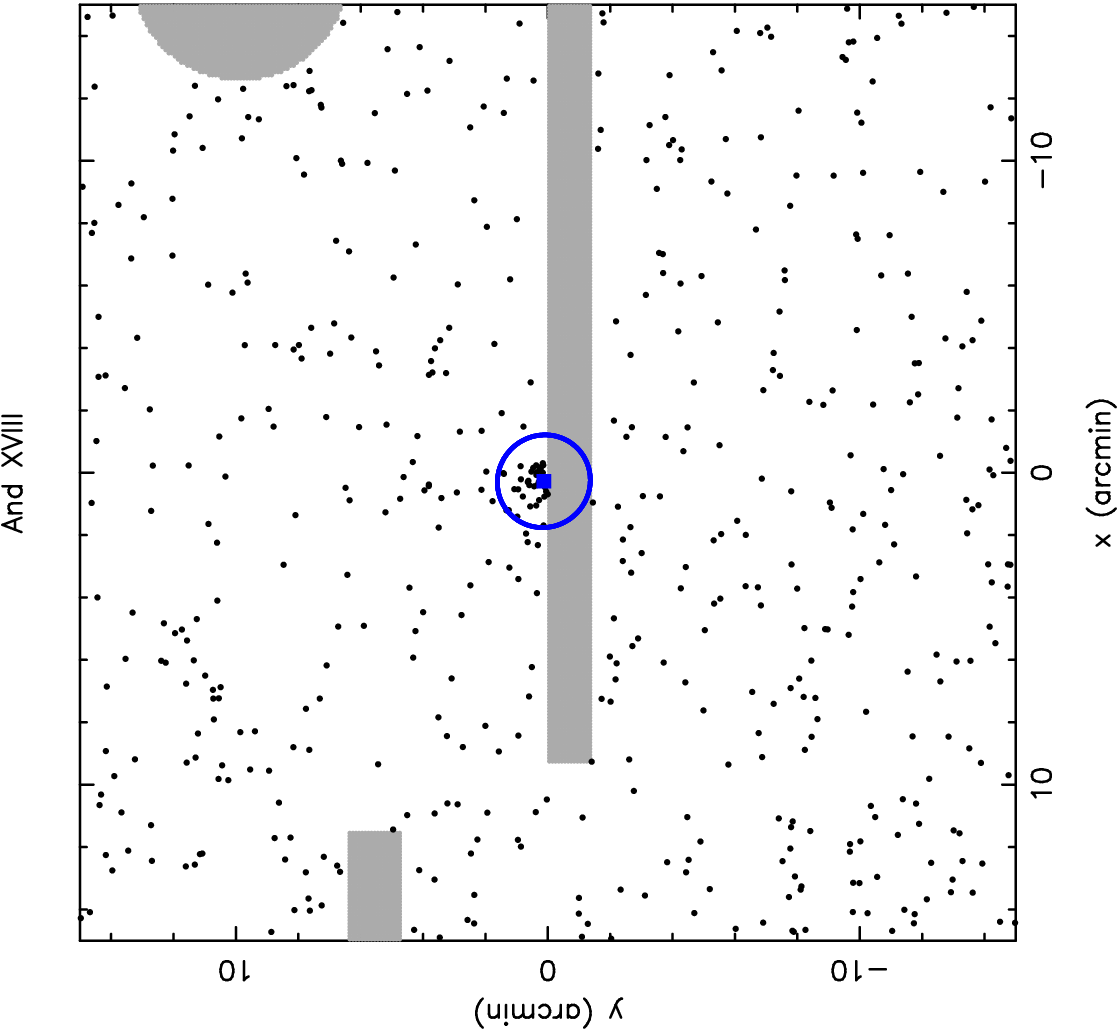}
\caption{\label{map_XVII_XVIII}Same as Figure~\ref{map_I_II} but for And~XVII and And~XVIII.}
\end{center}
\end{figure*}

\begin{figure*}
\begin{center}
\includegraphics[width=0.47\hsize,angle=270]{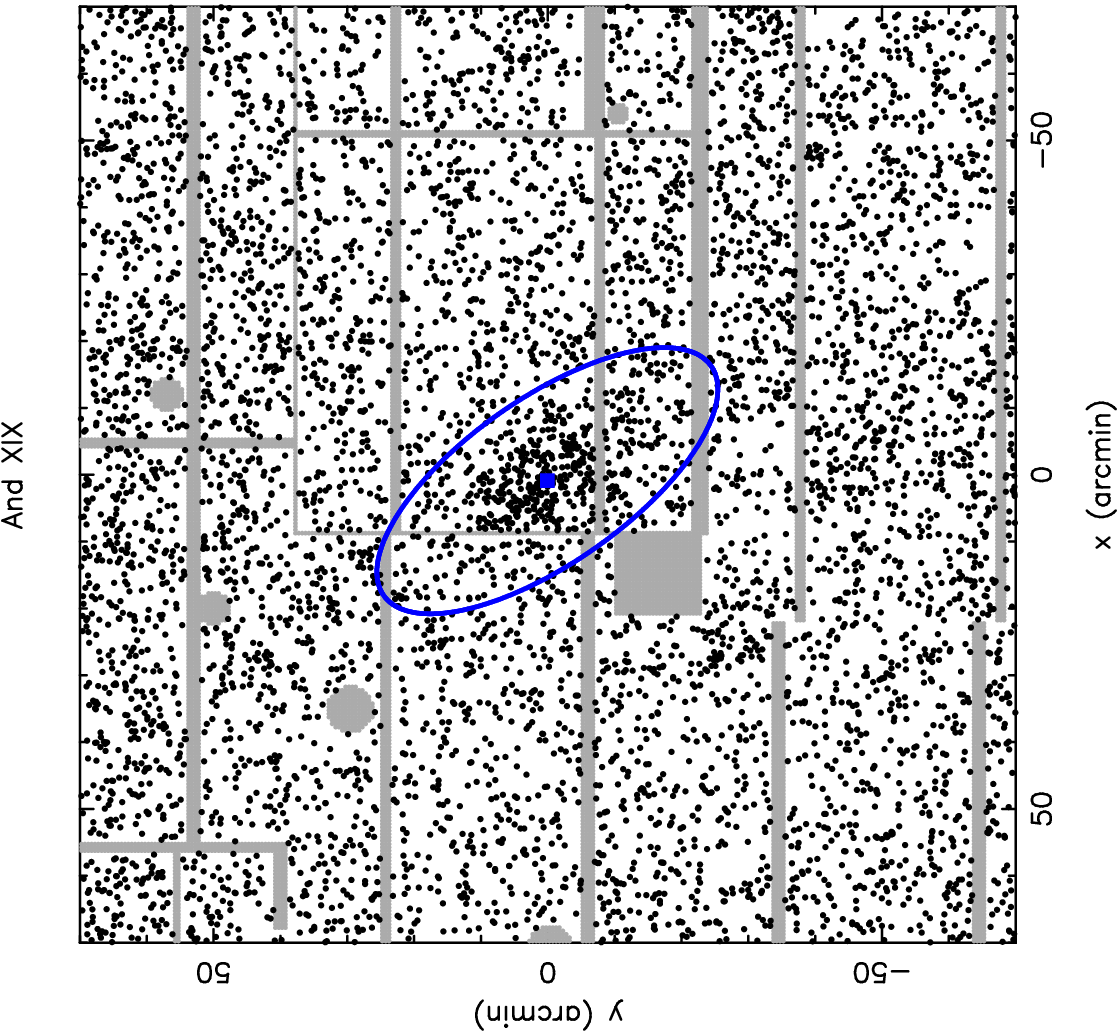}
\hspace{1cm}
\includegraphics[width=0.47\hsize,angle=270]{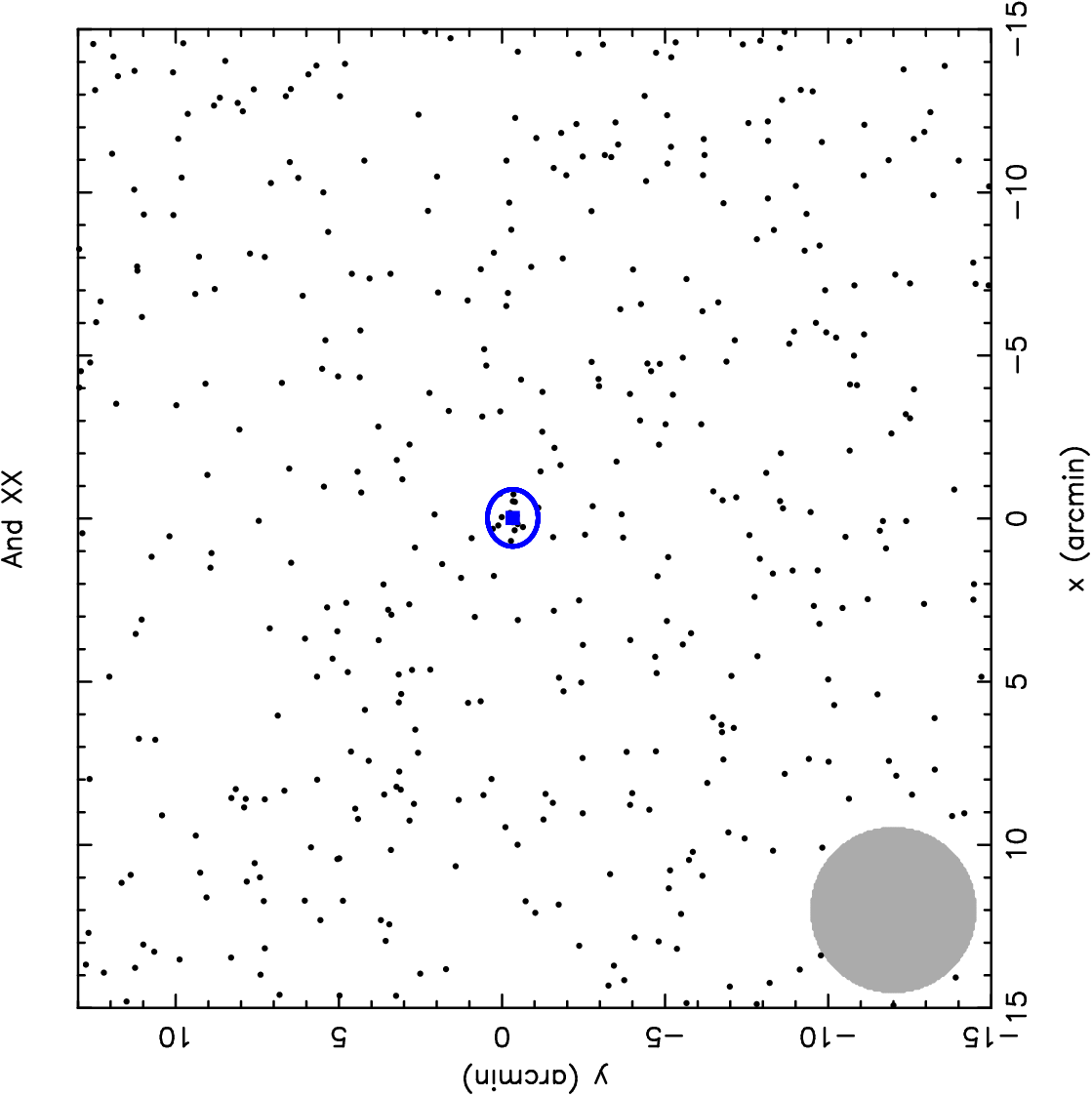}
\caption{\label{map_XIX_XX}Same as Figure~\ref{map_I_II} but for And~XIX and And~XX.}
\end{center}
\end{figure*}

\begin{figure*}
\begin{center}
\includegraphics[width=0.47\hsize,angle=270]{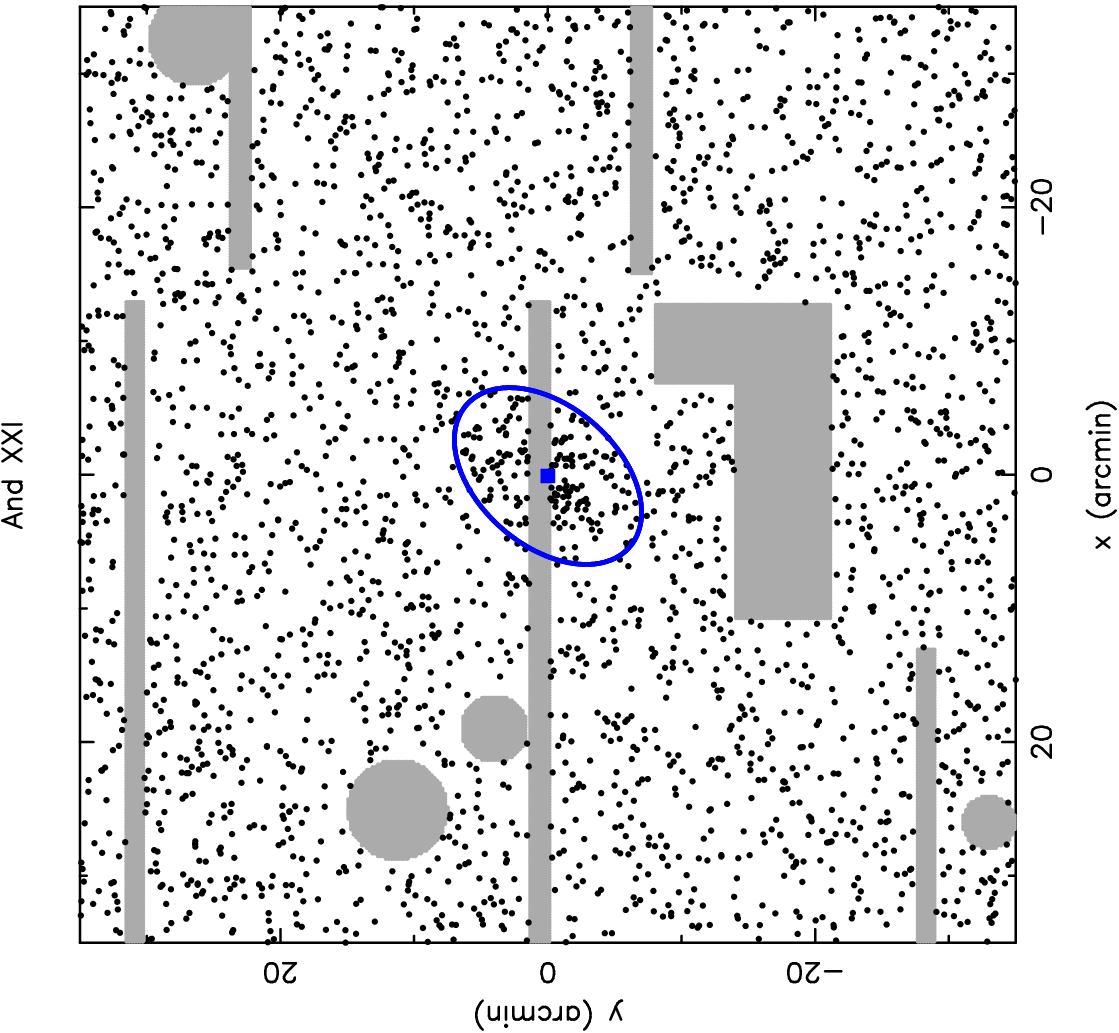}
\hspace{1cm}
\includegraphics[width=0.47\hsize,angle=270]{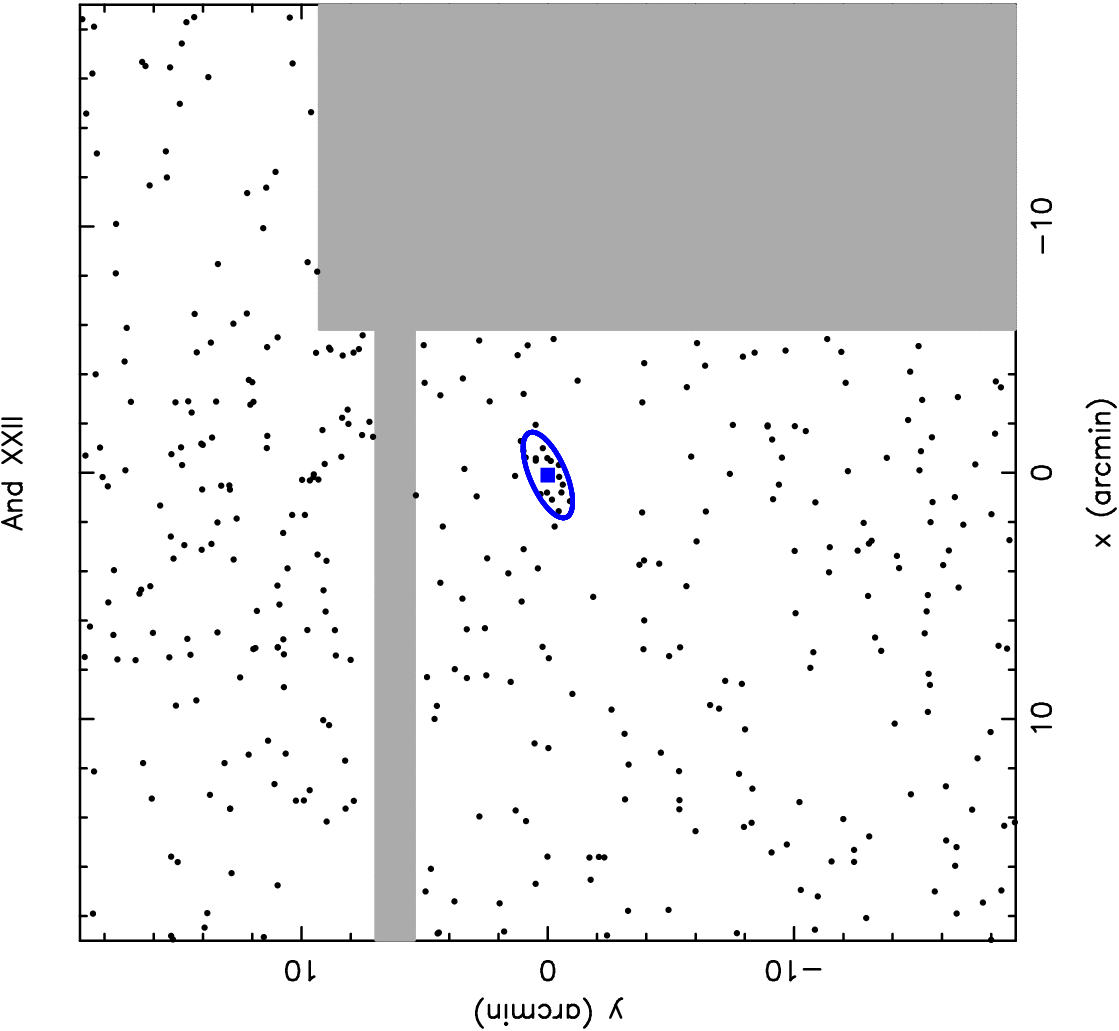}
\caption{\label{map_XXI_XXII}Same as Figure~\ref{map_I_II} but for And~XXI and And~XXII.}
\end{center}
\end{figure*}

\begin{figure*}
\begin{center}
\includegraphics[width=0.47\hsize,angle=270]{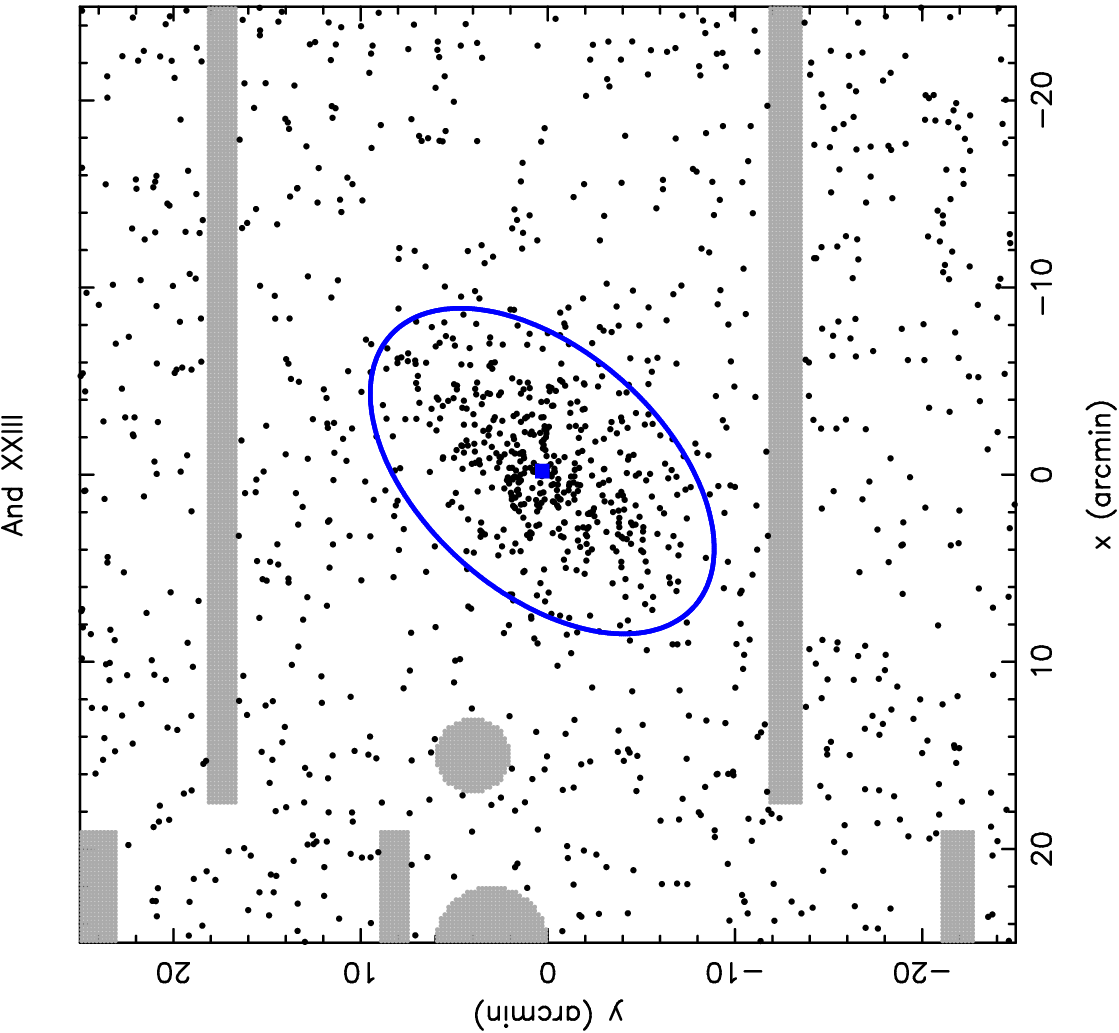}
\hspace{1cm}
\includegraphics[width=0.47\hsize,angle=270]{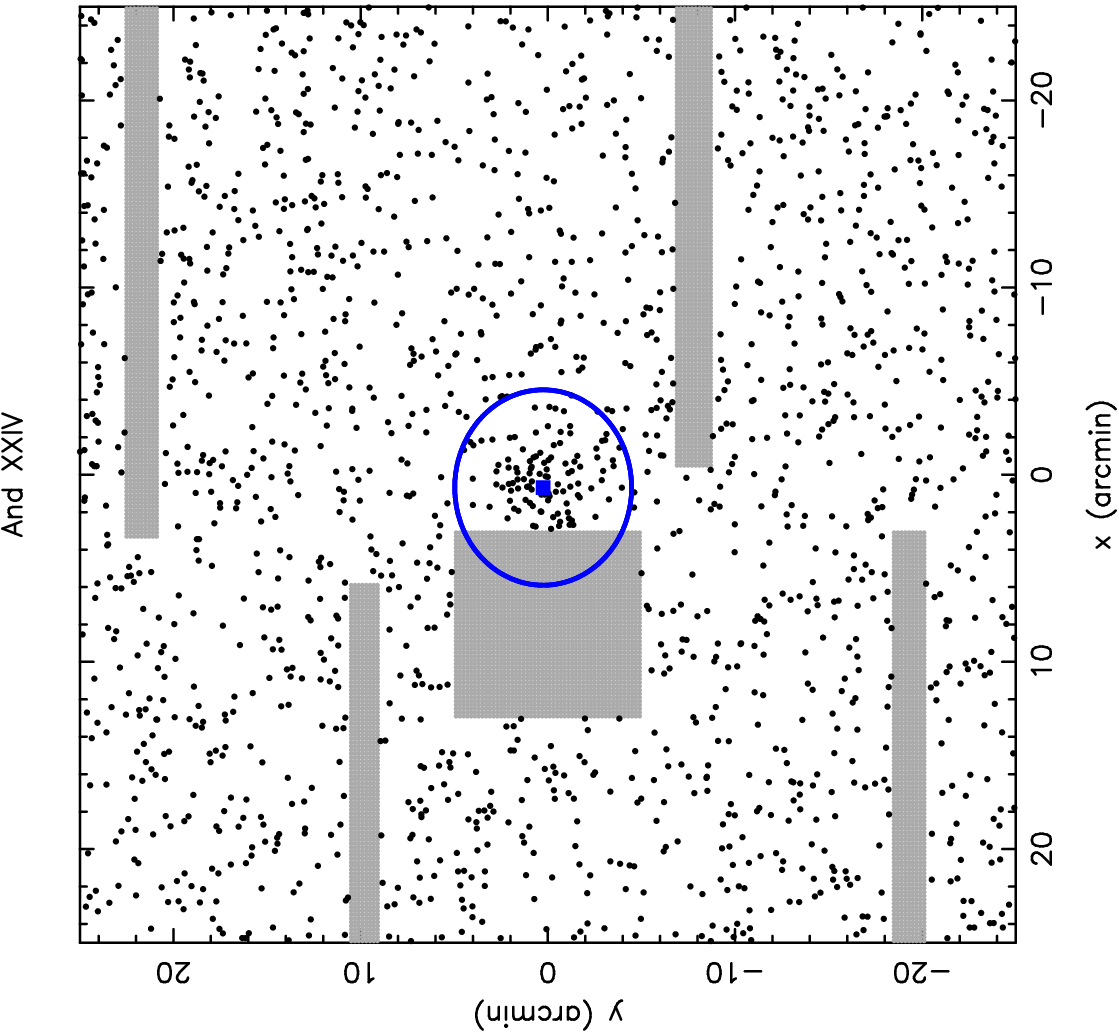}
\caption{\label{map_XXIII_XXIV}Same as Figure~\ref{map_I_II} but for And~XXIII and And~XXIV.}
\end{center}
\end{figure*}

\begin{figure*}
\begin{center}
\includegraphics[width=0.47\hsize,angle=270]{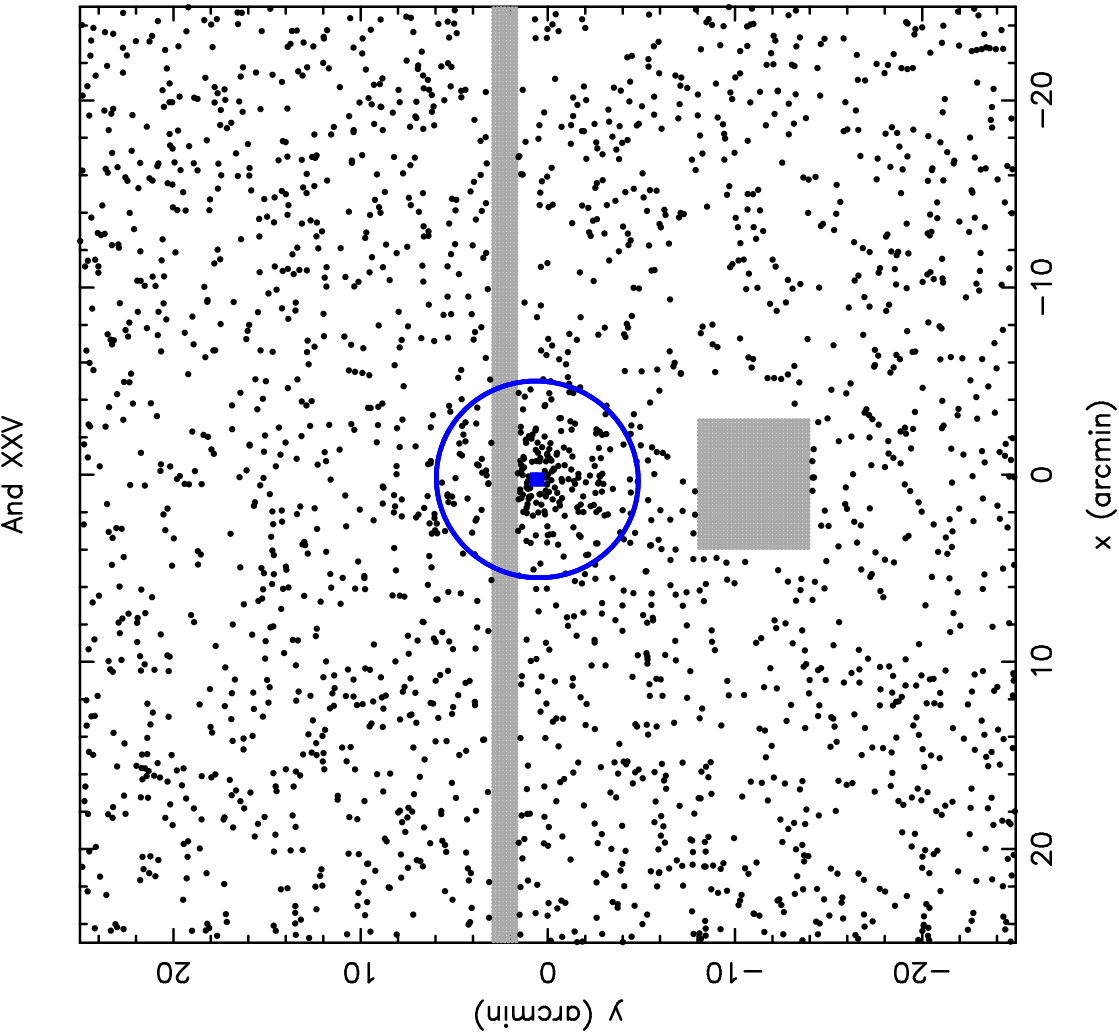}
\hspace{1cm}
\includegraphics[width=0.47\hsize,angle=270]{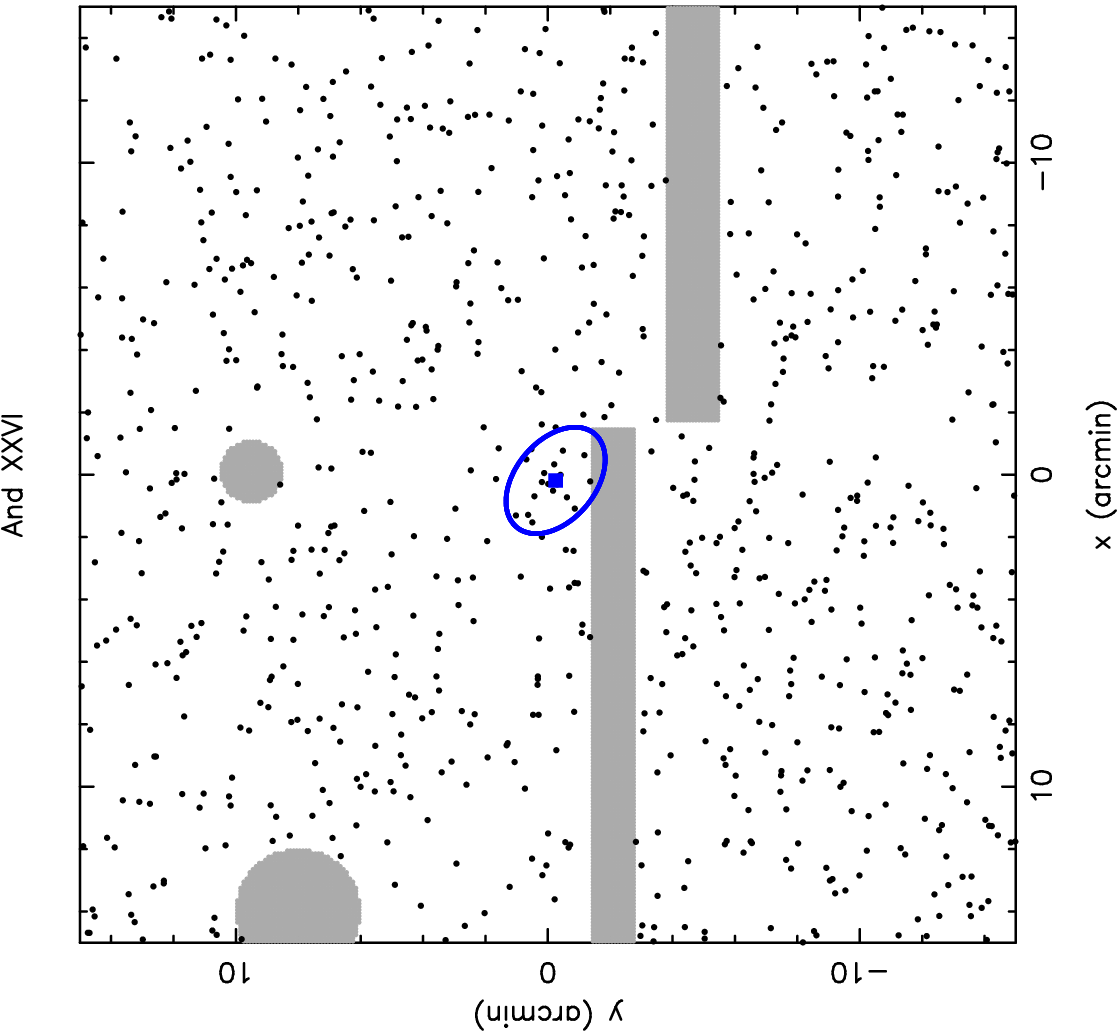}
\caption{\label{map_XXV_XXVI}Same as Figure~\ref{map_I_II} but for And~XXV and And~XXVI.}
\end{center}
\end{figure*}

\begin{figure}
\begin{center}
\includegraphics[width=0.47\hsize,angle=270]{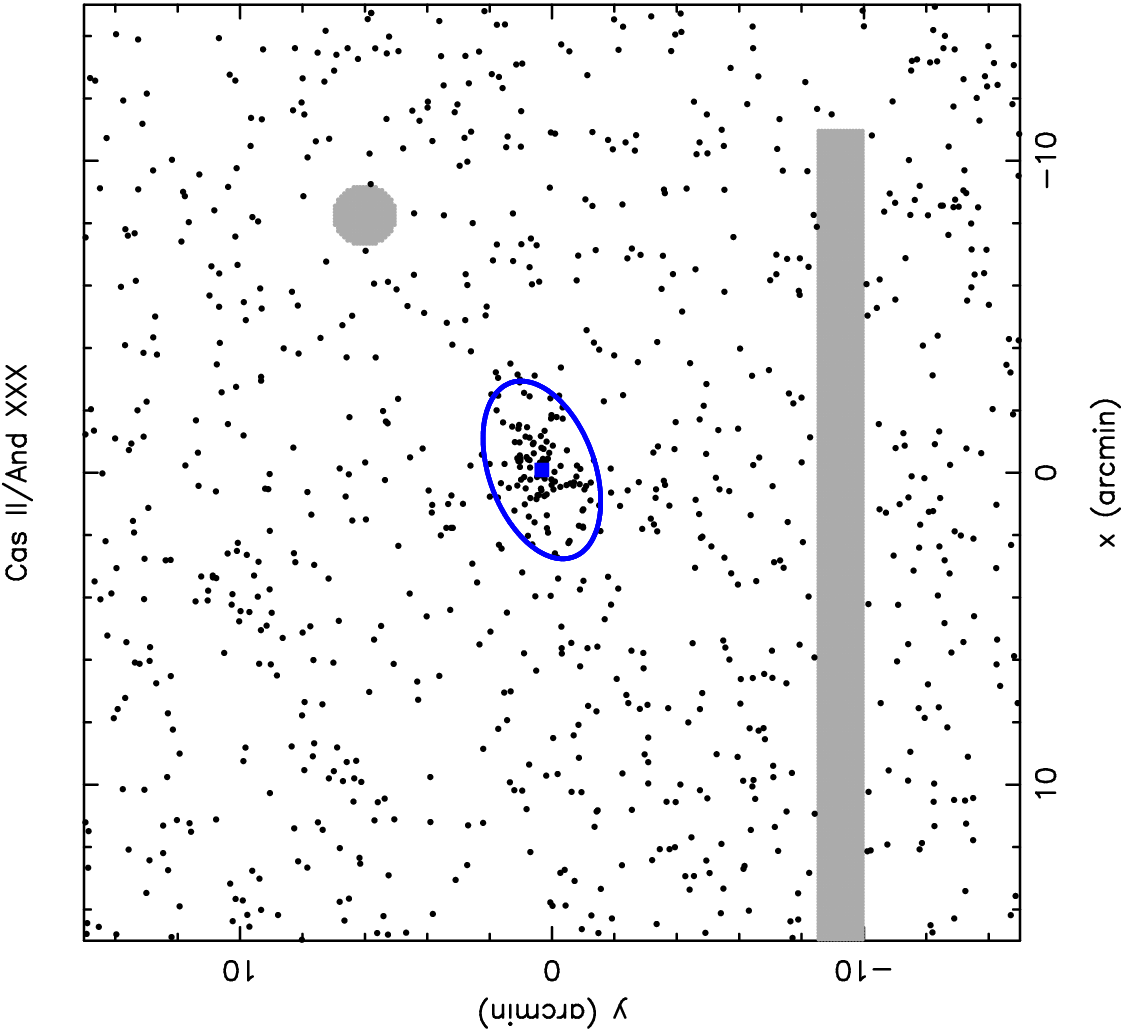}
\caption{\label{map_XXX}Same as Figure~\ref{map_I_II} but for Cas~II/And~XXX.}
\end{center}
\end{figure}

Figures~\ref{map_I_II} to \ref{map_XXX} show the stellar distribution $\mathcal{D}_n$ that was used for each dwarf galaxy. Each map corresponds to the region $\mathcal{A}$ of a given system, with masked out regions indicated in gray. The favored model, determined from the modes of the structural parameters' distributions is shown as the blue ellipses that represents the $2r_h$ region. The blue square locates the favored centroid $(x_0,y_0)$.

\label{maps}

\section{Chains}
\label{chains}

A sampling of the chains generated by the algorithm are available online in electronic tabular form for the 23 dwarf spheroidal galaxies discussed in the paper. Each table contains 500 randomly selected drawings from each chain. Table~\ref{table:online_table} provides an excerpt of the And~I chain for illustration. For each line, corresponding to a single drawing, columns (1) and (2) list the right ascension and declination of the centroid, column (3) gives the ellipticity $\epsilon$ and column (4) the position angle $\theta$. Columns (5) and (6) give the half-light radius $r_h$ in arcminutes and then in parsecs, as determined using a distance modulus value (listed further) randomly drawn from the relevant chain provided by \citet{aconn12}. Column (7) lists the apparent magnitude $m_{V,0}$ while column (8) lists absolute magnitude $M_{V,0}$, once again determined using the same distance modulus value as above. Column (9) gives the logarithm of the luminosity of the dwarf galaxy, while column (10) lists the central surface brightness in units of magnitudes per arcsec$^2$. Finally, column (11) gives the distance modulus value that was using to determine the relevant parameters. Note that the distance modulus values are drawn from the \citet{aconn12} chains and were not derived in this paper.

\begin{table}
\caption{\label{table:online_table}Excerpt from the And~I chain online table}
\begin{tabular}{ccccccccccc}
\hline
$\alpha$ (deg) & $\delta$ (deg) & $\epsilon$ & $\theta$ (deg) & $r_h$ ($'$) & $r_h$ (\pc) & $m_{V,0}$ & $M_{V,0}$ & $\log_{10}(L_V/\lsun)$ & $\mu_0$ (mag/arcsec$^2$) & distance modulus\\
(1) & (2) & (3) & (4) & (5) & (6) & (7) & (8) & (9) & (10) & (11)\\
\hline
 11.414190 & +38.037327 & 0.29 &    30.2 &  3.87 &    800 & 12.9 & $-11.3$ &          6.42 &              25.26 &    24.26\\
 11.415506 & +38.036674 & 0.31 &    28.6 &  3.97 &    848 & 13.1 & $-11.3$ &          6.40 &              25.41 &    24.33\\
 11.412939 & +38.037632 & 0.25 &    23.4 &  3.71 &    796 & 13.2 & $-11.2$ &          6.37 &              25.46 &    24.34\\
 11.415737 & +38.036972 & 0.25 &    33.7 &  3.75 &    801 & 13.4 & $-10.9$ &          6.27 &              25.72 &    24.33\\
 11.413170 & +38.035038 & 0.30 &    31.7 &  4.06 &    843 & 13.0 & $-11.3$ &          6.41 &              25.40 &    24.27\\
 11.413185 & +38.034157 & 0.31 &    26.2 &  4.02 &    847 & 13.2 & $-11.1$ &          6.32 &              25.61 &    24.30\\
 11.415273 & +38.037071 & 0.32 &    33.6 &  4.07 &    877 & 13.0 & $-11.3$ &          6.43 &              25.40 &    24.35\\
 11.416187 & +38.035320 & 0.25 &    28.4 &  3.78 &    858 & 13.1 & $-11.3$ &          6.42 &              25.47 &    24.46\\
 11.414933 & +38.037369 & 0.25 &    30.4 &  3.72 &    814 & 13.3 & $-11.1$ &          6.34 &              25.56 &    24.38\\
 11.414639 & +38.036373 & 0.24 &    26.1 &  3.87 &    865 & 13.3 & $-11.1$ &          6.34 &              25.73 &    24.43\\
\end{tabular}
\end{table}

\end{document}